%% file: main.tex
\documentclass[a4paper,11pt]{article}
\pdfoutput=1 

\usepackage{jcappub} 

\usepackage{xcolor}
\usepackage{hyperref}
\usepackage[noabbrev]{cleveref}
\usepackage{xspace}
\usepackage{verbatim}

\usepackage{fontawesome5}
\usepackage{orcidlink}
\usepackage{makecell}

\newcommand{\Planck}{{\it Planck}\xspace}
\newcommand{\mpgadget}{\texttt{MP-Gadget}\xspace}
\newcommand{\lacempg}{\texttt{lace-mpg}\xspace}
\newcommand{\lacelyssa}{\texttt{lace-lyssa}\xspace}
\newcommand{\lyssa}{\texttt{lyssa}\xspace}

\newcommand{\deltastar}{\ensuremath{\Delta^2_\star}\xspace}
\newcommand{\nstar}{\ensuremath{n_\star}\xspace}

\newcommand{\pone}{\ensuremath{P_{\rm 1D}}\xspace}
\newcommand{\lya}{Ly$\alpha$\xspace}
\newcommand{\lyaf}{Ly$\alpha$ forest\xspace}

\newcommand{\iMpc}{\ \text{Mpc}^{-1}}
\newcommand{\ikms}{\ensuremath{\mathrm{km}^{-1}\mathrm{s}}\xspace}
\newcommand{\kms}{\ensuremath{\mathrm{km}\,\mathrm{s}^{-1}}\xspace}
\newcommand{\kstarval}{\ensuremath{k_\star = 0.009\,\mathrm{km}^{-1}\mathrm{s}}\xspace}

\newcommand{\lyasiii}{\ensuremath{\rm Ly\alpha-Si\textsc{iii}}\xspace}
\newcommand{\lyasii}{\ensuremath{\rm Ly\alpha-Si\textsc{ii}}\xspace}
\newcommand{\siisii}{\ensuremath{\rm Si\textsc{ii}-Si\textsc{ii}}\xspace}
\newcommand{\siisiii}{\ensuremath{\rm Si\textsc{ii}-Si\textsc{iii}}\xspace}

\newcommand{\siia}{\ensuremath{\rm Si\textsc{ii}a}\xspace}
\newcommand{\siib}{\ensuremath{\rm Si\textsc{ii}b}\xspace}
\newcommand{\sii}{\ensuremath{\rm Si\textsc{ii}}\xspace}
\newcommand{\siii}{\ensuremath{\rm Si\textsc{iii}}\xspace}

\newcommand{\wsiia}{\ensuremath{\rm Si\textsc{ii}a(1190.42)}\xspace}
\newcommand{\wsiib}{\ensuremath{\rm Si\textsc{ii}b(1193.28)}\xspace}
\newcommand{\wsiii}{\ensuremath{\rm Si\textsc{iii}(1206.52)}\xspace}

\title{Cosmological analysis of the DESI DR1 Ly$\alpha$ 1D power spectrum}

\input{authors}

\abstract{
    We present the cosmological analysis of the one-dimensional Lyman-$\alpha$ flux power spectrum from the first data release of the Dark Energy Spectroscopic Instrument (DESI). We capture the dependence of the signal on cosmology and intergalactic medium physics using an emulator trained on a cosmological suite of hydrodynamical simulations, and we correct its predictions for the impact of astrophysical contaminants and systematics, many of these not considered in previous analyses. We employ this framework to constrain the amplitude and logarithmic slope of the linear matter power spectrum at $k_\star=0.009\,\mathrm{km^{-1}s}$ and redshift $z=3$, obtaining $\Delta^2_\star=0.379\pm0.032$ and $n_\star=-2.309\pm0.019$ \href{https://github.com/igmhub/cobaya_lya_p1d}{\faGithub}. The robustness of these constraints is validated through the analysis of mocks and a large number of alternative data analysis variations, with cosmological parameters kept blinded throughout the validation process. We then combine our results with constraints from DESI BAO and temperature, polarization, and lensing measurements from \textit{Planck}, ACT, and SPT-3G to set constraints on $\Lambda$CDM extensions. While our measurements do not significantly tighten the limits on the sum of neutrino masses from the combination of these probes, they sharpen the constraints on the effective number of relativistic species, $N_\mathrm{eff}=3.02\pm0.10$, the running of the spectral index, $\alpha_\mathrm{s}=0.0014\pm0.0041$, and the running of the running, $\beta_\mathrm{s}=-0.0006\pm0.0048$, by a factor of 1.18, 1.27, and 1.90, respectively. We conclude by outlining the improvements needed to fully reach the level of confidence implied by these uncertainties.
}

\begin{document}

\maketitle

\input{introduction_journal}
\input{data_journal}
\input{emulator_journal}
\input{contaminants_journal}
\input{likelihood_journal}
\input{results_journal}
\input{external_journal}
\input{discussion_journal}
\input{conclusions_journal}

\appendix
\crefname{section}{Appendix}{Appendices}

\renewcommand{\thefigure}{\thesection\arabic{figure}}
\renewcommand{\thetable}{\thesection\arabic{table}}

\setcounter{figure}{0} 
\setcounter{table}{0} 
\input{app_covariance_journal}

\setcounter{figure}{0} 
\setcounter{table}{0} 
\input{app_compressed_params_journal}

\setcounter{figure}{0} 
\setcounter{table}{0} 
\input{app_lyssa_emulator_journal}

\setcounter{figure}{0} 
\setcounter{table}{0} 
\input{app_nuisance_journal}

\setcounter{figure}{0} 
\setcounter{table}{0} 
\input{app_asns_journal}

\setcounter{figure}{0} 
\setcounter{table}{0}
\input{app_pivot_journal}

\input{acknow}
\input{affiliations}

\bibliographystyle{JHEP.bst}
\bibliography{main, desi}

\end{document}

%% file: authors.tex
\author[1]{J.~Chaves-Montero\orcidlink{0000-0002-9553-4261}}
\author[1,2]{A.~Font-Ribera\orcidlink{0000-0002-3033-7312}}
\author[3]{P.~McDonald\orcidlink{0000-0001-8346-8394}}
\author[4]{E.~Armengaud\orcidlink{0000-0001-7600-5148}}
\author[4]{D.~Chebat\orcidlink{0009-0006-7300-6616}}
\author[5,6]{C.~Garcia-Quintero\orcidlink{0000-0003-1481-4294}}
\author[7,8,9,10]{N.~G.~Kara\c{c}ayl{\i}\orcidlink{0000-0001-7336-8912}}
\author[11]{C.~Ravoux\orcidlink{0000-0002-3500-6635}}
\author[1]{S.~Satyavolu}
\author[12]{N.~Sch\"oneberg\orcidlink{0000-0002-7873-0404}}
\author[12,13]{M.~Walther\orcidlink{0000-0002-1748-3745}}
\author[3]{J.~Aguilar}
\author[14]{S.~Ahlen\orcidlink{0000-0001-6098-7247}}
\author[3]{S.~Bailey\orcidlink{0000-0003-4162-6619}}
\author[15,16]{D.~Bianchi\orcidlink{0000-0001-9712-0006}}
\author[17]{D.~Brooks}
\author[3]{T.~Claybaugh}
\author[3]{A.~Cuceu\orcidlink{0000-0002-2169-0595}}
\author[18]{A.~de la Macorra\orcidlink{0000-0002-1769-1640}}
\author[17]{P.~Doel}
\author[3,19]{S.~Ferraro\orcidlink{0000-0003-4992-7854}}
\author[20,21]{J.~E.~Forero-Romero\orcidlink{0000-0002-2890-3725}}
\author[22,23,24]{E.~Gaztañaga\orcidlink{0000-0001-9632-0815}}
\author[25]{S.~Gontcho A Gontcho\orcidlink{0000-0003-3142-233X}}
\author[26]{A.~X.~Gonzalez-Morales\orcidlink{0000-0003-4089-6924}}
\author[27]{G.~Gutierrez}
\author[3]{J.~Guy\orcidlink{0000-0001-9822-6793}}
\author[28]{C.~Hahn\orcidlink{0000-0003-1197-0902}}
\author[29,4]{H.~K.~Herrera-Alcantar\orcidlink{0000-0002-9136-9609}}
\author[8,9,7]{K.~Honscheid\orcidlink{0000-0002-6550-2023}}
\author[30]{M.~Ishak\orcidlink{0000-0002-6024-466X}}
\author[31]{R.~Joyce\orcidlink{0000-0003-0201-5241}}
\author[31]{S.~Juneau\orcidlink{0000-0002-0000-2394}}
\author[32]{D.~Kirkby\orcidlink{0000-0002-8828-5463}}
\author[3]{A.~Kremin\orcidlink{0000-0001-6356-7424}}
\author[17]{O.~Lahav}
\author[7]{C.~Lamman\orcidlink{0000-0002-6731-9329}}
\author[3]{M.~Landriau\orcidlink{0000-0003-1838-8528}}
\author[4]{J.M.~Le~Goff}
\author[33]{L.~Le~Guillou\orcidlink{0000-0001-7178-8868}}
\author[34,35]{A.~Leauthaud\orcidlink{0000-0002-3677-3617}}
\author[3]{M.~E.~Levi\orcidlink{0000-0003-1887-1018}}
\author[36,1]{M.~Manera\orcidlink{0000-0003-4962-8934}}
\author[8,10,7]{P.~Martini\orcidlink{0000-0002-4279-4182}}
\author[31]{A.~Meisner\orcidlink{0000-0002-1125-7384}}
\author[37,1]{R.~Miquel}
\author[38]{J.~Moustakas\orcidlink{0000-0002-2733-4559}}
\author[23]{S.~Nadathur\orcidlink{0000-0001-9070-3102}}
\author[26,39]{G.~Niz\orcidlink{0000-0002-1544-8946}}
\author[4,3]{N.~Palanque-Delabrouille\orcidlink{0000-0003-3188-784X}}
\author[40,41,42]{W.~J.~Percival\orcidlink{0000-0002-0644-5727}}
\author[43]{F.~Prada\orcidlink{0000-0001-7145-8674}}
\author[44]{I.~P\'erez-R\`afols\orcidlink{0000-0001-6979-0125}}
\author[45]{G.~Rossi}
\author[46]{E.~Sanchez\orcidlink{0000-0002-9646-8198}}
\author[3]{D.~Schlegel}
\author[47,48]{M.~Schubnell}
\author[49]{H.~Seo\orcidlink{0000-0002-6588-3508}}
\author[3]{J.~Silber\orcidlink{0000-0002-3461-0320}}
\author[31]{D.~Sprayberry}
\author[4]{T.~Tan\orcidlink{0000-0001-8289-1481}}
\author[48]{G.~Tarl\'{e}\orcidlink{0000-0003-1704-0781}}
\author[31]{B.~A.~Weaver}
\author[4]{C.~Y\`{e}che\orcidlink{0000-0001-5146-8533}}
\author[3]{R.~Zhou\orcidlink{0000-0001-5381-4372}}
\author[50]{H.~Zou\orcidlink{0000-0002-6684-3997}}

\emailAdd{jchaves@ifae.es}
\emailAdd{afont@ifae.es}

%% file: introduction_journal.tex
\section{Introduction} 
\label{sec:intro}

The Lyman-$\alpha$ (\lya) forest refers to absorption lines in the spectra of high redshift quasars resulting from \lya absorption by intergalactic neutral hydrogen along the quasar sightlines (see \cite{mcquinn2016EvolutionIntergalacticMedium} for a review). Therefore, the \lyaf is a powerful tool to study the clustering of matter from redshift $z=2$ to 6, beyond the reach of other probes such as galaxy clustering and weak gravitational lensing for the current generation of wide-field surveys. Traditionally, cosmological analyses of the \lyaf rely on either three-dimensional correlations of the \lya transmission field to measure baryonic acoustic oscillations (BAO \cite{busca2013LyaBAODR9, slosar2013MeasurementBaryonAcoustica, delubac2015LyaBAODR11, bautista2017MeasurementBaryonAcoustic, desainteagathe2019BaryonAcousticOscillations, dumasdesbourboux2020CompletedSDSSIVExtended}) or correlations along the line of sight of each individual quasar --- the one-dimensional \lyaf flux power spectrum (\pone \cite{McDonald2006, p1d_PalanqueDelabrouille2013, p1d_Chabanier2019}) --- to set constraints on the amplitude and shape of the matter power spectrum from quasi- to non-linear scales.

The precision of cosmological constraints from \pone analyses has improved significantly in recent years, driven by the steady increase in the number of \lyaf measurements. For two decades, the Sloan Digital Sky Survey (SDSS \cite{york2000_sdss}) and its successors provided the bulk of these data: SDSS measured $\simeq12\,000$ \lyaf over eight years, the Baryon Oscillation Spectroscopic Survey (BOSS \cite{Dawson2013}) expanded this number to $\simeq160\,000$ over five years, and the extended Baryon Oscillation Spectroscopic Survey (eBOSS \cite{Dawson2016}) further increased it to $\simeq210\,000$ over six years. This number rose dramatically with the advent of the Dark Energy Spectroscopic Instrument survey (DESI \cite{DESI2016a.Science}). Remarkably, DESI measured $\simeq55\,000$ \lya forests \cite{Karacayli2024_edr, Ravoux2023} within the Early Data Release (EDR \cite{DESI2023b.KP1.EDR}), which contains observations from the six-month Survey Validation period (SV \cite{DESI2023a.KP1.SV}) and the first two months of the main survey. In this work, we analyze \pone measurements \cite{Karacayli2025_p1d_dr1, Ravoux2025} from the first DESI data release (DR1 \cite{DESI2024.I.DR1}), which includes 450\,000 \lya forests obtained during the first year of main survey observations combined with the SV data.

Within the $\Lambda$CDM model and several of its extensions, there is no significant loss of constraining power when compressing the \pone cosmological information into the amplitude and logarithmic slope of the linear power spectrum at the pivot scale \kstarval and redshift $z_\star=3$ \cite{McDonald2006, Pedersen2023},
\begin{align}
  \Delta^2_\star& = \frac{k_\star^3 P_\mathrm{lin}(k_\star, z_\star)}{2\pi^2},\\
  n_\star& = \frac{\mathrm{d}\log P_\mathrm{lin}(k, z)}{\mathrm{d}\log k}\bigg|_{k_\star, \, z_\star},
  \label{eq:nstar}
\end{align}
where $P_\mathrm{lin}$ is the linear matter power spectrum. The pivot scale and redshift adopted here are optimal for the analysis of SDSS DR2 \pone measurements \cite{McDonald2006}; in \cref{app:pivot}, we derive the corresponding optimal values for DESI DR1. We define the pivot wavelength in velocity units because \pone is naturally measured in velocity space, and converting velocity to comoving units would require assuming a specific expansion history, which would make the value of the compressed parameters dependent on cosmology \cite{Pedersen2021, Pedersen2023}. The motivation for using this parameterization is that \lya measurements probe redshifts during which the universe is practically Einstein de-Sitter, and thus the majority of the cosmological information resides in the power spectrum\footnote{At fixed physical matter density, the expansion rate and logarithmic growth rate at $z=2.2$ differ by less than 1\% between cosmologies with dimensionless Hubble constant $h=0.67$ and $h=0.74$.} \cite{Croft1998, McDonald2000, emuparam_Gnedin, Croft2002, Viel2004}.

The main goal of this work is to analyze \pone measurements from DESI DR1 to measure the value of \deltastar and \nstar; compressed parameters hereafter. To this end, we train an emulator on \pone predictions from a suite of cosmological hydrodynamical simulations \cite{Pedersen2021}, which captures the dependence of this statistics on both cosmology and the physics of the intergalactic medium (IGM). We then present a new model to correct these predictions for metal contamination, high column density contamination, and systematic effects, with the full analysis pipeline being an updated version of the publicly available likelihood code \texttt{cup1d}\footnote{\url{https://github.com/igmhub/cup1d}} \cite{Pedersen2021, Pedersen2023}. We validate the robustness of our cosmological constraints through the analysis of \pone mocks based on hydrodynamical simulations run with different codes and resolutions, as well as alternative data analyses varying the type of \pone measurement, covariance matrix, emulator, and modeling assumptions. We optimized the model and validated the analysis under blinded conditions to avoid experimenter bias.

The Standard Model of particle physics and the $\Lambda$CDM cosmological model successfully describe a wide range of laboratory and observational measurements with high precision; however, there are fundamental open questions such as the origin of neutrino masses, the nature of the dark sector, and the mechanism seeding large-scale structure. Our \pone measurements probe the linear matter power spectrum on scales far smaller than those accessible to cosmic microwave background and galaxy clustering analyses; consequently, the combination of all these probes provides a powerful means of detecting any physical effect that modify the power spectrum on small scales. In particular, we combine our measurements with temperature, polarization, and lensing data from \Planck \cite{planck2018_like, Planck2018, Carron2022_planck_lensing}, ACT DR6 \cite{Qu2024_act_lensing, Madhavacheril2024_act_lensing, Louis2025_actdr6like, Calabrese2025_actdr6cosmo}, and SPT-3G D1 \cite{Camphuis2025_spt3gd1, Ge2025_sptlensing, Qu2025_cpalensing} (CMB-SPA \cite{Camphuis2025_spt3gd1}) and DESI-DR2 BAO measurements \cite{DESI.DR2.BAO.lya, DESI.DR2.BAO.cosmo} to measure the sum of neutrino masses, the effective number of relativistic species, the running of the spectral index, and the running of the running. We also combine our measurements with DESI full-shape \cite{DESI2024.VII.KP7B} and Big Bang Nucleosynthesis \cite{schoneberg2024_bbn} constraints to set upper limits on the sum of neutrino masses independent of CMB measurements. On the other hand, we find that our measurements are not precise enough to improve upon existing constraints on the dark energy equation of state.

Measuring the aforementioned quantities has important physical implications. Neutrino oscillation experiments are sensitive only to mass-squared differences; therefore, precise constraints on the sum of neutrino masses would enable the measurement of individual masses up to the uncertainty associated with the mass ordering \cite{Navas2024_rev_particle_phy}. A detection of an effective number of relativistic species exceeding the Standard Model prediction would indicate the presence of new light particles \cite{Antel2023_reportneff}, while evidence for a nonzero running of the scalar spectral index, or of the running of the running, would place stringent constraints on the shape of the inflationary potential \cite{Lorenzoni2024_inflation, Martin2024_inflation}.

The structure of this paper is as follows. In \cref{sec:data}, we describe DESI DR1 \pone measurements and the hydrodynamical simulations used to train our emulator and validate the analysis. The construction of the emulator is detailed in \cref{sec:emulator}, and the modeling of astrophysical contaminants and instrumental systematics is presented in \cref{sec:contaminants}. In \cref{sec:model}, we put together a model for analyzing \pone measurements and validate it using \pone mocks. In \cref{sec:results}, we present our constraints on cosmology, together with an extensive suite of robustness tests. We combine our measurements with external data to set constraints on $\Lambda$CDM extensions in \cref{sec:extensions}, discuss several avenues to further strengthen the robustness of future \pone analyses in \cref{sec:discussion}, and summarize our main findings and conclude in \cref{sec:conclusions}.

%% file: data_journal.tex
\section{Measurements and simulations}
\label{sec:data}

We first describe the DESI DR1 \pone measurements in \cref{sec:data_obs}, and we then introduce the simulations used to train the \pone emulator and to validate the analysis pipeline in \cref{sec:data_sim}.


\subsection{DESI DR1 measurements}
\label{sec:data_obs}

DESI is a robotic, multi-fiber spectrograph mounted on the Mayall 4-meter telescope at Kitt Peak National Observatory \cite{DESI2016b.Instr, DESI2022.KP1.Instr}. It can obtain spectra for nearly 5000 sources per exposure thanks to its high-precision focal plane \cite{FocalPlane.Silber.2023}, corrector \cite{Corrector.Miller.2023}, and fiber system \cite{FiberSystem.Poppett.2024} together with a sophisticated survey operation planning \cite{SurveyOps.Schlafly.2023}. DESI spectra cover wavelengths from the near UV to the near infrared (3600 to 9800 $\mathrm{\AA}$) with spectral resolution ranging from 2000 to 5000 \cite{Spectro.Pipeline.Guy.2023}, which enables measurements of the \lyaf in quasars at $z>2.1$.

DESI uses three automated classification algorithms to identify and compute the redshift of quasars \cite{QSO.TS.Chaussidon.2023}, which are pre-selected through a dedicated targeting strategy \cite{LS.Overview.Dey.2019, QSOPrelim.Yeche.2020}. The primary tool is the template-fitting code \texttt{redrock} \cite{RedrockQSO.Brodzeller.2023}, which classifies sources as stars, galaxies, or quasars and estimates their redshifts. A visual inspection campaign during the SV phase revealed that \texttt{redrock} misses some quasars \cite{VIQSO.Alexander.2023} and, to mitigate this problem and improve completeness, DESI decided to employ two additional classification algorithms \cite{Farr20_QN, Napolitano2023_mgii}. Finally, DESI applies selection cuts to increase the purity of the DR1 sample \cite{Karacayli2025_p1d_dr1}, which contains more than 450\,000 quasars at $z>2.1$.

Some absorption features in the \lyaf are not produced by intergalactic gas, and it is useful to correct for these before computing \pone. Broad absorption line features, caused by quasar outflows, are first identified \cite{KP6s9-Martini, Filbert2024_bal} and then masked \cite{Karacayli2025_p1d_dr1}. Likewise, damped \lya systems with high column densities (HCD; see \cref{sec:cont_hcd}) are identified by three independent algorithms \cite{Ho2021_dlas, Wang2022, Brodzeller2025_desiDLA} and subsequently modeled and masked in the forests \cite{Karacayli2025_p1d_dr1}. Absorption lines from the Milky Way and strong atmospheric emission lines are also masked \cite{Ravoux2023}.

Using this dataset, DESI measured \pone from $z=2.2$ to 4.2 in steps of $\Delta z=0.2$ using two methods\footnote{The QMLE and FFT measurements analyzed in this work are available at \url{https://zenodo.org/records/16943723} and \url{https://zenodo.org/records/17100543}, respectively.}: a quadratic maximum likelihood estimator (QMLE; \cite{Karacayli2025_p1d_dr1}) and Fast Fourier Transforms (FFT; \cite{Ravoux2025}). In principle, the two approaches should yield consistent results; however, differences in the selection of the quasar sample, the treatment of systematics, and the estimation of noise lead to discrepancies. To minimize the impact of systematic uncertainties, we focus on the analysis QMLE measurements since these are not affected by masking systematics like FFT-based estimates \cite{Ravoux2025, Lokken2025_masking}. Our baseline analysis further restricts the sample to Ly$\alpha$ forests with an average signal-to-noise ratio per pixel exceeding 3 in the \lyaf region rather than the $\mathrm{SNR}>1$ criterion adopted in the fiducial QMLE measurements presented in \cite{Karacayli2025_p1d_dr1}. This choice reduces the quasar sample to 62\,807 sources, increasing statistical uncertainties by approximately 25\% but reducing systematic uncertainties by a factor of 2.7 owing to the substantial mitigation of noise-related systematics \cite{Karacayli2025_p1d_dr1}. The two QMLE datasets nonetheless remain fully consistent within their uncertainties.

\begin{figure}
    \centering
    \includegraphics[width=\linewidth]{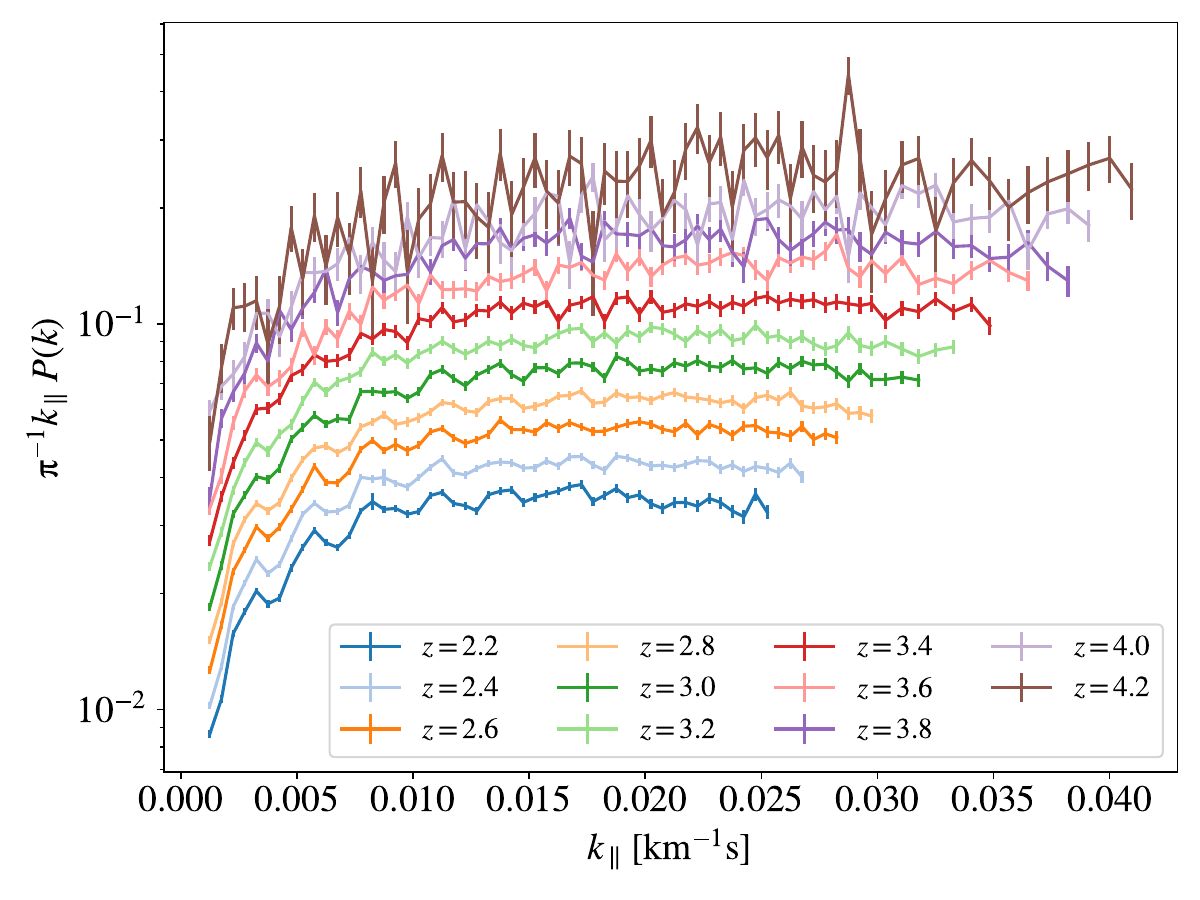}
    \caption{DESI DR1 measurements from the QMLE estimator using forests with an average signal-to-noise ratio greater than 3 per pixel in the \lyaf region. As indicated in the legend, each line displays the results at a different redshift. Error bars display the square root of the diagonal elements of the full covariance matrix, which contain terms accounting for statistical, systematic, and emulator errors.
    }
    \label{fig:p1d_data}
\end{figure}

In \cref{fig:p1d_data}, each line displays QMLE measurements from $\mathrm{SNR}>3$ forests at a different redshift. We use measurements within the wavenumber range $10^{-3}<k_\parallel[\ikms]<0.5 \pi R_z$ \cite{karacayli2025_p1d_validation}, where $R_z = c\,(1+z)^{-1} \Delta\lambda_\mathrm{DESI}/\lambda_\mathrm{Ly\alpha}$ is the pixel width in velocity units, and $\Delta\lambda_\mathrm{DESI} = 0.8\,\mathrm{\AA}$. The lower and upper limits are set by the strength of continuum fitting errors and the spectrograph resolution, respectively. As expected, the amplitude of the signal increases significantly with redshift, reflecting the increasing opacity of the intergalactic medium. This trend is less apparent in the three highest-redshift bins due to statistical errors resulting from the limited number of $\mathrm{SNR}>3$ forests at those redshifts. 

We can also see that the measurements exhibit periodic oscillations as a function of wavenumber. These features result from the absorption of quasar light by other elements besides hydrogen that are collectively referred to as ``metals.'' The \pone measurements shown in \cref{fig:p1d_data} are already corrected for contamination from metal lines with rest-frame wavelengths much longer than $\lambda_\mathrm{Ly\alpha}$ by subtracting the power spectrum measured in a sideband spanning from 1268 to $1380\,\mathrm{\AA}$. Contamination from lines with wavelengths below $\lambda \approx 1300\,\mathrm{\AA}$ cannot be handled in this way since they overlap with \lya and cannot be observed independently. The oscillations seen in the data are precisely due to these metal lines (see \cref{sec:cont_metals}).

There are three main sources of uncertainty affecting \pone studies: statistical, systematic, and emulator errors. The first two affect \pone measurements, while the last refers to inaccuracies in model predictions due to the number, volume, and resolution of the hydrodynamical simulations employed in the training of a \pone emulator (see \cref{sec:emulator_performance}). An important difference between the QMLE and FFT measurements is that only the first estimate statistical correlations of measurements from different redshift bins \cite{Karacayli2025_p1d_dr1, Ravoux2025}. We can subdivide the systematic errors affecting QMLE measurements into four categories \cite{Karacayli2024_edr, Karacayli2025_p1d_dr1}: incomplete masking of HCD and BAL features, incorrect characterization of either continuum fitting or the spectrograph resolution, and noise calibration errors. Of these, we omit the contributions from residual HCD contamination and uncertainties in the spectrograph resolution to the systematic covariance matrix, as both effects are explicitly marginalized over in our analysis (see \cref{sec:model_single}).

Our treatment of systematic and emulator errors differs from that adopted in some of the previous \pone studies of SDSS measurements \cite{PalanqueDelabrouille2013, p1d_Chabanier2019, palanque-delabrouille2020HintsNeutrinoBounds, Walther2025_lyssa}. First, we model each of the previous systematic errors as fully correlated across the DESI wavenumbers by incorporating these into the covariance matrix by taking their outer product, while previous studies added systematic errors to the diagonal of the covariance matrix. The motivation for this revised treatment is that each systematic error tends to coherently increase or decrease \pone, rather than behaving as random fluctuations \citep{Ravoux2025, Karacayli2025_p1d_dr1}. However, this is an approximation and future studies would need to compute these correlations directly. On the other hand, we assume that different systematic errors are uncorrelated, following the same approach as previous studies. Second, even though emulator errors can dominate the error budget, the aforementioned analyses did not account for these during inference. In our study, emulator and statistical errors are of the same order up to $z\simeq2.6$.

In \cref{sec:model_single}, we show that our model reproduces the \pone measurements accurately at all redshifts except $z = 3.0$, 3.6, and 4.0, for which the fit probability falls below 1\%. The residuals at these redshifts, however, show no significant wavenumber-dependent structure, suggesting either that our model fails to capture a redshift-specific physical or systematic effect, or that the measurement uncertainties are underestimated. Recent CCD image simulations indicate that DESI DR1 \pone measurements may be affected by percent-level large-scale biases not included in the systematic covariance matrix \cite{karacayli2025_p1d_validation}. These biases appear most significant at $z = 3.0$, although similar studies are not yet available for $z = 3.6$ and 4.0. To account for possible systematic effects and improve the fit probability, we increase the statistical uncertainty of all wavenumbers and redshifts by 5\%.

\begin{figure}
    \centering
    \includegraphics[width=\linewidth]{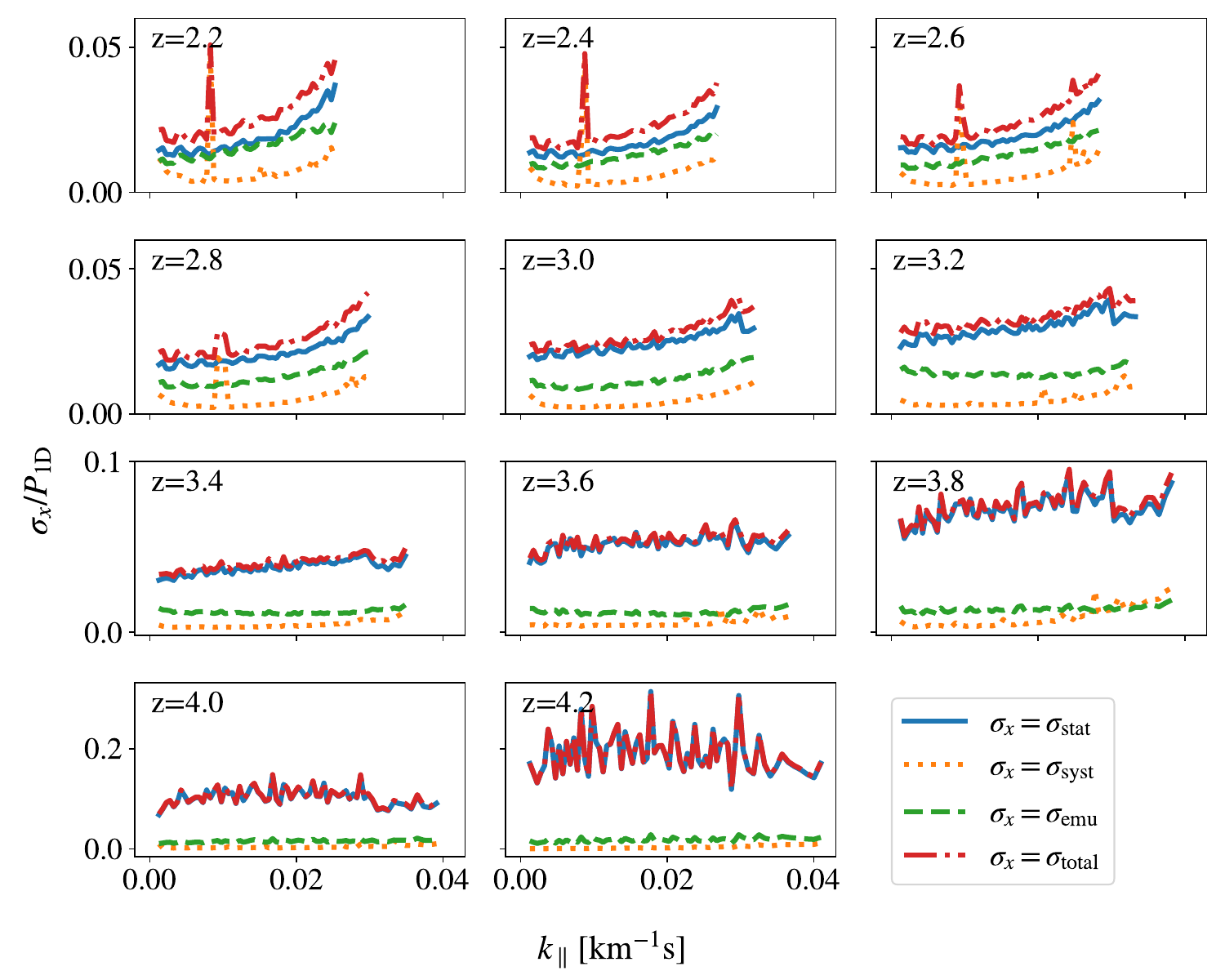}
    \caption{Noise to signal ratio of \pone measurements. The red dot-dashed lines depict the ratio for the total error, while the blue solid, orange dotted, and green dashed lines do so for the statistical, systematic, and emulator error components. Each panel shows the results for the redshift indicated at the bottom right.
    }
    \label{fig:p1d_errors}
\end{figure}

We display the contribution of the different sources of uncertainty to total \pone errors in \cref{fig:p1d_errors}. The red lines show the ratio between the square root of the diagonal elements of the total \pone covariance matrix and the measured \pone values, while the blue, orange, and green lines represent the statistical, systematic, and emulator contributions, respectively. As we can see, systematic uncertainties dominate the error budget only at a few wavenumbers at low redshift, which is caused by noise systematics \cite{Karacayli2025_p1d_dr1}. However, it is important to note that including the contributions from residual HCD contamination and uncertainties in the characterization of the spectrograph resolution to the covariance matrix would cause systematic errors to dominate the total uncertainty on both large and small scales at most redshifts \cite{Ravoux2025, Karacayli2025_p1d_dr1}. Emulator and statistical uncertainties are comparable up to $z = 2.8$, beyond which statistical errors dominate the error budget. The decrease in statistical errors at large wavenumbers for $z \ge 3$ arises from a change in the $k$-binning scheme.

To test the robustness of our cosmological constraints, in \cref{sec:results} we also analyze alternative \pone measurements from the QMLE estimator including low-SNR quasars (fiducial in \cite{Karacayli2025_p1d_dr1}) and from the FFT estimator applied to forests with average $\mathrm{SNR}>3$ per pixel in the \lyaf region. These samples differ in both signal-to-noise ratio and sensitivity to systematics compared to the fiducial measurements, but we expect the resulting cosmological constraints to be mutually consistent.


\subsection{Simulated data}
\label{sec:data_sim}

In this section, we describe the hydrodynamical simulations from which we draw \pone measurements to train and validate our emulator. For computational efficiency, these simulations do not include prescriptions for star formation or for stellar and AGN feedback \cite{viel2004ConstraintsPrimordialPower} since these effects have a small impact on the low column density absorption systems probed by the \lyaf \cite{bolton2017SherwoodSimulationSuite, chabanier2020ImpactAGNFeedback}. We summarize the properties of these simulation in \cref{tab:simulations}.

\begin{table}[]
    \centering
    \begin{tabular}{crrrcl}
    Name & Box size & Elements & Resolution & Code & Details\\
    \hline
    \multicolumn{6}{c}{Suites}\\
    \hline
    \mpgadget & 67.5 Mpc   & $768^3$  & 87.9 kpc & \mpgadget    & 30 f\&p sims \\
    \lyssa    & 120.0 Mpc  & $4096^3$ & 29.3 kpc & \texttt{Nyx} & 18 sims\\
    \hline
    \multicolumn{6}{c}{Individual}\\
    \hline
    \texttt{mpg-central}   & 67.5 Mpc  & $768^3$  & 87.9 kpc & \mpgadget    & \makecell[l]{f\&p, at the center of \\the \mpgadget suite} \\
    \texttt{mpg-seed}      & 67.5 Mpc  & $768^3$  & 87.9 kpc & \mpgadget    & \makecell[l]{Same but other \\ initial phases} \\
    \texttt{lyssa-central} & 118.8 Mpc & $4096^3$ & 29.0 kpc & \texttt{Nyx} & \makecell[l]{At the center of \\the \lyssa suite} \\
    \texttt{lyssa-seed}    & 118.8 Mpc & $4096^3$ & 29.0 kpc & \texttt{Nyx} &  \makecell[l]{Same but other \\ initial phases} \\
    \texttt{sherwood}      & 59.0 Mpc  & $2048^3$ & 28.8 kpc & \texttt{P-Gadget3} & \\
    \hline
    \end{tabular}
    \caption{Summary of the hydrodynamical simulations employed in this work. The resolution is defined as $L/N^{1/3}$, where $L$ is the box size and $N$ is the number of particles for the SPH codes (\mpgadget and \texttt{P-Gadget3}) and the number of grid cells for \texttt{Nyx}. However, note that the converge of \pone measurements for SPH and grid codes may be different at the same resolution \cite{hydro_Chabanier2023}. The term f\&p stands for fixed and paired initial conditions \cite{angulo2016CosmologicalNbodySimulations, anderson2019CosmologicalHydrodynamicSimulations, fixedpaired_Villaescusa}.}
    \label{tab:simulations}
\end{table}


\subsubsection{\mpgadget simulations}
\label{sec:data_sim_lace}

The first suite consists of 30 fixed and paired \cite{angulo2016CosmologicalNbodySimulations, anderson2019CosmologicalHydrodynamicSimulations, fixedpaired_Villaescusa} cosmological hydrodynamical simulations \cite{Pedersen2021} run using an implementation \cite{feng2018MpGadgetMpGadgetTag} of the \texttt{TreeSPH} code \mpgadget \cite{Feng2016_bluetides, emugp_bird2019, bird2022ASTRIDSimulationGalaxy} based on a massively scalable version of the cosmological structure formation code \texttt{P-Gadget3} \cite{Gadget_Springel, dimatteo2012_pgadget3, Khandai2015_pgadget3}. Each simulation tracks the evolution of $768^3$ cold dark matter particles along with an equal number of baryonic particles inside a simulation box of size $L=67.5$ Mpc. The limited resolution of these simulations could lead to an underestimation of \pone by approximately 10\% on the smallest scales probed by DESI at $z=3$, with differences increasing (decreasing) towards smaller (larger) scales and higher (lower) redshifts \cite{bolton2017SherwoodSimulationSuite}. However, as shown in \cref{sec:model_validation}, inaccuracies due to the limited resolution of the simulations do not bias cosmological constraints.

All the simulations begin at $z=99$ from initial conditions generated with \texttt{MP-GenIC} \cite{Bird2020} using baryon and dark matter particles initialized on offset grids with species-specific transfer functions calculated employing \texttt{CLASS} \cite{lesgourgues2011CosmicLinearAnisotropy, Blas2011_class}, displacements computed using the Zel’dovich approximation, and same distribution of Fourier phases. This configuration is accurate over the range of scales and redshifts accessed by DESI measurements \cite{Bird2020, Khan2024_lyaic}. We refer the reader to \cite{Pedersen2021, cabayol-garcia2023NeuralNetworkEmulator} for more details about these simulations.

\pone measurements are sensitive to the small-scale amplitude and logarithmic slope of the linear power spectrum, as well as the ionization, thermal state, and pressure of the intergalactic medium. In order to explore these quantities, the simulations of the suite use a flat $\Lambda$CDM cosmology with compressed parameters within the ranges $\deltastar\in[0.25,\,0.45]$ and $\nstar\in[-2.35,\,-2.25]$ while varying one parameter modifying the midpoint of the hydrogen reionization and two controlling the thermal history of intergalactic gas. These parameters were selected according to a Latin hypercube design to sample the space of interest efficiently \cite{mckay1979ComparisonThreeMethods}. Given that \pone measurements are only weakly sensitive to other cosmological parameters, all simulations use the same value of the Hubble parameter, $H_0=67\,\mathrm{km\,s^{-1}Mpc^{-1}}$, physical cold dark matter density, $\omega_\mathrm{cdm}=0.12$, and physical baryon density, $\omega_\mathrm{b}=0.022$). As shown in \cite{Pedersen2023}, emulators trained on these simulations yield accurate cosmological constraints when applied to \pone predictions from simulations with different values of the fixed parameters as well as for some extensions of the $\Lambda$CDM model. 

To capture the cosmological and astrophysical dependence of \pone, we optimize a Gaussian process emulator (see \cref{sec:emulator}) using the \pone simulation measurements presented in \cite{cabayol-garcia2023NeuralNetworkEmulator}, which were extracted from simulation snapshots between $z=2$ and 4.5 in steps of $\Delta z=0.25$ using \texttt{FSFE}\footnote{\url{https://github.com/sbird/fake_spectra}} \cite{bird2017FSFEFakeSpectra}. The transmitted flux fraction was computed along $768^2$ uniformly distributed lines of sight along each simulation axis with line-of-sight resolution of 0.05 Mpc, which is sufficient to resolve the thermal broadening scale. The \pone predictions were then obtained by averaging the Fourier transforms of all flux skewers.

In addition to the 30 simulations used to train the emulator, we employ two additional simulations to validate our pipeline. The first is the \texttt{mpg-central} simulation, a fixed and paired simulation with cosmological and astrophysical parameters at the center of the hypercube. The second is the \texttt{mpg-seed} simulation, which we use to assess the impact of cosmic variance since it shares the same parameters as the \texttt{mpg-central} simulation but employs different initial conditions. Naively, one would expect cosmic variance to induce $\simeq10\%$ variations in \pone at the smallest wavenumbers of DESI measurements given the simulation volume \cite{hydro_Lukic2015, bolton2017SherwoodSimulationSuite, hydro_Walther2021}. However, the fixed-and-paired technique substantially suppresses this source of uncertainty \cite{anderson2019CosmologicalHydrodynamicSimulations} in a manner that is difficult to quantify analytically \cite{maion2022fpvariance}.


\subsubsection{\texttt{Lyssa} simulations}
\label{sec:data_sim_lyssa}

We use \pone predictions from \lyssa, a cosmological suite of high-resolution hydrodynamical simulations, to validate our analysis pipeline in \cref{sec:model_validation} and to train an alternative \pone emulator in \cref{app:lace-lyssa}. The \lyssa suite is fully described in \cite{Walther2025_lyssa}; we provide a brief summary of its main characteristics below.

The \lyssa simulations were run using the adaptive mesh hydrodynamical code \texttt{Nyx}\footnote{\url{https://github.com/AMReX-Astro/Nyx}} \cite{Almgren+14, Lukic+15, Sexton+21}, which evolves the hydrodynamical fluid on an Eulerian grid together with $N$-body particles tracing the dark matter. The suite comprises 18 simulations of $120\,\mathrm{Mpc}$ on a side with $4096^3$ hydrodynamical cells and the same number of dark matter particles, and a ``fiducial'' simulation at approximately the center of the parameter space (\texttt{lyssa-central} hereafter) described in \cite{hydro_Walther2021}. The resolution of these simulations is approximately a factor of 3 better than the one of the \mpgadget simulations. The simulations were initialized at $z=99$ and took snapshots in intervals of $\Delta z=0.2$ between $z=2.2$ and 4.6, with two additional snapshots at $z=5.0$ and 5.4. The initial conditions of the simulations were generated using \texttt{2lpt-ic} \cite{Crocce2006_icsims} and produced particles following the total matter transfer function at the initial redshift. The baryonic mass of the particles was deposited on a regular mesh using a cloud-in-cell scheme.

The \lyssa suite explores cosmological parameters $A_\mathrm{Ly\alpha}$ and $n_\mathrm{Ly\alpha}$ --- closely related to \deltastar and \nstar ---, the physical matter density $\omega_\mathrm{m}$, and the dimensionless Hubble parameter $h$ according to a Latin hypercube while holding fixed the physical density of baryons to the value $\omega_\mathrm{b}=0.02233$. For each of the redshift and simulations discussed above, the code \texttt{gimlet} (last described in \cite{friesen+2016}) was used to rescale the gas temperature and effective optical depth \cite{hydro_Walther2021}, extract flux skewers, and measure \pone. In this way, the code extracted $4096^2$ flux skewers along each simulation axis with line-of-sight resolution of 0.03 Mpc.


\subsubsection{\texttt{Sherwood} simulations}
\label{sec:data_sim_other}

We also validate our methodology using \pone predictions from the 40-2048 simulation of the \texttt{sherwood} suite; see \cite{bolton2017SherwoodSimulationSuite} for a full description. The \texttt{sherwood} simulations were run using the Tree-PM SPH code \texttt{P-Gadget-3}, which is an updated version of \texttt{Gadget-2} \cite{Gadget_Springel}. The 40-2048 simulation evolved $2048^3$ gas and the same number of baryonic particles within a simulation box of $L=40\,h^{-1}\mathrm{Mpc}$ on a side while assuming a \Planck 2013 cosmology \cite{planck2013_cosmo}. The resolution of this simulation is approximately a factor of 3 better than the one of the \mpgadget simulations. The simulation was initialized at $z=99$ on a regular grid using the \texttt{N-GenIC} code and transfer functions generated by \texttt{CAMB} \cite{lewis2000EfficientComputationCosmic}, and took snapshots in intervals of $\Delta z=0.1$ between $z=2$ and 5. At each redshift, \pone is measured by extracting 5000 flux skewers with line-of-sight resolution of $0.02\,h^{-1}\mathrm{Mpc}$. Therefore, the resolution of the \lyssa and \texttt{sherwood} simulations is very similar.

%% file: emulator_journal.tex
\section{Emulator}
\label{sec:emulator}

In this section, we construct an emulator to model the dependence of \pone on cosmology and IGM physics. We use training data similar to that employed in previous emulator developments \cite{Pedersen2021, Pedersen2023, cabayol-garcia2023NeuralNetworkEmulator, chavesmontero2025forestflow}, but adopt a different internal parameterization and network architecture. We summarize the emulation strategy in \cref{sec:emulator_strategy}, describe the implementation in \cref{sec:emulator_implementation}, and evaluate the emulator performance in \cref{sec:emulator_performance}.


\subsection{Strategy}
\label{sec:emulator_strategy}

Following the same approach as in \cite{Pedersen2021, Pedersen2023, cabayol-garcia2023NeuralNetworkEmulator, chavesmontero2025forestflow}, we emulate \pone as a function of the small-scale amplitude and logarithmic slope of the linear power spectrum and four parameters capturing the ionization, temperature, and pressure of the intergalactic medium. We refer to those works for a full description of the parameterization; in what follows, we briefly discuss it.

We summarize the cosmological information contained by \pone using two parameters: $\Delta^2_{\mathrm{p}}(z)$ and $n_{\mathrm{p}}(z)$, which represent the amplitude and logarithmic slope of the linear matter power spectrum at the pivot scale $k_\mathrm{p} = 0.7\iMpc$ and redshift $z$, respectively. These parameters are computed following the method described in \cite{Pedersen2021}. The pivot scale is defined in comoving rather than velocity units because the \pone $k$-binning is uniform in comoving coordinates across different cosmologies, which facilitates emulator training. As discussed in \cref{sec:model}, emulator predictions are subsequently converted from comoving to velocity units during cosmological inference, since DESI measurements are expressed in velocity units.

We capture the dependence of \pone upon the properties of the IGM using the mean transmitted flux fraction $\bar{F}(z)=\exp{(-\tau_\mathrm{eff})}$, where $\tau_\mathrm{eff}$ is the effective optical depth, the normalization ($T_0$) and slope ($\gamma$) of the temperature-density relation, $T=T_0\Delta_\mathrm{b}^{\gamma-1}$, where $\Delta_\mathrm{b}$ is the baryon overdensity, and the pressure smoothing scale $k_\mathrm{F}$. Rather than using $T_0$ as input, our emulator adopts $\sigma_\mathrm{T} = \sigma_\mathrm{T,0} \sqrt{10^{-4} T_0[K]} (1+z) H^{-1}(z)$, where the value of the normalization is $\sigma_\mathrm{T,0}=9.1\,\kms$. The factor $(1+z) H^{-1}(z)$ is required to convert the broadening scale from velocity to comoving units, thus keeping the internal units of the emulator consistent.

Even though the emulator parameters are defined at specific redshifts, redshift itself is not an input to our emulator. As a result, the emulator can generate predictions at redshifts not included in the training set \cite{Pedersen2021, cabayol-garcia2023NeuralNetworkEmulator, chavesmontero2025forestflow}. It is important to also highlight that this emulation strategy enables extracting accurate constraints even when analyzing simulations with $\Lambda$CDM parameters outside of the training set or contemplating $\Lambda$CDM extensions such as massive neutrinos, non-zero running of the spectral index, and curvature \cite{Pedersen2023, cabayol-garcia2023NeuralNetworkEmulator}.


\subsection{Implementation}
\label{sec:emulator_implementation}

Hydrodynamical simulations are far more computationally expensive than gravity-only simulations, severely limiting both their volume and number of realizations in cosmological suites. The restriction in the volume of the simulations poses a particular challenge for constructing a \pone emulator since a naive implementation would learn spurious large-scale fluctuations arising from the small number of long-wavelength modes in each simulation, potentially biasing cosmological inference \cite{Pedersen2023}. To mitigate this issue, our emulator predicts the coefficients of a smooth \pone model rather than the power spectrum at individual $k$ values. This smooth model is constructed as the product of a mean flux–dependent baseline and correction terms that capture cosmology- and IGM-induced variations.

We construct the base model using \texttt{forestflow} \cite{chavesmontero2025forestflow}, a cosmological emulator for the three-dimensional \lya flux power spectrum ($P_\mathrm{3D}$) trained on \pone measurements from the \mpgadget simulations. The main advantage of \texttt{forestflow} is that it predicts the \lya linear biases ($b_\delta$ and $b_\eta$) and the parameters of a smooth functional form describing small-scale deviations of $P_\mathrm{3D}$ from linear theory; therefore, its predictions agree with the \lya generalization of the Kaiser model on large scales \cite{kaiser1987ClusteringRealSpace}, which mitigates the impact of cosmic variance. On the other hand, even though \texttt{forestflow} provides \pone predictions through direct integration of $P_\mathrm{3D}$, we do not employ it for inference because it is not currently optimized for fast \pone analyses.

Using \texttt{forestflow}, we make \pone predictions using as input the cosmological and IGM parameters of the 11 snapshots of the \texttt{mpg-central} simulation (see \cref{sec:data_sim_lace}), for which \texttt{forestflow} achieves an accuracy better than 0.6\% \cite{chavesmontero2025forestflow}. We then interpolate these \pone predictions as a function of the mean transmitted flux of each snapshot, $\bar{F}(z_i)$, to obtain the base model $\pone^\mathrm{base}(k_\parallel,\,\bar{F})$. Then, we build a smooth model that combines the base model with parametric corrections
\begin{align}
    \label{eq:psmooth}
    \pone^\mathrm{smooth} &= \exp\left[g(k_\parallel/k^\mathrm{max}_\parallel, \mathbf{a})\right] ~ \pone^\mathrm{base}(k_\parallel,\,\bar{F}),\\
    g(x, \mathbf{a}) &= \sum_{i=0}^4 \frac{a_i}{1 + e^{2^{i-1} x}}
\end{align}
where $a_i$ are five free parameters designed to capture the dependence of \pone on cosmology and IGM physics besides that of the baseline model upon mean flux. The motivation for using this functional form is that, for a null value of all the free parameters, the smooth model agrees with the baseline model, and this with the \lya generalization of the Kaiser model on large scales. 

\begin{figure}
    \centering
    \includegraphics[width=0.495\linewidth]{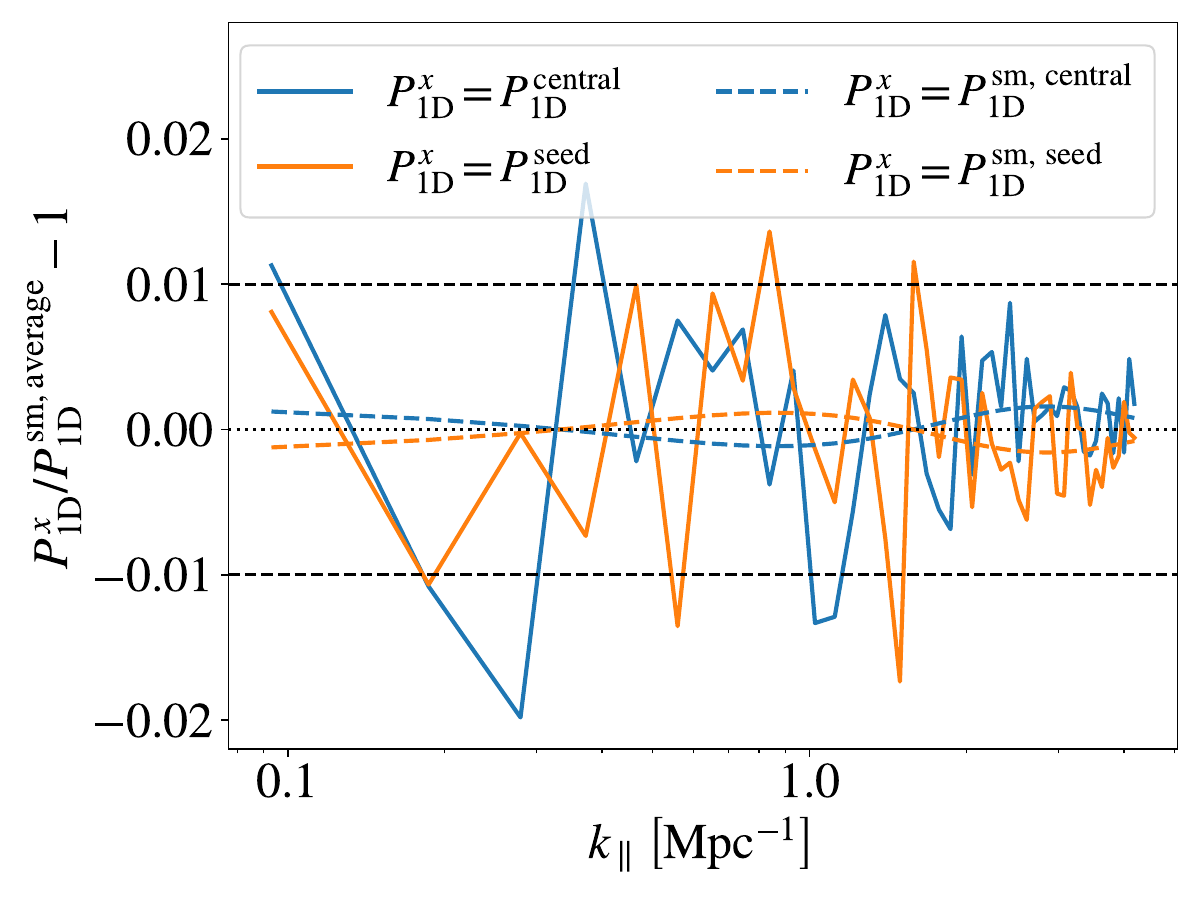}
    \includegraphics[width=0.495\linewidth]{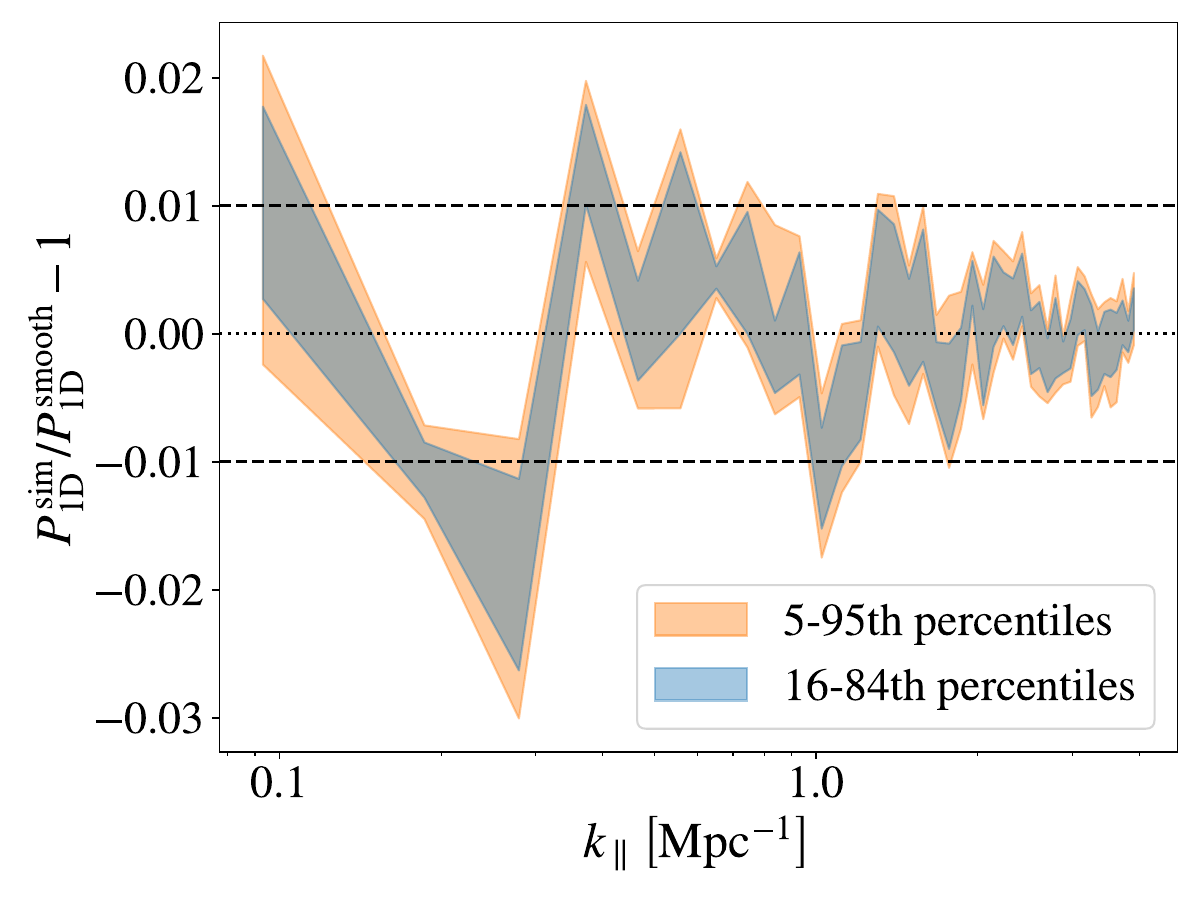}
    \caption{
    Performance of the \pone smooth model from \cref{eq:psmooth} in reproducing \pone predictions from the \mpgadget simulations. In the left panel, we show the residual between different samples and the mean of the best-fitting smooth models to the \texttt{mpg-central} and \texttt{mpg-seed} simulations at $z = 3$. The blue and orange solid lines correspond to residuals for \texttt{mpg-central} and \texttt{mpg-seed}, respectively, while the blue and orange dashed lines do so for their respective best-fitting smooth models.
    In the right panel, the blue and orange shaded areas show the 16 to 84th and the 5 to 95th percentile regions, respectively, of the relative difference between \pone predictions from all the \mpgadget simulations in the training set and the best-fitting smooth model to each of these.}
    \label{fig:emulator_smooth}
\end{figure}

In the left panel of \cref{fig:emulator_smooth}, we evaluate the accuracy of the smooth model in capturing \pone predictions from the \texttt{mpg-central} and \texttt{mpg-seed} simulations at $z = 3$. We show the residual between different samples and the mean of the best-fitting smooth models to each of the two simulations, which is expected to be less affected by cosmic variance than the predictions from the two simulation or the best-fitting smooth model to each of these. The blue and orange solid lines show the residuals for \texttt{mpg-central} and \texttt{mpg-seed} predictions, respectively. As we can see, the difference between these two residuals reaches 2.5\% for some scales, which is attributable to cosmic variance since the only distinction between the \texttt{mpg-central} and \texttt{mpg-seed} simulations is their initial distribution of Fourier phases. The blue and orange dashed lines display the residuals for the best-fitting smooth models to the \texttt{mpg-central} and \texttt{mpg-seed} simulations, respectively. The difference between the dashed lines is approximately an order of magnitude smaller than between the solid lines, indicating that cosmic variance has a negligible impact on the smooth model fits.

In the right panel of \cref{fig:emulator_smooth}, we evaluate the performance of the smooth model in reproducing \pone predictions from all the \mpgadget simulations. The largest discrepancies appear at the same wavenumbers as for the \texttt{mpg-central} simulation, indicating that a) the smooth model performs similarly across redshifts and cosmologies and b) all training simulations use the same initial Fourier phases. Aside from this noise pattern, the smooth model reproduces the data with accuracy better than 1\% on scales not dominated by cosmic variance. This demonstrates that the smoothing procedure accurately captures \pone predictions without introducing scale-dependent residuals and it mitigates cosmic variance fluctuations that would otherwise enter the emulator training set and propagate into emulator predictions \cite{Pedersen2023}.

In order to get the input data to our emulator, we compute the best-fitting value of the free parameters $\mathbf{a}$ to measurements from each snapshot of the \mpgadget simulations by minimizing the mean squared difference between the smooth model and the simulation data. The minimization is performed over scales ranging from the fundamental mode of the simulations up to $k^\mathrm{max}_\parallel = 4.25\,\iMpc$, which exceeds the maximum wavenumber accurately measured by DESI, using the Nelder–Mead algorithm implemented in the \texttt{scipy}'s \texttt{minimize} package \cite{scipy:2020}. Then, we train a Gaussian process emulator, \lacempg hereafter, to predict the value of the free parameters $\mathbf{a}$ as a function of the cosmological and IGM parameters described in the previous section: $\Delta^2_\mathrm{p}$, $n_\mathrm{p}$, $\bar{F}$, $\sigma_T$, $\gamma$, and $k_\mathrm{F}$. 

We implement the emulator using the \texttt{GaussianProcessRegressor} routine of the package \texttt{scikit-learn} \cite{Pedregosa2011_scikitlearn} with an anisotropic Matern 1/2 kernel, and optimize it using the \texttt{L-BFGS-B} algorithm from the \texttt{scipy}'s \texttt{minimize} package \cite{scipy:2020}. To reduce training and evaluation time, we divide the input data into two slightly overlapping $\bar{F}$ bins containing 1\,000 elements each and train a separate Gaussian process on each subsample. When evaluating the \lacempg emulator at a given $\bar{F}$, we use the Gaussian process corresponding to the bin whose central $\bar{F}$ value is closest. If $\bar{F}$ lies within the overlap region between the two bins, we take the average of the predictions from the two Gaussian processes, which typically agree at the 0.1\% level within this region.


\subsection{Performance}
\label{sec:emulator_performance}

\begin{figure}
    \centering
    \includegraphics[width=0.475\linewidth]{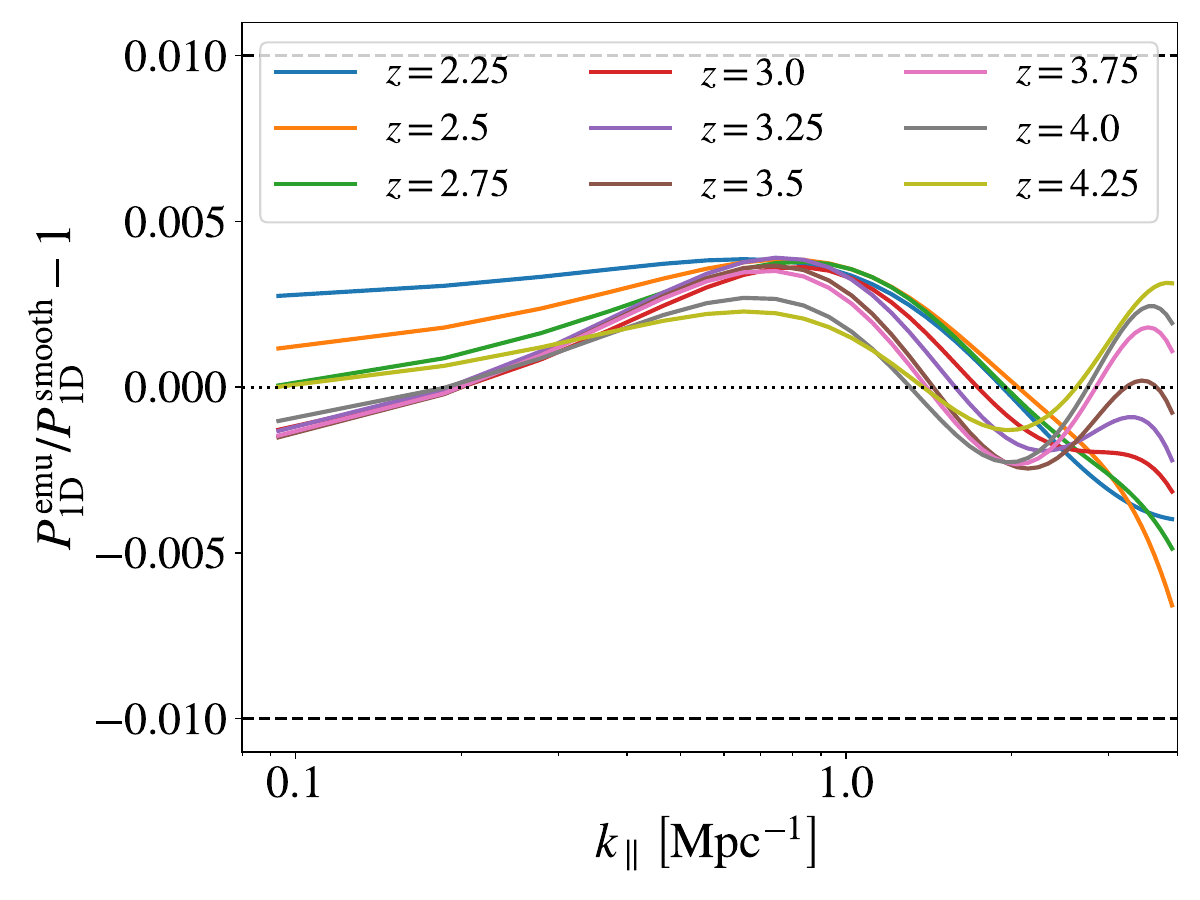}
    \includegraphics[width=0.475\linewidth]{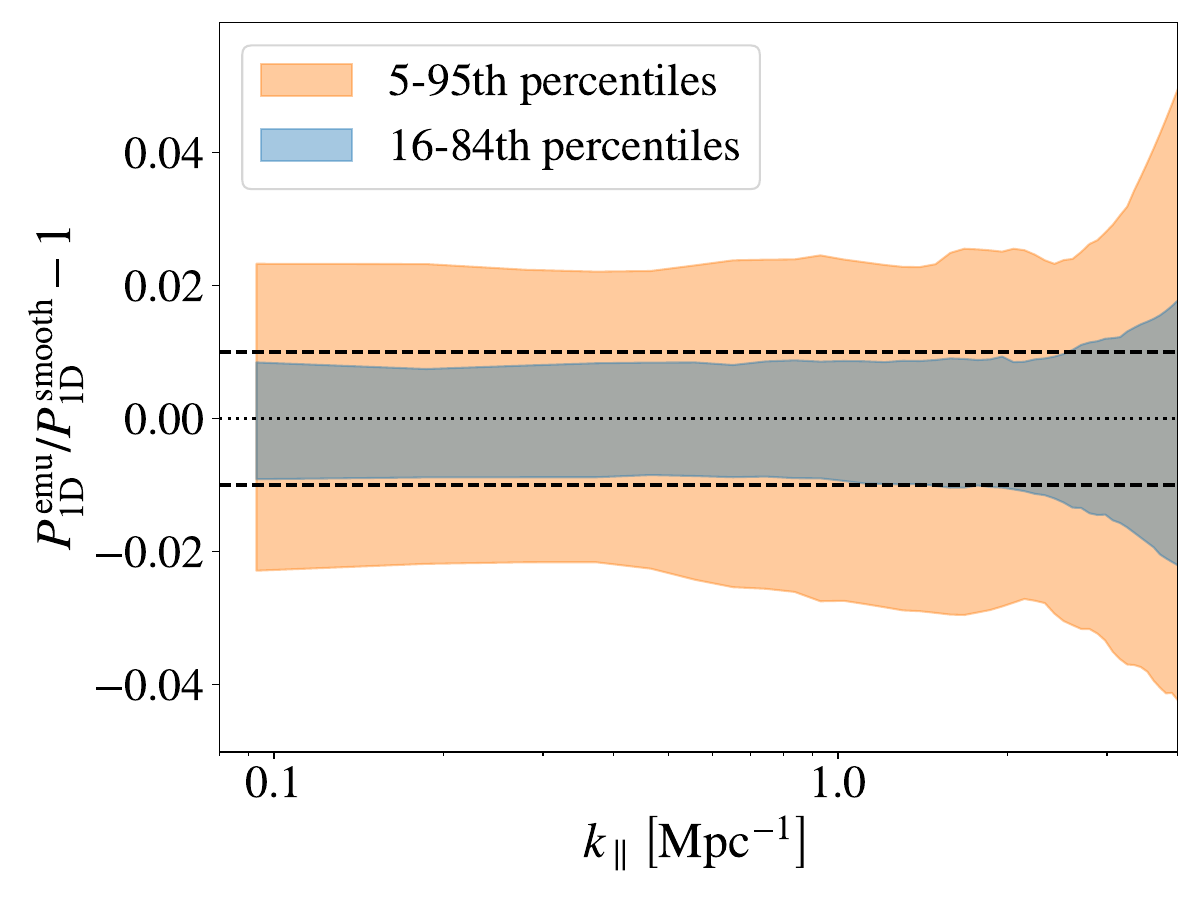}
    \caption{Performance of \lacempg in recovering smoothed \pone predictions from simulations outside of the training set. In the left panel, we show the relative difference between our emulator and smoothed predictions from the \texttt{mpg-seed} simulation. In the right panel, the blue and orange shaded areas show the 16 to 84th and the 5 to 95th percentile regions, respectively, of 30 leave-one-out tests. The accuracy of the emulator is better than 1\% at $1\sigma$ for most scales.
    }
    \label{fig:emulator_performance}
\end{figure}

In this section, we assess the performance of the \lacempg emulator in recovering smoothed \pone measurements from the \mpgadget simulations. The left panel of \cref{fig:emulator_performance} shows that the emulator reproduces smoothed predictions from the \texttt{mpg-seed} simulation with better than 1\% accuracy across all redshifts and scales. The results for the \texttt{mpg-central} simulation are similar, indicating that cosmic variance has a negligible impact on emulator predictions. However, since both simulations lie near the center of the parameter space spanned by the training set --- where machine-learning methods generally perform best --- this level of precision cannot be assumed across the full parameter space. 

To obtain a more reliable assessment of the performance of \lacempg across the parameter space, we carry out 30 leave-one-out tests. In each case, we train a new emulator using all simulations except one, and then evaluate its accuracy against the excluded simulation. The average performance over the 30 tests is shown in \cref{fig:emulator_performance}. On average, we find that the accuracy is better than 1\% across the parameter space. As expected, the emulator performs best near the center of the parameter space, where the density of training points is highest, and its accuracy degrades with increasing distance from the nearest training point. However, in the leave-one-out setup, each test evaluates the emulator at a point that is, by construction, farther from the training set than would be the case for the full emulator (since one simulation has been removed). Consequently, these tests likely provide a conservative estimate of emulator errors.

In \cref{app:emu_covariance}, we describe how we quantify the impact of emulator uncertainties on the analysis of \pone measurements. During inference, we add these uncertainties to the statistical and systematic errors described in \cref{sec:data_obs}.

%% file: contaminants_journal.tex
\section{Astrophysical contaminants and systematics}
\label{sec:contaminants}

A complete characterization of the impact of cosmology, IGM physics, HCD systems, and metal lines on \pone would require extremely high-resolution hydrodynamical simulations with full subgrid physics and a broad range of ionization, thermal, and pressure histories. However, it is currently infeasible to run a cosmological suite of simulations that simultaneously samples all these properties while maintaining the volume and resolution required for \pone studies. Instead, we model the impact of HCD and metal contamination on the \lya–only predictions from the \lacempg emulator. To do so, we adopt parameterized corrections for metal contamination based on \cite{Karacayli2023_doublet, Karacayli2025_p1d_dr1} in \cref{sec:cont_metals} and for HCD contamination following \cite{Rogers2018a} in \cref{sec:cont_hcd}, with modifications tailored to the analysis of DESI data. We also model the impact of uncertainties in the DESI spectrograph resolution on \pone measurements in \cref{sec:cont_res}, and the combination of all these corrections in \cref{sec:cont_full}.


\subsection{Metal contamination}
\label{sec:cont_metals}

The \lyaf results from the absorption of quasar light by neutral hydrogen, and similarly, other ions create forests that leave their imprint in \pone. Modeling the absorption by metals is challenging due to the large number of ions that can contaminate \pone measurements; as a result, the preferred solution is to remove the metal contamination directly by subtracting the power spectrum of a side band free from neutral hydrogen absorption \cite{McDonald2006, p1d_PalanqueDelabrouille2013, Ravoux2023, Karacayli2024_edr}. However, it is not possible to subtract the contamination from absorption lines with wavelengths smaller than $\lambda \approx 1300\,\mathrm{\AA}$ following this procedure since the resulting absorptions are mixed with the \lyaf. In what follows, we model the impact of absorption lines with such wavelengths on the \lyaf.

Intergalactic gas containing neutral hydrogen and metals at redshift $z$ absorbs light at observed-frame wavelengths $\lambda_\alpha (1+z)$ and $\lambda_x (1+z)$, respectively, where $\lambda_x$ is the rest-frame wavelength of a metal line $x$. If metal absorption is mistakenly interpreted as \lya absorption, it will be assigned an incorrect redshift $z_\alpha$, related to the true redshift $z_x$ of the metal line by $(1 + z_x) = (\lambda_x / \lambda_\alpha)(1 + z_\alpha)$. This misassignment causes spurious correlations at a characteristic velocity separation $\Delta v_{x\alpha} = c \, \log(\lambda_x / \lambda_\alpha)$, where $c$ is the speed of light. Consequently, metal contamination produces an oscillatory pattern in \pone with frequency $2\pi / \Delta v_{x\alpha}$. Note that the previous derivation also applies to pairs of metal lines.

It is thus useful to analyze the one-dimensional correlation function, $\xi(v)=\langle \delta(v) \delta(v+\Delta v) \rangle$, to detect possible metal contamination. As shown in fig.~7 of \cite{Karacayli2025_p1d_dr1}, there are six strong peaks in the correlation function that approximately correspond to the velocity difference between the line pairs \wsiia-\wsiib, \lya-\wsiii, \wsiib-\wsiii, \wsiia-\wsiii, \lya-\wsiib, and \lya-\wsiia. We provide the velocity separation of these lines in \cref{tab:metals}. Two more peaks appear at a velocity differences comparable to the maximum length of the forests considered in our analysis, $\Delta v=1.8\times10^4\,\kms$, and thus the frequency of the corresponding \pone oscillations is too low to be detected. 

We can split the impact of metal contamination on \pone into the terms $C_\mathrm{Ly\alpha-\siii}$, $C_\mathrm{Ly\alpha-\sii}$, $C_\mathrm{\sii-\siii}$, and $C_\mathrm{\sii-\sii}$, which capture the impact of the \lyasiii pair, the two \lyasii pairs, the two \siisiii pairs, and the \siisii pair, respectively. In what follows, we model the different-ion pairs (first three terms) in \cref{sec:cont_metal_diff} and same-ion pairs (last term) in \cref{sec:cont_metal_same}.

\begin{table}
\centering
\begin{tabular}{cccc}
    Line 1 [$\mathrm{\AA}$] & Line 2 [$\mathrm{\AA}$] & $\Delta v\,[\mathrm{km}\,\mathrm{s}^{-1}]$ & $\Delta k_\parallel\,[\ikms]$ \\ \hline
    \siia 1190.42 & \siib 1193.28 &  719.39 & 0.0087\\
    \lya  1215.67 & \siii 1206.52 & 2264.99 & 0.0028\\
    \siib 1193.28 & \siii 1206.52 & 3308.02 & 0.0019\\
    \siia 1190.42 & \siii 1206.52 & 4027.41 & 0.0016\\
    \lya  1215.67 & \siib 1193.28 & 5573.01 & 0.0011\\
    \lya  1215.67 & \siia 1190.42 & 6292.40 & 0.0010\\
\end{tabular}
\caption{Wavelength, velocity difference, and frequency of the oscillatory pattern imprinted on \pone by the line pairs responsible for the metal contamination.}
\label{tab:metals}
\end{table}


\subsubsection{Different ion pairs}
\label{sec:cont_metal_diff}

The simplest approach to model the contamination of the \lya-metal pairs is to assume that the density contrast of a metal is equal to the \lya density contrast up to an overall normalization \cite{McDonald2006, Karacayli2025_p1d_dr1}:
\begin{equation}
    \label{eq:delta_metals}
    \delta_\mathrm{F} = \delta(v) + A_{\siii} \delta(v + \Delta v_{\lyasiii}) + A_{\sii} [\delta(v + \Delta v_\mathrm{Ly\alpha-\siib}) + r\,\delta(v + \Delta v_\mathrm{Ly\alpha-\siia})],
\end{equation}
where $r=(\mathfrak{f}_{\siia}\lambda_{\siia}) / (\mathfrak{f}_{\siib}\lambda_{\siib})$ scales the strength of the \sii lines under the optically thin limit, the values of the oscillator strengths are $\mathfrak{f}_{\siia}=0.277$ and $\mathfrak{f}_{\siib}=0.575$ \cite{NIST_database}, $A_x = (1-\bar{F})^{-1} f_x$ scales the amplitude of the lines by the mean flux, and $f_x$ are free parameters.

We continue by taking the Fourier transform of the previous expression while considering all possible \lya-metal combinations,
\begin{equation}
    P_\mathrm{1D}^\mathrm{w/ metals}=(1 + C_\mathrm{Ly\alpha-\siii} + C_\mathrm{Ly\alpha-\sii} + C_\mathrm{\sii-\siii})\pone,
\end{equation}
where $\pone$ refers to the uncontaminated power spectrum. The term $C_\mathrm{Ly\alpha-\siii}$ accounts for the contamination by \lya-\wsiii,
\begin{equation}
    \label{eq:lya-siii}
    C_\mathrm{Ly\alpha-\siii}= A^2_{\siii} + 2 A_{\siii} \cos\left(k_\parallel \Delta v_{\lyasiii}\right),
\end{equation}
the term $C_\mathrm{Ly\alpha-\sii}$ models the contamination by \lya-\wsiib and \lya-\wsiia,
\begin{equation}
    \label{eq:lya-sii}
    C_\mathrm{Ly\alpha-\sii}= A^2_{\sii} \left(1 + r^2\right) + 2 A_{\sii} \left[\cos\left(k_\parallel \Delta v_\mathrm{Ly\alpha-\siib}\right) + r \cos\left(k_\parallel \Delta v_\mathrm{Ly\alpha-\siia}\right)\right],
\end{equation}
and the term $C_\mathrm{\sii-\siii}$ captures the contamination by \wsiib-\wsiii and \, \wsiia-\wsiii,
\begin{equation}
    \label{eq:sii-siii}
    C_\mathrm{\sii-\siii} = 2 A_{\siii} A_{\sii} \left[\cos\left(k_\parallel \Delta v_{\siii-\siib}\right) + r \cos\left(k_\parallel \Delta v_{\siii-\siia}\right)\right].
\end{equation}

However, these expressions are not complex enough to describe the impact of different-ion contamination on DESI measurements (see \cref{sec:model_single}). We find it critical to account for two additional effects: deviations from the optically-thin limit and the progressive decorrelation between the small-scale distribution of neutral hydrogen and silicon. We model the first by substituting $r$ by $r$ times the free parameter $r_{\siia}$ in the previous equations, as well as by multiplying the RHS of \cref{eq:sii-siii} by the free parameter $r_{\siisiii}$. On the other hand, we capture the decorrelation by multiplying the second term in the RHS of \cref{eq:lya-siii,,eq:lya-sii} by $D_\mathrm{Si\,III}$ and $D_\mathrm{Si\,II}$, respectively, where $D_x=2 - 2/\left[1 + \exp(-k_\parallel\,/k_x)\right]$ and $k_x$ is a free parameter controlling the scale at which the distributions decouple. We do not model the decorrelation between different silicon ions.

\begin{figure}
    \centering
    \includegraphics[width=0.75\linewidth]{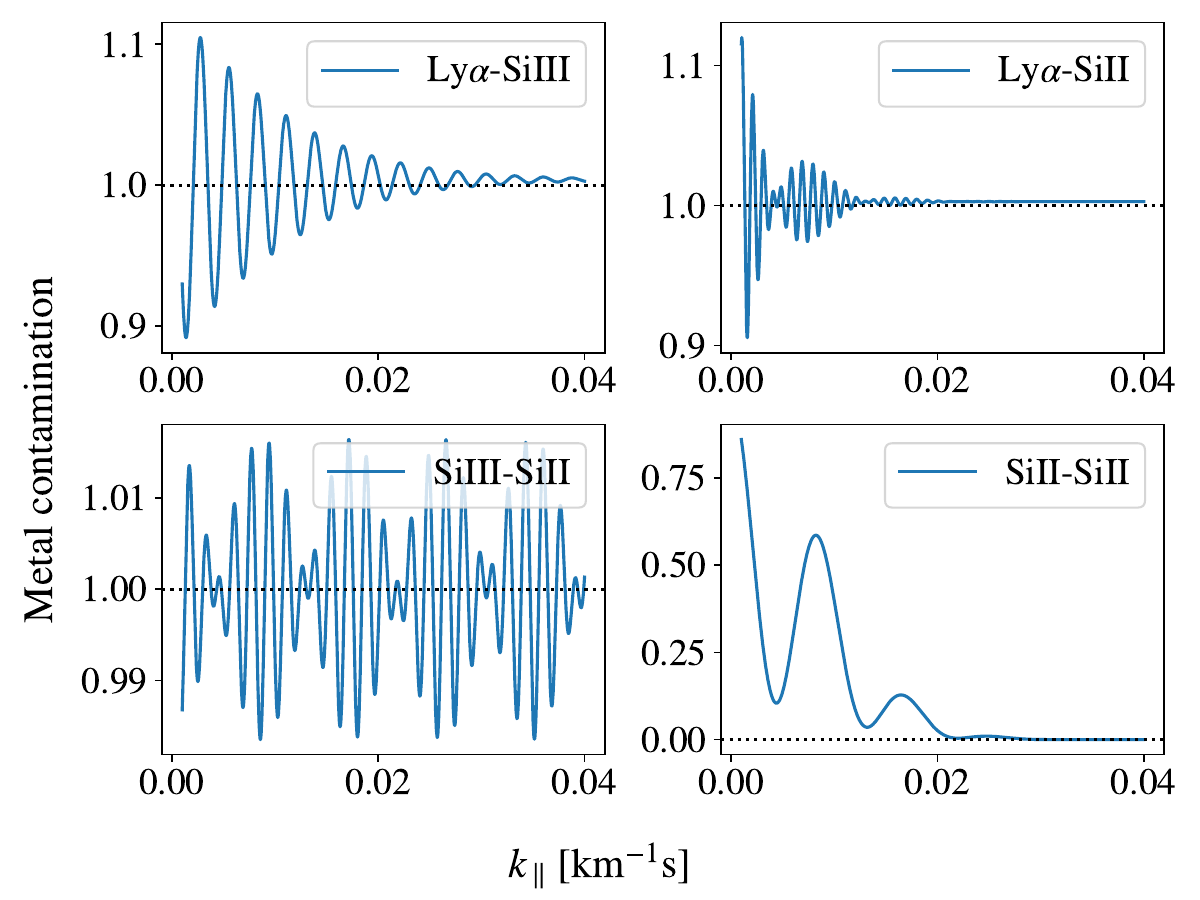}
    \caption{Illustrative example of the impact of metal contamination on \pone. Each panel displays the contribution of a different line pair, as indicated at the top right of each panel. The contribution of the \siisii pair is additive, while those of the other line pairs are multiplicative.}
    \label{fig:metal_illustrate}
\end{figure}

In total, our model has 6 free parameters: $f_{\siii}$, $f_{\sii}$, $k_{\siii}$, $k_{\sii}$, $r_{\siia}$, and $r_{\siisiii}$. In the top-left, top-right, and bottom-left panels of \cref{fig:metal_illustrate}, we display illustrative examples for the metal contamination arising from \lyasiii, \lyasii, and \siisiii, respectively, produced using the best-fitting value of the parameters at $z=2.2$ from the baseline analysis (see \cref{sec:results}). We can readily see that the \lyasiii and \lyasii terms are increasingly damped as we move towards smaller scales due to the progressively greater de-correlation between the distribution of neutral hydrogen and silicon. We can also see that the \lyasiii term oscillates with a single frequency, while the \lyasii and \siisiii terms have complex oscillatory patterns owing to the combined oscillations caused by \wsiia and \wsiib.

\subsubsection{Same ion pairs}
\label{sec:cont_metal_same}

Instead of assuming that the \wsiia–\wsiib cross-correlation is a scaled version of the \lyaf autocorrelation, as in the previous section, we model it by approximating the transmission profile of the doublet as a pair of Gaussian functions. As shown in \cite{Karacayli2023_doublet}, this approach describes accurately the correlation function of same-ion pairs. In this way, the contribution of same-ion pairs to \pone predictions from our emulator becomes additive, while the contributions of different-ion pairs explained above were multiplicative.

In the optically-thin limit, one of the Gaussians is a factor of $r$ weaker than the other, $K(v)=G(v-\Delta v_{\siisii}/2) + r G(v+\Delta v_{\siisii}/2)$, and the corresponding power spectrum is given by 
\begin{equation}
    C_{\sii-\sii} = f_{\siisii} \left[1 + r^2 + 2 r \cos\left(k_\parallel \Delta v_{\siisii}\right)\right] \exp\left(- k^2_\parallel/k^2_{\siisii}\right),
\end{equation}
where the exponential term accounts for the thermal broadening of the lines, which suppresses the small-scale power. In total, this model has one parameter controlling the strength of the contamination and another the scale of the smoothing: $f_{\siisii}$ and $k_{\siisii}$. In the bottom-right panel of \cref{fig:metal_illustrate}, we display an illustrative example of the contribution of \siisii for the best-fitting value of the model parameters at $z=2.2$ from the baseline analysis (see \cref{sec:results}).


\subsection{High-column density systems}
\label{sec:cont_hcd}

Low-density intergalactic gas is responsible for the majority of the \lya absorption, while HCD systems near galaxies account for only a small fraction of this. Nevertheless, the impact of HCD contamination on \pone measurements is substantial because the largest systems --- commonly known as damped Lyman-$\alpha$ systems (DLAs) --- exhibit extended damping wings that reach far beyond the location of the absorbing gas, contaminating the \lya forest even on relatively large scales \cite{McDonald2005b, McQuinn2011, FontRibera2012a, Rogers2018a, dMdB2020, DESI2024.IV.KP6}. In this section, we model the effect of HCD contamination on \pone.

Following \cite{Rogers2018a}, we classify HCDs into four groups by column density: Lyman limit systems (LLSs; $1.6\times10^{17} < N_\mathrm{H I}[\mathrm{cm}^{-2}] < 10^{19}$), sub-damped \lya systems (sub-DLAs; $10^{19} < N_\mathrm{H I}[\mathrm{cm}^{-2}] < 2\times10^{20}$), small DLAs ($2\times10^{20} < N_\mathrm{H I}[\mathrm{cm}^{-2}] < 10^{21}$), and large DLAs ($N_\mathrm{H I} > 10^{21} \mathrm{cm}^{-2}$). In principle, DLAs are modeled and corrected before measuring DESI DR1 \pone. However, the DESI DLA finder is $\simeq70\%$ complete for systems with $N_\mathrm{H I} > 2\times 10^{20}\mathrm{cm}^{-2}$ in spectra with average $\mathrm{SNR}>2$ per pixel within the 1420-1480$\mathrm{\AA}$ rest-frame interval \cite{Y3.lya-s2.Brodzeller.2025}, and the completeness drops rapidly for systems with lower column densities. As a result, DESI DR1 \pone measurements are contaminated by LLSs and sub-DLAs, as well as some fraction of DLAs not identified by the algorithm.

We model the impact of HCD contamination on \pone using an empirically-motivated multiplicative correction based on \cite{Rogers2018a}
\begin{equation}
    \label{eq:hcd}
    C_\mathrm{HCD} = 1 + f_\mathrm{norm}^\mathrm{HCD} + \sum_i f^\mathrm{HCD}_i\, \left[a_i(z) e^{b_i(z)k_\parallel}-1\right]^{-2},
\end{equation}
where $i$ indexes each HCD category, the free parameters $f^\mathrm{HCD}_i$ control the contamination from each system, the parameter $f_\mathrm{norm}^\mathrm{HCD}$ acts as an overall normalization factor, and the redshift evolution of the remaining terms follows $a(z)=a_0[(1+z)/3]^{a_1}$ and $b(z)=b_0[(1+z)/3]^{b_1}$, where $a_0$, $a_1$, $b_0$, and $b_1$ are fixed to the values reported by \cite{Rogers2018a} and gathered in \cref{tab:hcd_rogers}. The value of the $f^\mathrm{HCD}_i$ parameters reported in \cite{Rogers2018a} corresponds to the fraction of line-of-sights in the highest resolution simulation from the Illustris project \cite{Vogelsberger2014_illustris, Nelson2015_illustris} containing at least a particular type of HCD system, and thus these values would be different for a larger simulation. Consequently, we treat these quantities as free parameters of our model, and we allow them to have a flexible redshift evolution instead of the one found in \cite{Rogers2018a} for a particular simulation and cosmology.

\begin{table}[]
    \centering
    \begin{tabular}{ccccc}
    Category & $a_0$ & $a_1$ & $b_0\,[\mathrm{km\,s}^{-1}]$ & $b_1$\\
    \hline
    LLS       & 2.2001 & 0.0134 & 36.449 & -0.0674 \\
    Sub-DLA   & 1.5083 & 0.0994 & 81.388 & -0.2287 \\
    Small DLA & 1.1415 & 0.0937 & 162.95 &  0.0126 \\
    Large DLA & 0.8633 & 0.2943 & 429.58 & -0.4964 \\
    \end{tabular}
    \caption{Value of the fixed parameters describing the impact of HCDs on \pone \cite{Rogers2018a}.}
    \label{tab:hcd_rogers}
\end{table}

\begin{figure}
    \centering
    \includegraphics[width=0.75\linewidth]{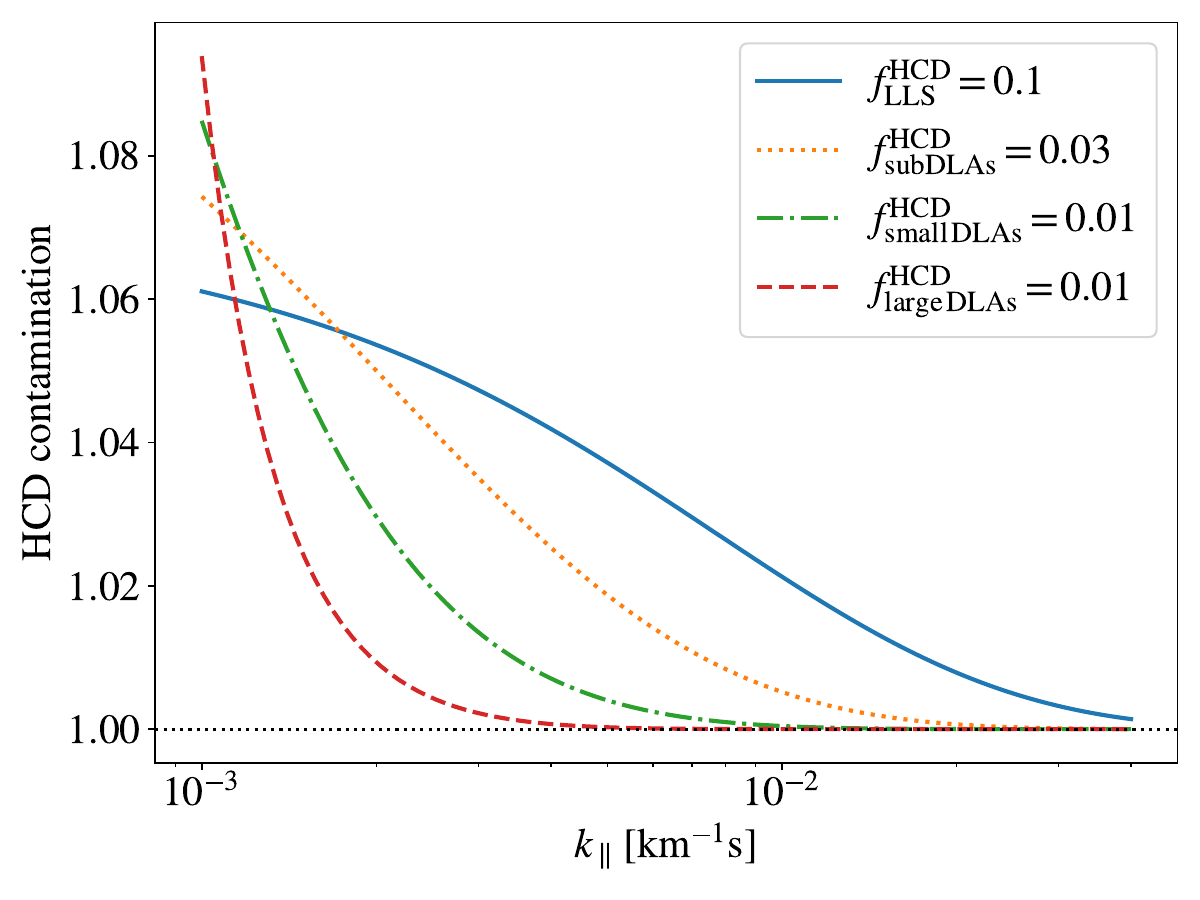}
    \caption{Illustrative example of the impact of HCD contamination on \pone. Lines show the contribution of each of the four terms of the HCD model for a different value of the free parameters, as indicated in the legend. The large DLA term increases very rapidly with decreasing wavenumber, while the increase of the LLS term is very modest.}
    \label{fig:hcd_illustrate}
\end{figure}

In \cref{fig:hcd_illustrate}, we illustrate the impact of HCD contamination on \pone using the $f^\mathrm{HCD}_i$ values listed in the legend. We observe that higher column density systems mainly affect large scales, while lower column density systems have a stronger impact on small scales. Since the DESI DLA finder algorithm identifies most of the high column density systems, the DLA parameters will primarily reflect the efficiency of this algorithm. We show the results while assuming $f_\mathrm{norm}^\mathrm{HCD}=0$, a parameter that we omit in the baseline analysis because it is strongly degenerate with any rescaling of the mean flux \citep{Rogers2018a}.

\subsection{Spectrograph resolution}
\label{sec:cont_res}

Accurate knowledge of the DESI spectrograph resolution is essential for reliably extracting \pone on small scales, where residual uncertainties in its characterization constitute the leading source of systematic error in this regime \cite{Karacayli2025_p1d_dr1, Ravoux2025}. Instead of including this error in the systematic covariance matrix, we marginalize over it using a multiplicative correction \cite{McDonald2006, Karacayli2024_edr},
\begin{equation}
    C_\mathrm{res} = 1 + 2 f_\mathrm{res}\, R_z^2\, k^2_\parallel,
\end{equation}
where $f_\mathrm{res}$ is a free parameter, $R_z = c\,(1+z)^{-1} \Delta\lambda_\mathrm{DESI}/\lambda_\mathrm{Ly\alpha}$ is the pixel width in velocity units, and $\Delta\lambda_\mathrm{DESI} = 0.8\,\mathrm{\AA}$. The value of the resolution bias is expected to be of the order of a few per cent \cite{Karacayli2025_p1d_dr1, karacayli2025_p1d_validation}.

Even though we expect the resolution bias to change slowly with redshift, we use a different parameter at each redshift when analyzing DESI data in order to capture the impact of systematic effects such as the transition of the \lya line between the blue and red channels of the DESI spectrograph \cite{Spectro.Pipeline.Guy.2023}.


\subsection{Combination}
\label{sec:cont_full}

Throughout this section, we have introduced parametric models to account for the impact of metal contamination, HCD contamination, and uncertainties in the characterization of the DESI spectrograph resolution on \pone measurements. Combined, these corrections modify the emulator’s \lya-only \pone prediction, $P^\mathrm{emu}_\mathrm{1D}$, as follows:
\begin{equation}
    P^\mathrm{cont}_\mathrm{1D} = \left[(1 + C_\mathrm{Ly\alpha-\siii} + C_\mathrm{Ly\alpha-\sii} + C_\mathrm{\sii-\siii})\, C_\mathrm{HCD}\, P^\mathrm{emu}_\mathrm{1D} + C_\mathrm{\sii-\sii}\right]C_\mathrm{res},
\end{equation}
where the full model includes 8 free parameters for the metals contamination, 5 for the HCD contamination, and 1 for resolution systematics. We checked that the results are essentially the same when $C_\mathrm{HCD}$ also multiples $C_\mathrm{\sii-\sii}$ in the previous expression. Together with the 6 input parameters of the \pone emulator, the model includes a total of 20 free parameters per redshift.

It is important to note that, rather than evaluating the model directly on the same $k$-binning as DESI measurements, we employ a binning eight times finer and subsequently rebin the results. This procedure mitigates biases that would otherwise arise from the artificial smoothing of the oscillatory corrections introduced by metal contamination.

%% file: likelihood_journal.tex
\section{Model}
\label{sec:model}

In \cref{sec:emulator}, we presented an emulator capturing the dependence of \pone on cosmology and the physics of the intergalactic medium. In \cref{sec:contaminants}, we introduced models for additional effects not captured by the emulator: metal contamination, HCD contamination, and resolution systematics. In this section, we combine these components into a unified framework, which is an updated version of publicly available likelihood code \texttt{cup1d}\footnote{\url{https://github.com/igmhub/cup1d}} \cite{Pedersen2021, Pedersen2023}. We construct a model to fit measurements at individual redshifts (\cref{sec:model_single}), extend it for the joint analysis of multiple redshift bins (\cref{sec:model_multiple}), and validate it using \pone mocks from distinct hydrodynamical simulations (\cref{sec:model_validation}).


\subsection{Single redshift}
\label{sec:model_single}

In this section, we assess whether the combination of our emulator and the corrections for astrophysical and systematic contaminants are sufficiently flexible to describe DESI DR1 measurements at individual redshifts. We perform the fits while assuming the \textit{Planck} 2018 $\Lambda$CDM cosmology \cite{Planck2018} since reasonable variations in cosmological parameters lead to $\chi^2$ shifts too small to affect the conclusions of single-redshift analyses.

To evaluate the model at redshift $z$, we first compute the amplitude and logarithmic slope of the linear matter power spectrum at $k_\mathrm{p} = 0.7\,\mathrm{Mpc}^{-1}$ and the target redshift $z$, $\Delta^2_\mathrm{p}(z)$ and $n_\mathrm{p}(z)$, which serve as the input cosmological parameters to our emulator (see \cref{sec:emulator_strategy}). Then, after evaluating the emulator, we convert its predictions from comoving to velocity units, since both the remaining components of the model and the DESI \pone measurements are expressed in velocity units. For these calculations, we use the \texttt{camb} Boltzmann solver \cite{lewis2000EfficientComputationCosmic} following the approach described in \cite{Pedersen2021}.

Regarding the IGM parameters entering the emulator, rather than fitting directly their values, we constrain deviations from the values predicted by the \texttt{mpg-central} simulation: $\Delta \tau_\mathrm{eff}$, $\Delta \sigma_T$, $\Delta \gamma$, and $\Delta k_\mathrm{F}$. This approach is motivated by the fact that the IGM parameter space is highly non-trivial and the \texttt{mpg-central} values lie near its center, making it easier to define meaningful parameter ranges for the fit \cite{Pedersen2021}. Specifically, we adopt flat priors spanning the full range of values covered by the \mpgadget simulations. Note that constraint deviations from the value of $\sigma_T$ and $k_\mathrm{F}$ in velocity units so that they remain independent of the Hubble expansion used (see also \cref{sec:intro}).

Regarding the parameters controlling astrophysical contaminants and systematics, we use broad flat priors that encompass the best-fitting values obtained in all the analyses presented in this work. We fix the value of $f_\mathrm{norm}^\mathrm{HCD}$ to zero, since it is strongly degenerate with the parameters scaling the mean flux. In \cref{sec:results_robust}, we perform alternative analyses allowing this parameter to vary, and we find that doing so has practically no impact on the inferred cosmological constraints. A full description of the model parameters and their respective variation ranges is given in \cref{tab:parameters}.

\begin{table}[]
    \centering
    \begin{tabular}{ccl}
        Parameter            & Range         & Description \\
        \hline
        \multicolumn{3}{c}{Cosmology}\\
        \hline
        $A_\mathrm{s}$         & $[1.0,\,3.4]\times 10^{-9}$ & \makecell[l]{Amplitude of primordial curvature perturbations\\ at $k_\mathrm{s}=0.05\,\iMpc$} \\
        $n_\mathrm{s}$         & $[0.7,\,1.3]$ & Primordial spectral index at $k_\mathrm{s}=0.05\,\iMpc$ \\
        \hline
        \multicolumn{3}{c}{IGM (deviations from \texttt{mpg-central} simulation)}\\
        \hline
        $\log \Delta \tau_{\rm eff}$        & [-0.2, 0.2]   & Optical depth\\
        $\Delta \sigma_T$       & [0.8, 1.3]    & Normalization of the temperature-density relation\\
        $\Delta \gamma$         & [0.8, 1.3]    & Slope of the temperature-density relation\\
        $\Delta k_\mathrm{F}$   & [0.8, 1.3]    & Pressure smoothing scale\\
        \hline
        \multicolumn{3}{c}{Metal contamination}\\
        \hline
        $\log f_{\lyasiii}$  & [-11, -2]     & Strength of \lyasiii contamination\\
        $\log k_{\lyasiii}$  & [2, 7]        & Damping scale of \lyasiii contamination\\
        $\log f_{\lyasii}$   & [-11, -2]     & Strength of \lyasii contamination\\
        $\log k_{\lyasii}$   & [2, 7]        & Damping scale of \lyasii contamination\\
        $\log r_{\siia}$   & [-1, 3]        & Deviations from the optically-thin limit for \sii\\
        $\log r_{\siisiii}$   & [-1, 3]        & Deviations from the optically-thin limit for \siisiii\\
        $\log f_{\siisii}$   & [-3, 3]       & Strength of \siisii contamination\\
        $\log k_{\siisii}$   & [0, 7]        & Damping scale of \siisii contamination\\
        \hline
        \multicolumn{3}{c}{HCD contamination}\\
        \hline
        $\log f_\mathrm{LLS}^\mathrm{HCD}$       & [-11, -0.03] & Strength of LLS contamination\\
        $\log f_\mathrm{sub-DLA}^\mathrm{HCD}$   & [-11, -0.03] & Strength of sub-DLA contamination\\
        $\log f_\mathrm{small DLA}^\mathrm{HCD}$ & [-11, -0.03] & Strength of small DLA contamination\\
        $\log f_\mathrm{large DLA}^\mathrm{HCD}$ & [-11, -0.03] & Strength of large DLA contamination\\
        $f_\mathrm{norm}^\mathrm{HCD}$ & Fixed to 0 & Overall normalization\\
        \hline
        \multicolumn{3}{c}{Spectrograph resolution}\\
        \hline
        $f_\mathrm{res}$    & [-0.02, 0.02] & Spectrograph resolution bias
    \end{tabular}
    \caption{Range and brief description of the free parameters of our model. The range of variations in $A_\mathrm{s}$ and $n_\mathrm{s}$ translate into correlated priors in the compressed parameters, with the projections onto these parameters being $\deltastar\in[0.1, 1.3]$ and $\nstar\in[-2.6, -2.0]$. The IGM parameters capture variations around the ionization, thermal, and pressure histories predicted by the \texttt{mpg-central} simulation. The damping scales are in units of $\ikms$.}
    \label{tab:parameters}
\end{table}

We compute the best-fitting model to \pone measurements at a particular redshift by minimizing 
\begin{equation}
    \chi^2_z = \left[\pone^\mathrm{z-data}-\pone^{\mathrm{z-model}}(z, \theta)\right]^\mathrm{T}  C^{-1}_z \left[\pone^\mathrm{z-data}-\pone^{\mathrm{z-model}}(z, \theta)\right],
\end{equation}
where we carry out the minimization using the Nelder-Mead routine from the \texttt{scipy} package \cite{scipy:2020}, $\pone^\mathrm{z-data}$ and $\pone^\mathrm{z-model}$ refer to the data and model vectors, respectively, $\theta$ indicates the free parameters of the model, and $C^{-1}_z$ is the inverse of the covariance at redshift $z$, which includes terms accounting for statistical, systematic, and emulator errors (see \cref{sec:data_obs}).

\begin{figure}
    \centering
    \includegraphics[width=0.9\linewidth]{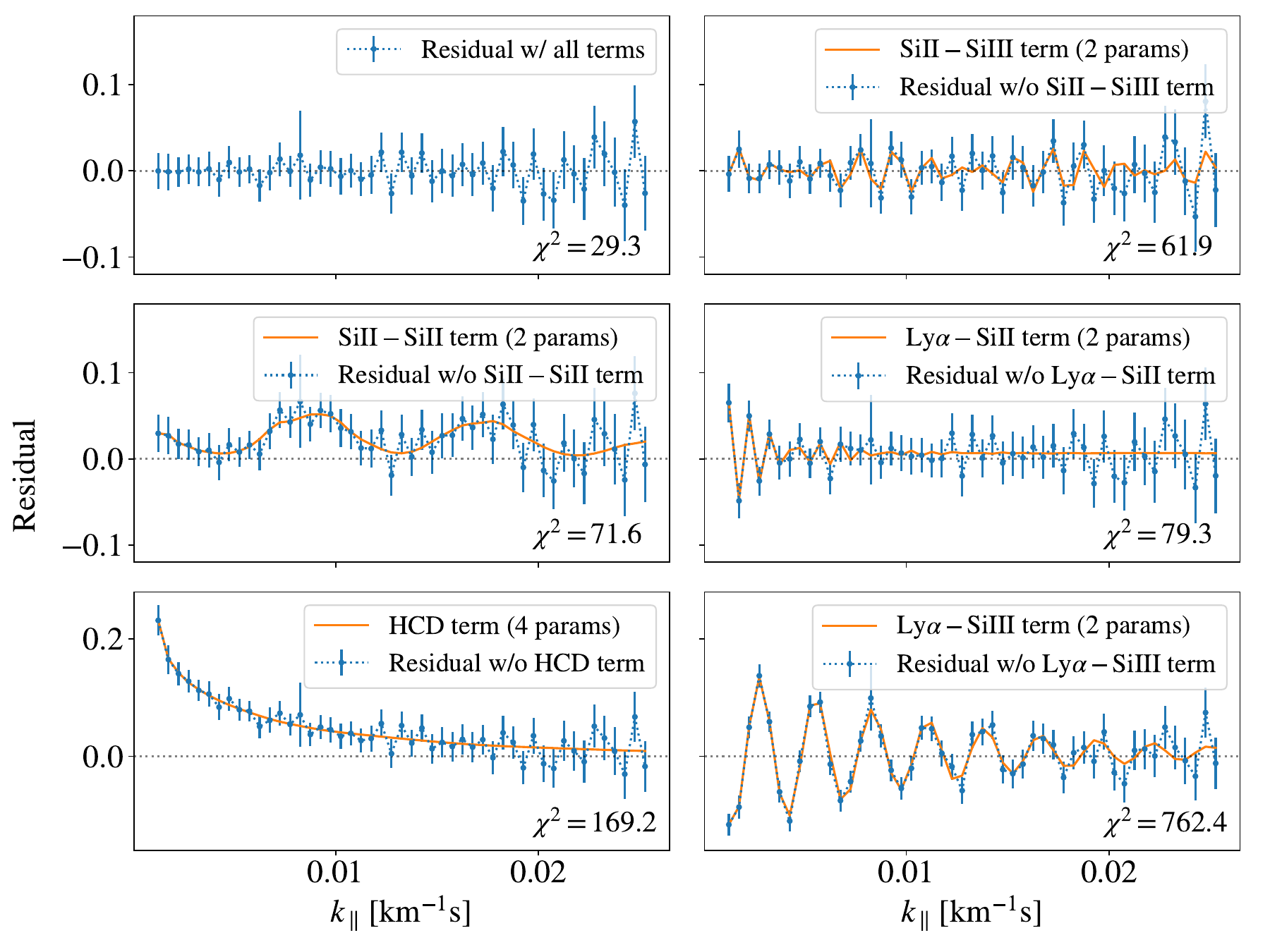}
    \caption{
    Contribution of the HCD and metal terms to the best-fitting model to DESI DR1 measurements at $z=2.2$. The top-left panel shows the residual between the data and the best-fitting model including all components. In the remaining panels, data points show the residuals obtained after switching off the term indicated in the legend, while the orange line depicts the term removed. Error bars show the square root of the diagonal of the total covariance matrix, which includes statistical, systematic, and emulator contributions. We report the $\chi^2$ of the residuals at the bottom-right corner of each panel.
    }
    \label{fig:contaminants_z2.2}
\end{figure}

In \cref{fig:contaminants_z2.2}, we dissect the contributions of the HCD and metal components of the best-fitting model to DESI measurements at $z=2.2$. The top-left panel shows the residual between the data and the full model. In the remaining panels, the points show residuals obtained after switching off a single term of the model, while the orange line indicates the term removed. The top-right panel highlights the \siisiii contribution, which displays a complex oscillatory pattern arising from the combined \siia--\siii and \siib--\siii oscillations. The middle-left panel focuses on the \siisii component, responsible for the longest-period oscillations, while the middle-right panel isolates the \lyasii term, which captures the strong, rapid oscillations on the largest scales. The bottom-left panel examines the impact of HCD contamination, dominated by the LLS term and affecting primarily large scales. The bottom-right panel shows the \lyasiii contribution, which accurately reproduces the strongest oscillations.

Previous \pone cosmological analyses neither modeled the scale-dependent damping of the \lyasii and \lyasiii terms nor included the \siisii and \siisiii components, whereas we find all of these ingredients essential for achieving a good fit. Specifically, we carry out alternative fits to the $z=2.2$ data while neglecting, one at a time, the \siisiii, HCD, \siisii, \lyasii, and \lyasiii terms, finding that the $\chi^2$ of the fits increases by 10.7, 21.2, 29.0, 46.5, 632.5, respectively, while not accounting for any of these contaminants increases it by 735.1. Note that the increase in $\chi^2$ obtained when refitting the data while neglecting the HCD term is substantially smaller than that resulting from simply switching off the HCD component of the best-fitting model. This difference arises because, in the fit neglecting the HCD term, the IGM parameters adjust significantly to partially absorb the effects of unmodeled HCD contamination.

\begin{figure}
    \centering
    \includegraphics[width=1\linewidth]{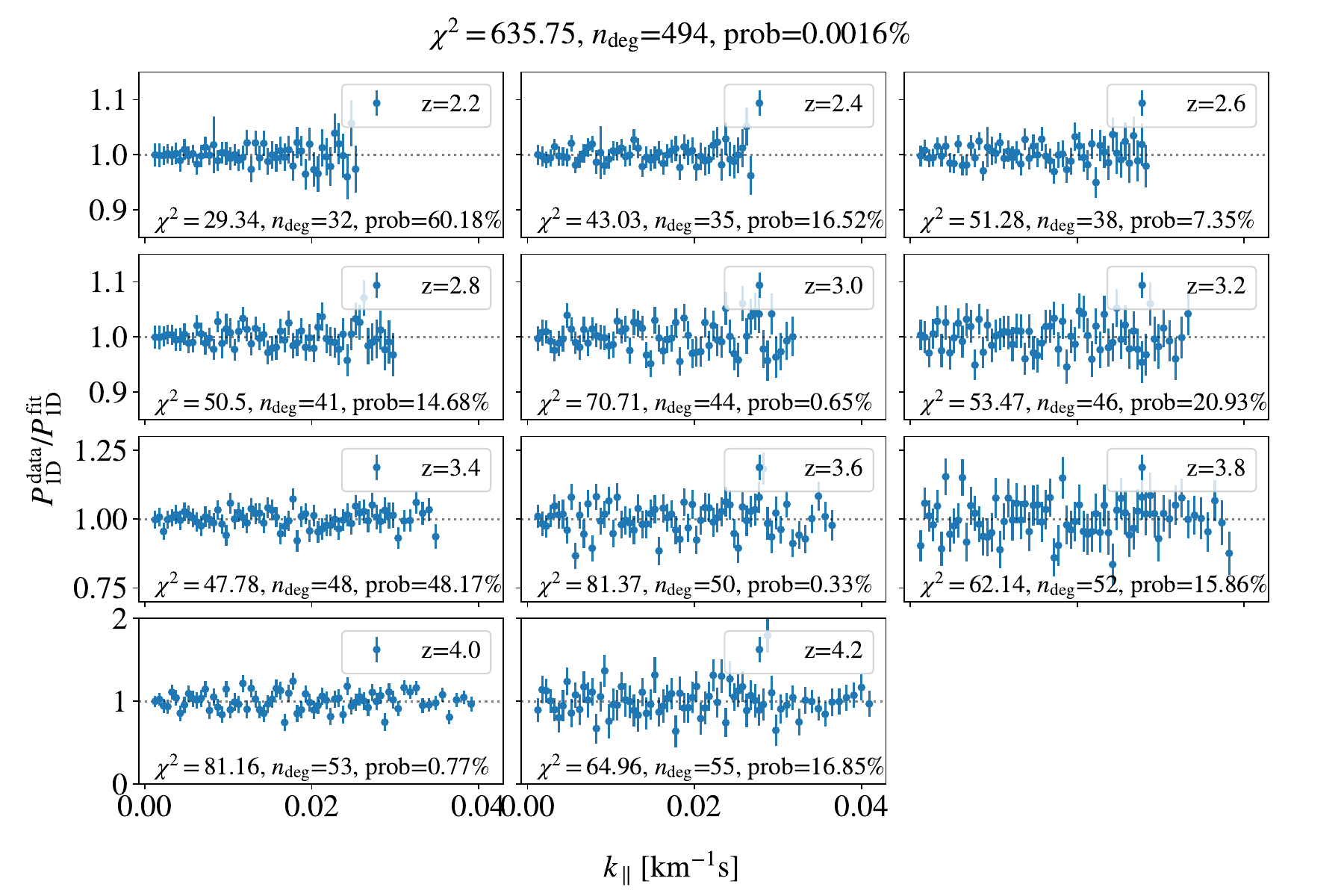}
    \caption{Ratio between DESI measurements and the best-fitting model to each redshift. Each panel displays the results for the redshift indicated at the top right. At the bottom left of each panel, we show the $\chi^2$, number of degrees of freedom, and probability of the fit to the corresponding redshift. At the top of the figure, we quote these quantities for the combination of the individual fits when not considering correlations among redshift bins. Error bars represent the square root of the diagonal elements of the total covariance matrix, which accounts for statistical, systematic, and emulator uncertainties. Note that we do not inflate statistical error bars by 5\% in order to carry out this fit, while we do inflate them throughout the remainder of this work.
    }
    \label{fig:z_at_time}
\end{figure}

In each panel of \cref{fig:z_at_time}, we show the ratio between the \pone measurements at the redshift indicated in the top-right corner and the corresponding best-fitting model. We display the results when not inflating statistical error bars by 5\% (see \cref{sec:data_obs}). As we can see, our model provides a reasonable description of the data across all redshifts but $z=3.0$, 3.6, and 4, for which the fit probabilities are 0.65, 0.33 and 0.77\%, respectively. However, the residuals at these redshifts show no clear wavenumber-dependent structure, indicating that either a) a physical effect specific to these redshifts is not captured by our model, or b) statistical, systematic, or emulator uncertainties are underestimated at these redshifts. A recent validation of DESI DR1 measurements reported evidence for a possible systematic bias at $z = 3$ \cite{karacayli2025_p1d_validation}, but no analogous studies are available for $z = 3.6$ or 4. 

Owing to the poor fits at the previous redshifts, the sum of the $\chi^2$ values across all redshift bins is $\chi^2=636$ for 494 degrees of freedom, corresponding to a fit probability of $1.6\times10^{-5}$, despite individual bins typically exhibiting fit probabilities above 10\%. We increase the statistical uncertainty of each $k$-bin at all redshifts by 5\% to account for a likely underestimation of the errors and to obtain a more reasonable overall fit probability. With this adjustment, the total $\chi^2$ decreases to 577 and the corresponding probability increases to 0.58\%.


\subsection{Multiple redshifts}
\label{sec:model_multiple}

As discussed in \cref{sec:intro}, we summarize the cosmological information encoded in \pone by measuring the amplitude and logarithmic slope of the linear power spectrum at $z_\star=3$ and \kstarval, denoted as $\Delta^2_\star$ and $n_\star$, respectively. For brevity, we refer to these as ``compressed parameters.'' Before starting the inference procedure, we compute their values for the \Planck~2018 $\Lambda$CDM cosmology \cite{Planck2018}: Hubble parameter of $H_0=67.66\,\kms$, physical density of cold dark matter and baryons of $\Omega_\mathrm{cdm}h^2=0.119$ and $\Omega_\mathrm{b}h^2=0.0244$, respectively, and amplitude of primordial curvature perturbations and spectral index of $A_\mathrm{s}=2.105\times10^{-9}$ and $n_\mathrm{s}=0.9665$ at $k_\mathrm{s}=0.05\,\iMpc$. During the inference, we vary the parameters controlling the primordial power spectrum within the ranges $A_\mathrm{s} \in [1.0, 3.4]\times 10^{-9}$ and $n_\mathrm{s} \in [0.7, 1.3]$, hold fixed the Hubble parameter and the physical densities to the \Planck values, and scale the value of the compressed parameters from the \Planck cosmology to the target cosmology following \cite{Pedersen2023}. We hold fixed some cosmological parameters during inference to speed up calculations since the scaling of the compressed parameters does not require calling a Boltzmann solver for each new cosmology and the Universe is effectively Einstein–de Sitter at the redshifts relevant to the \lya forest. We validate the accuracy of this approach \cref{sec:results_robust}, see also \cite{Pedersen2023}.

Rather than fitting the IGM parameters independently at each redshift, we constrain their values at four logarithmically spaced nodes between $z=2.2$ and $4.2$, $z_\mathrm{node}=2.20$, 2.73, 3.38, and 4.20, since we expect the redshift evolution of $\tau_\mathrm{eff}$, $\gamma$, and $k_\mathrm{F}$ to be smooth, with the main potential complication arising for $T_0$ due to the heating of the IGM caused by He II reionization \cite{LaPlante2017_heiireiosims, Upton_Sanderbeck2020_heii}. In principle, four nodes could not be enough to resolve this feature, but we conducted an alternative analysis using six redshift nodes in \cref{sec:results} and found an excellent agreement with the baseline analysis. We remind the reader that these parameters capture deviations from the values predicted by the \texttt{mpg-central} simulation. We obtain the value of the IGM parameters at each of the redshifts of DESI data by linearly interpolating the values at the nodes. This approach results in a low-resolution representation of the redshift evolution of the IGM parameters. 

Similarly, we constrain the metal and HCD parameters using redshift nodes at $z=2.2$ and 4.2, and then linearly interpolate between these to get predictions for the redshifts of DESI data. This is equivalent to assuming a single power-law redshift evolution for these parameters. On the other hand, we marginalize over spectrograph resolution uncertainties by introducing one free parameter per redshift bin (see \cref{sec:cont_res}). In total, the model comprises 53 free parameters: 2 for cosmology, 16 for IGM evolution, 24 for HCD and metal contamination, and 11 for resolution systematics.

We compute the best-fitting model to \pone measurements from multiple redshifts by minimizing
\begin{equation}
    \label{eq:chi2}
    \chi^2 = \left[ \pone^\mathrm{data} - \pone^\mathrm{model} (\tilde{\theta}) \right]^\mathrm{T}  C^{-1} \left[ \pone^\mathrm{data} - \pone^\mathrm{model} (\tilde{\theta}) \right],
\end{equation}
where $\pone^\mathrm{data}$ and $\pone^\mathrm{model}$ are vectors containing measurements and model predictions from multiple redshifts, respectively, $\tilde{\theta}$ stands for the free parameters of the multi-redshift model, and $C^{-1}$ is the inverse of the full covariance matrix.

\begin{figure}
    \centering
    \includegraphics[width=1\linewidth]{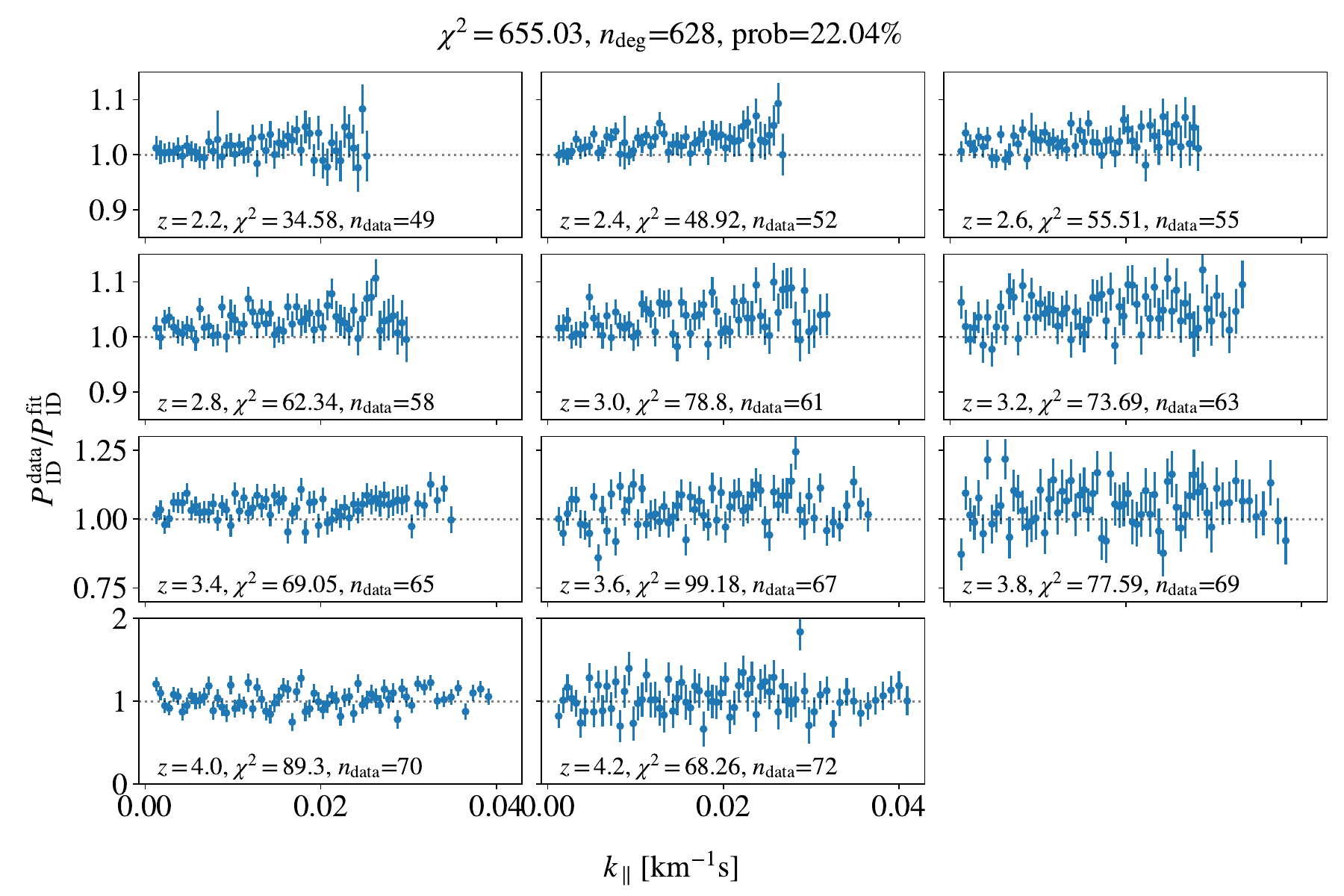}
    \caption{Same as \cref{fig:z_at_time} but for the joint analysis of all DESI DR1 measurements using the multi-redshift model. As we can see, the model provides an accurate description of the data, with a fit probability of 22\%. Despite the good overall fit, model predictions at low redshift are systematically lower than the data by about one percent. This discrepancy arises from the strong off-diagonal terms in the emulator covariance matrix. We remind the reader that we inflate statistical error bars by 5\% in order to carry out this fit.
    }
    \label{fig:global_fit}
\end{figure}

In \cref{fig:global_fit}, we show the ratio between DESI measurements and the best-fitting model to these. As we can see, the model describes the data accurately, with no clear redshift- or wavenumber-dependent structure in the residuals. The percent-level systematic offset between model predictions and measurements at some redshifts is explained by the off-diagonal terms of the covariance matrix, which are dominated by the emulator covariance. Reducing the number of free parameters from 187 in the combination of all the single-redshift analyses to 53 in the multi-redshift analysis increases the $\chi^2$ from 577 to 655, but the associated probability goes from 0.58 to 22.0\% due to the 134 fewer degrees of freedom. Consequently, our multi-redshift model describes the data accurately.

\subsection{Validation}
\label{sec:model_validation}

In principle, the accuracy of cosmological inference may depend on the volume, resolution, or other specific characteristics of the simulations used to train the \lacempg emulator. To validate our model, we analyze mock \pone measurements generated from four simulations that differ in code, resolution, and initial conditions (see \cref{sec:data_sim}): \texttt{mpg-central}, \texttt{mpg-seed}, \texttt{lyssa-central}, and \texttt{sherwood}.

We generate \pone mocks using measurements from the aforementioned simulations between $z=2.2$ and 4.2, which corresponds to the redshift range of DESI DR1 data. The \texttt{mpg-central} and \texttt{mpg-seed} simulations provide nine snapshots within this range, from $z=2.25$ to 4.25 in steps of $\Delta z = 0.25$, while the \texttt{lyssa-central} and \texttt{sherwood} simulations include snapshots at all DESI DR1 redshifts. To mitigate cosmic variance, we first apply the smoothing procedure described in \cref{sec:emulator_implementation} to the simulated data. We then convert the smoothed measurements from comoving to velocity space and interpolate them to match the scales used in the DESI DR1 analysis. In the case of the \mpgadget mocks, we also interpolate from the native redshift of the simulations to the target redshift of the DESI DR1 data. The resulting \pone mocks are assigned the same covariance matrix as the one employed in the DESI DR1 analysis. These mocks do not include contaminants or systematics; however, we analyze them with the model presented in the previous section, which accounts for both.

We extract cosmological constraints using the publicly available affine-invariant Markov chain Monte Carlo (MCMC) ensemble sampler \texttt{emcee}\footnote{\url{https://emcee.readthedocs.io/en/stable/}} \cite{foremanmackey13}. At each step, \texttt{emcee} proposes new values for the model parameters within the priors gathered in \cref{tab:parameters}, evaluates the model, and computes the corresponding compressed parameters. The model prediction is then compared to mock measurements using a Gaussian likelihood, $L=\exp(-\chi^2/2)$, with $\chi^2$ given by \cref{eq:chi2}. For each target sample, we run \texttt{emcee} with 11\,648 independent chains of 3\,000 steps each, discarding the first 1\,500 steps of every chain as burn-in and retaining only 1 in 20 of the remaining steps. Each run takes roughly one node hour in the National Energy Research Scientific Computing Center (NERSC). We verify that this setup provides a robust sampling of the posterior distribution, and we apply the same configuration to analyze observational and mock data throughout the remainder of this work.

\begin{figure}
    \centering
    \includegraphics[width=0.495\linewidth]{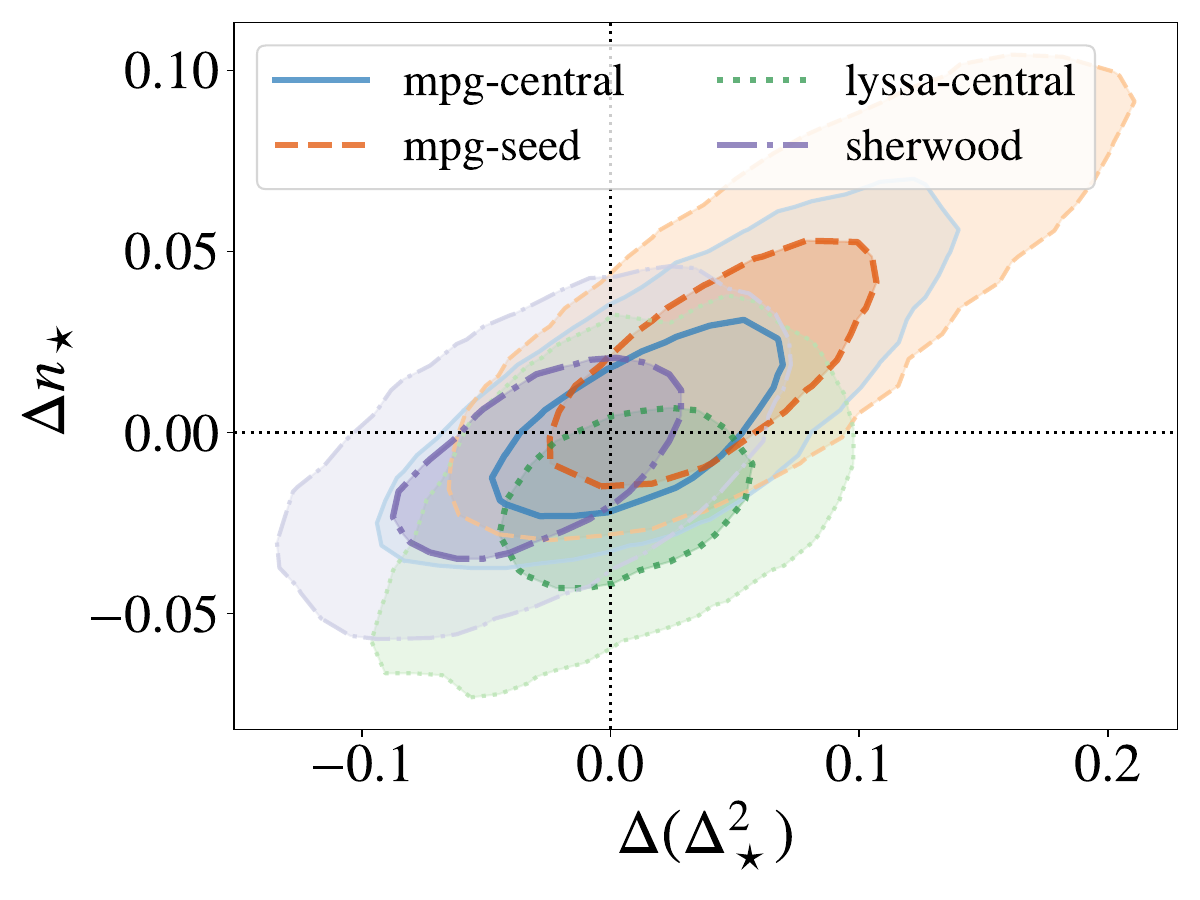}
    \includegraphics[width=0.495\linewidth]{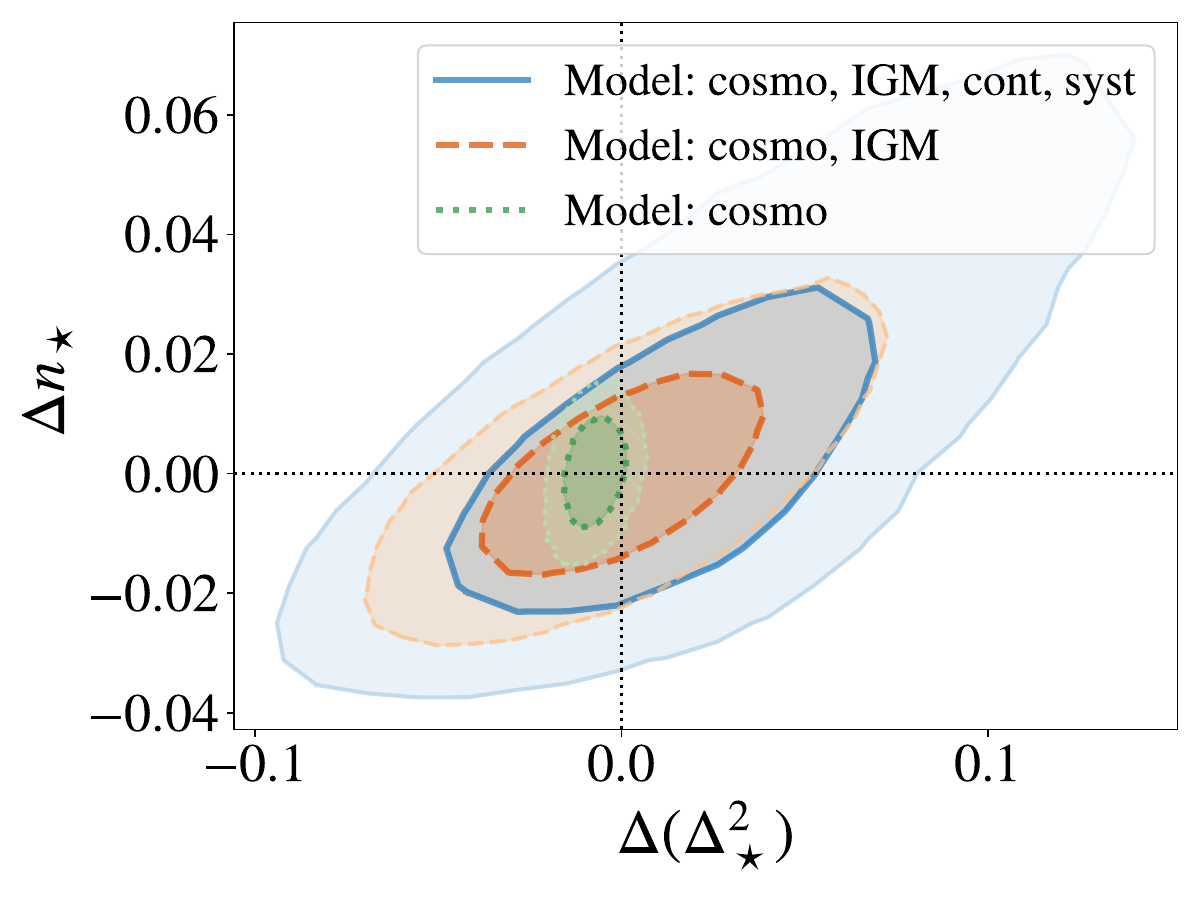}
    \caption{
    Validation of the analysis pipeline using \pone mocks derived from distinct hydrodynamical simulations. In the left panel, blue, orange, green, and purple contours display the results for the analysis of the \texttt{mpg-central}, \texttt{mpg-seed}, \texttt{lyssa-central}, and \texttt{sherwood} mocks, respectively, plotted after subtracting the true value of the compressed parameters for each mock. In the right panel, blue contours show the baseline analysis of \texttt{mpg-central}, while orange and green contours display, respectively, the results obtained when marginalizing only over IGM parameters and when varying only cosmological parameters. Inner and outer contours denote the 68 and 95\% credible regions, respectively.
    }
    \label{fig:validation}
\end{figure}

In the left panel of \cref{fig:validation}, we show constraints on the compressed parameters from the analysis of the four \pone mocks, with the true values subtracted. As expected, our model accurately recovers the true cosmology for the \texttt{mpg-central} mock, since the emulator was trained on the \mpgadget suite and this simulation is at the center of the parameter space. It also recovers the correct cosmology for the \texttt{mpg-seed} mock, confirming that cosmic variance has a limited impact on the emulator predictions --- the only difference between the \texttt{mpg-central} and \texttt{mpg-seed} simulations is the initial distribution of Fourier phases. More importantly, the model recovers compressed parameters consistent with the true values for the \texttt{lyssa-central} and \texttt{sherwood} mocks, which are based on simulations run with different codes and 3 times better resolution than the \mpgadget simulations. 

These results indicate that potential emulator inaccuracies arising from the limited resolution of the \mpgadget simulations or differences between SPH and grid codes \cite{hydro_Chabanier2023} do not bias the inferred cosmological constraints beyond the $1\sigma$ level. However, inaccuracies of this kind could be absorbed by the IGM, metal, HCD, and resolution parameters, thereby altering their physical interpretation. We verify that, in all mock analyses considered above, the best-fitting values of the metal and HCD parameters are consistent with zero at the $1\sigma$ level, as expected given that the \pone mocks were generated with these contributions set to zero. Nevertheless, a definitive assessment of the ability of \texttt{cup1d} to recover the IGM, metal, HCD, and resolution parameters without bias would require a large suite of mocks systematically varying all of these parameters, which is beyond the scope of this work. We therefore treat these quantities as nuisance parameters during inference, reporting their values with the caveat that some parameters may take values without a direct physical interpretation.

In the right panel of \cref{fig:validation}, we display the results for the baseline analysis of the \texttt{mpg-central} mock and two alternative analyses: marginalizing over IGM parameters while fixing contamination and systematic parameters to their true values (orange contours), and only varying cosmological parameters (green contours). Marginalizing over contaminants and systematic effects increases the uncertainties on \deltastar and \nstar by 1.5 and 1.7, respectively, while also marginalizing over IGM parameters increases the uncertainties by 4.6 and 1.8. Hence, informative priors on IGM and nuisance parameters would substantially enhance the constraining power of \pone measurements.

%% file: results_journal.tex
\section{Results}
\label{sec:results}

In this section, we use \texttt{cup1d} to extract cosmological constraints from the analysis of DESI DR1 \pone measurements. We first assess the robustness of these constraints through a set of alternative data analyses in \cref{sec:results_robust}. We then present the best-fitting constraints on the compressed parameters in \cref{sec:results_cosmo}, followed by those on the IGM, metal, HCD, and resolution parameters in \cref{sec:results_nuisance}.


\subsection{Robustness tests}
\label{sec:results_robust}

In the baseline analysis, we derive constraints from \pone measurements obtained with the QMLE estimator using spectra with $\mathrm{SNR}>3$ per pixel in the forest region (see \cref{sec:data_obs}). To do so, we employ a model that combines an emulator capturing the dependence of \pone on cosmology and IGM physics (see \cref{sec:emulator}) and corrections for the impact of metal contamination, HCD contamination, and resolution systematics (\cref{sec:model}). We adopt a covariance matrix that includes statistical, systematic, and emulator uncertainties, with the statistical component inflated by 5\% to account for a potential underestimation of the errors (see \cref{sec:model}). In this section, we carry out a suite of alternative analyses to evaluate the sensitivity of the baseline constraints to variations in the redshift range and type of \pone measurements, the covariance matrix, the emulator, the cosmological parameters fixed during inference, and the model ingredients. 

All analyses, including the model optimization described in previous sections, were carried out under blinded conditions to mitigate human bias. We implemented the blinding by adding random offsets to the best-fitting compressed parameters from each analysis, fixing the random seed to ensure consistency across all analysis variants. These offsets were drawn from Gaussian distributions with zero mean and standard deviations of 0.05 and 0.01 for \deltastar and \nstar, respectively, matching the uncertainties obtained in one analysis of eBOSS measurements \cite{p1d_Chabanier2019}. After validating our methodology using mocks, we found that the \nstar blinding was smaller than its statistical uncertainty. This motivated an additional blinding step, in which we shift the posterior of all analyses so that its center aligns with that of \Planck~2018 $\Lambda$CDM constraints. The results were unblinded only after the model passed all consistency tests, and no further changes were made thereafter.

We quantify the consistency between cosmological constraints from the baseline and alternative analyses as follows
\begin{equation}
    \label{eq:consistency}
    \Delta\chi^2_\mathrm{cosmo} = 
    \frac{1}{1-\rho_\mathrm{base}^2} 
    \left[\frac{\left[\Delta(\deltastar)\right]^2}{\tilde{\sigma}^2_{\Delta^2_\star}} 
    - 2\rho_\mathrm{base} \frac{\Delta(\deltastar) \Delta \nstar} {\tilde{\sigma}_{\Delta^2_\star} \tilde{\sigma}_{n_\star}} 
    + \frac{(\Delta \nstar)^2} {\tilde{\sigma}^2_{n_\star}}\right],
\end{equation}
where the subindex base indicates the results for the baseline configuration, $\Delta(\deltastar) = \Delta^2_\star - \Delta^2_{\star, \mathrm{base}}$ and $\Delta \nstar = n_\star - n_{\star, \mathrm{base}}$ refer to the difference in the value of the compressed parameters between the baseline and an alternative analysis, and $\tilde{\sigma}^2_x = \mathrm{max}(\sigma^2_x, \sigma^2_{x, \mathrm{base}})$ denotes the maximum of the error from the baseline and the alternative analyses. We take the maximum rather than adding the errors in quadrature since these are correlated. In two dimensions, a $\Delta\chi^2$ of 0.96, 2.30, and 6.18 correspond to 0.5, 1, and $2\,\sigma$, respectively. We gather the results for all variations in \cref{tab:variations_model}, and discuss these in what follows.

\begin{table}[]
    \centering
    \begin{tabular}{lclccc}
        Variation  & \multicolumn{1}{c}{$\Delta$(\deltastar)} & \multicolumn{1}{c}{$\Delta$\nstar} & $\Delta\chi^2_\mathrm{cosmo}$ & \multicolumn{1}{c}{$\chi^2_\mathrm{fit}$} & \multicolumn{1}{c}{Prob. fit}\\ 
        \hline
Fiducial                                    & $\;\;\,0.000^{+0.032}_{-0.033}$ & $\;\;\,0.000^{+0.019}_{-0.019}$ & 0.00  & 655.0  & 0.22                \\
Data: $z \leq 3.4$                          & $\;\;\,0.032^{+0.046}_{-0.040}$ & $\;\;\,0.014^{+0.030}_{-0.029}$ & 0.71  & 358.1  & 0.44                \\
Data: $z \geq 2.6$                          & $-0.047^{+0.042}_{-0.043}$      & $-0.005^{+0.023}_{-0.029}$      & 1.53  & 570.7  & 0.10                \\
Data: w/ low SNR                            & $\;\;\,0.032^{+0.035}_{-0.035}$ & $-0.004^{+0.023}_{-0.022}$      & 0.99  & 685.1  & 0.06                \\
Data: FFT                                   & $\;\;\,0.016^{+0.037}_{-0.032}$ & $\;\;\,0.033^{+0.023}_{-0.022}$ & 2.05  & 867.7  & $1.7\times10^{-7}$  \\
Cov: uncorr syst                            & $-0.006^{+0.026}_{-0.027}$      & $\;\;\,0.002^{+0.016}_{-0.015}$ & 0.03  & 636.7  & 0.40                \\
Cov: w/o 5\% err                            & $-0.005^{+0.029}_{-0.029}$      & $\;\;\,0.003^{+0.016}_{-0.016}$ & 0.01  & 719.5  & $6.5\times10^{-3}$  \\
Cov: w/o emu err                            & $-0.012^{+0.025}_{-0.025}$      & $\;\;\,0.018^{+0.017}_{-0.015}$ & 0.99  & 678.7  & 0.08                \\
Cov: emu diag                               & $-0.013^{+0.027}_{-0.028}$      & $\;\;\,0.019^{+0.019}_{-0.015}$ & 0.81  & 569.1  & 0.96                \\
Cov: emu block diag                         & $\;\;\,0.000^{+0.035}_{-0.034}$ & $\;\;\,0.021^{+0.025}_{-0.020}$ & 0.81  & 675.2  & 0.09                \\
Cov: emu infl 25\%                          & $\;\;\,0.005^{+0.029}_{-0.030}$ & $-0.002^{+0.017}_{-0.017}$      & 0.06  & 644.2  & 0.32                \\
Syst: 5\% eBOSS                             & $-0.002^{+0.029}_{-0.030}$      & $-0.008^{+0.017}_{-0.017}$      & 0.19  & 656.0  & 0.21                \\
Emulator: lace-lyssa                        & $\;\;\,0.045^{+0.056}_{-0.054}$ & $-0.013^{+0.018}_{-0.018}$      & 1.15  & 584.4  & 0.89                \\
Cosmo: $\omega_0\omega_a$CDM                & $-0.003^{+0.030}_{-0.029}$      & $\;\;\,0.003^{+0.016}_{-0.016}$ & 0.03  & 654.8  & 0.22                \\
Cosmo: $h=0.74$                             & $-0.006^{+0.029}_{-0.030}$      & $\;\;\,0.002^{+0.017}_{-0.017}$ & 0.01  & 655.0  & 0.22                \\
Cosmo: $\sum m_\nu=0.3$ eV                  & $-0.016^{+0.032}_{-0.033}$      & $\;\;\,0.008^{+0.020}_{-0.018}$ & 0.25  & 653.7  & 0.23                \\
Cosmo: high $\Omega_\mathrm{cdm}h^2$        & $\;\;\,0.001^{+0.028}_{-0.030}$ & $-0.001^{+0.017}_{-0.016}$      & 0.01  & 653.9  & 0.23                \\
Cosmo: low $\Omega_\mathrm{cdm}h^2$         & $-0.008^{+0.029}_{-0.029}$      & $\;\;\,0.005^{+0.017}_{-0.016}$ & 0.08  & 654.6  & 0.22                \\
IGM: $n_z=6$                                & $\;\;\,0.019^{+0.032}_{-0.030}$ & $-0.010^{+0.017}_{-0.018}$      & 0.89  & 651.6  & 0.18                \\
IGM: larger priors                          & $-0.014^{+0.033}_{-0.032}$      & $\;\;\,0.000^{+0.021}_{-0.020}$ & 0.17  & 654.3  & 0.23                \\
HCD: LLS $n_z=4$                            & $-0.004^{+0.030}_{-0.030}$      & $\;\;\,0.002^{+0.018}_{-0.017}$ & 0.01  & 654.7  & 0.21                \\
HCD: w/ $f_{\mathrm{const}}^{\mathrm{HCD}}$ & $\;\;\,0.004^{+0.032}_{-0.032}$ & $\;\;\,0.012^{+0.022}_{-0.019}$ & 0.33  & 654.4  & 0.22                \\
HCD: only DLAs                              & $\;\;\,0.082^{+0.020}_{-0.022}$ & $-0.052^{+0.012}_{-0.013}$      & 13.70 & 664.5  & 0.18                \\
HCD: BOSS                                   & $\;\;\,0.079^{+0.023}_{-0.025}$ & $-0.048^{+0.013}_{-0.014}$      & 12.43 & 698.9  & 0.04                \\
Metals: opt thin                            & $\;\;\,0.004^{+0.030}_{-0.031}$ & $\;\;\,0.003^{+0.018}_{-0.017}$ & 0.13  & 725.5  & $5.7\times10^{-3}$  \\
Metals: no H-Si decorr                      & $-0.041^{+0.034}_{-0.030}$      & $\;\;\,0.003^{+0.019}_{-0.019}$ & 1.41  & 1388.1 & $1.2\times10^{-58}$ \\
Metals: no SiII-SiII                        & $\;\;\,0.079^{+0.029}_{-0.026}$ & $-0.014^{+0.015}_{-0.016}$      & 6.70  & 747.6  & $1.0\times10^{-3}$  \\
Metals: BOSS                                & $\;\;\,0.063^{+0.031}_{-0.030}$ & $-0.002^{+0.016}_{-0.017}$      & 4.43  & 1539.6 & $7.1\times10^{-76}$ \\
Metals: Ma+2026                             & $\;\;\,0.074^{+0.028}_{-0.026}$ & $-0.019^{+0.014}_{-0.017}$      & 5.71  & 991.6  & $5.1\times10^{-18}$ \\
    \end{tabular}
    \caption{    
    Consistency of cosmological constraints between the baseline and alternative analyses. Columns 2 and 3 show differences in the compressed parameters relative to the baseline analysis, with the median as the central value and the 16th/84th percentiles as uncertainties. Column 4 quantifies the consistency between the cosmological constraints from each variation and the baseline, as defined in \cref{eq:consistency}, where $\Delta\chi^2_\mathrm{cosmo}\simeq2.30$, 6.18, and 11.83 correspond to consistency at the 1, 2, and $3,\sigma$ level, respectively. Columns 5 and 6 report the $\chi^2$ of the fit and its probability. Some of the variations use a different number of free parameters, so it is more informative to compare fit probabilities.
}
    \label{tab:variations_model}
\end{table}


\subsubsection{Measurements and covariance matrix}
\label{sec:results_robust_data}

\begin{figure}
    \centering
    \includegraphics[width=0.495\linewidth]{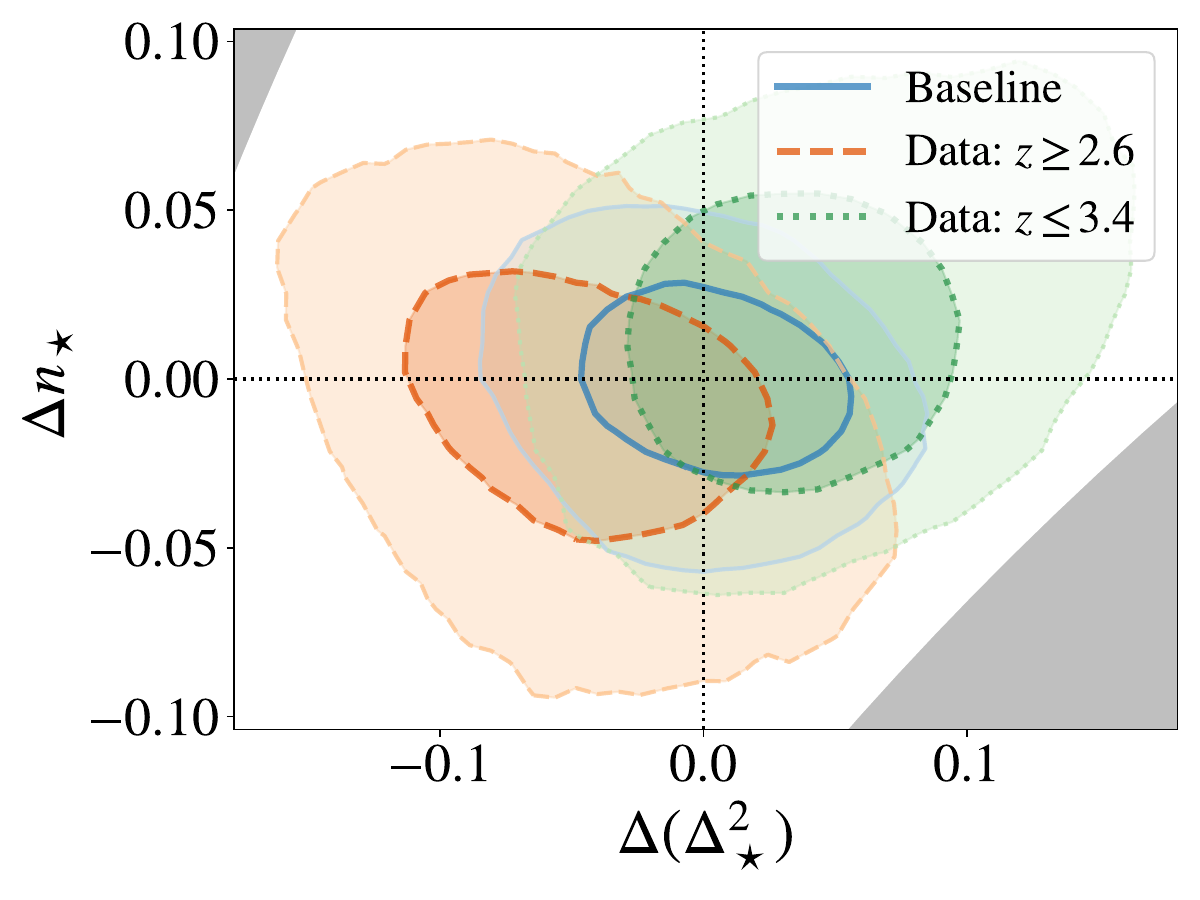}
    \includegraphics[width=0.495\linewidth]{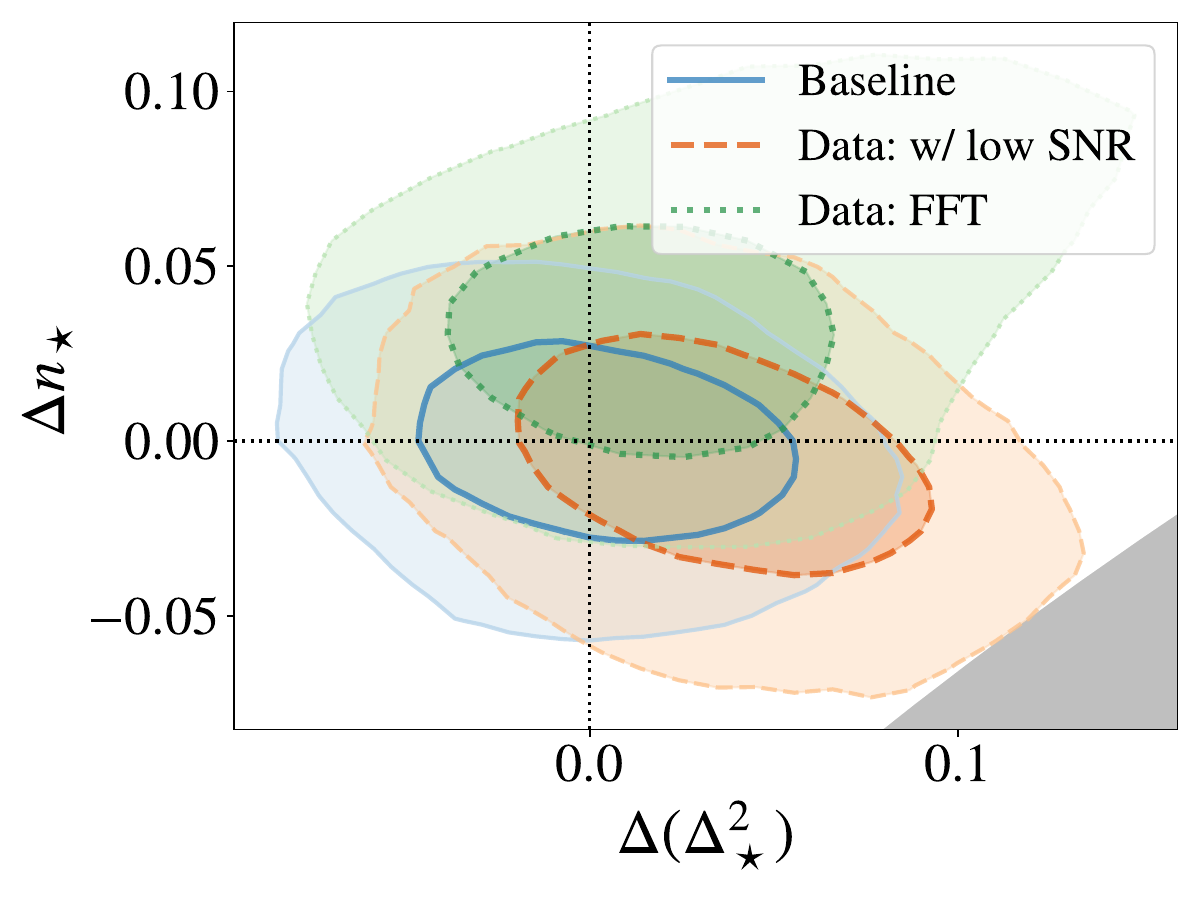}
    \caption{
    Dependence of cosmological constraints on the redshift range (left) and type of measurements (right). Blue contours show the results of the baseline analysis in both panels. In the left panel, the orange and green contours show the results when only considering measurements at $z\ge2.6$ or $z\leq3.4$, respectively. In the right panel, orange contours show the results for QMLE measurements including low-SNR quasars, while green contours do so for FFT measurements with high-SNR quasars. Inner and outer contours show the 68 and 95\% credible intervals, after subtracting the best-fitting values from the baseline analysis. The gray shaded are indicates the region of the parameter space excluded by the priors.
}
    \label{fig:variations_data}
\end{figure}

In the left panel of \cref{fig:variations_data}, we compare the baseline constraints (blue) with those obtained when restricting the analysis to \pone measurements at either $z \geq 2.6$ (orange) or $z \leq 3.4$ (green). The cosmological constraints from these two redshift splits and the baseline results are consistent at the $1\,\sigma$ level, which is noteworthy given that recent eBOSS re-analyses reported significant discrepancies between constraints derived with and without including \pone measurements from $z < 2.6$ \cite{Fernandez2024_priya, Walther2025_lyssa}. 

In the right panel, the orange contours show the results obtained when analyzing QMLE measurements from spectra with $\mathrm{SNR}>0.3$ per pixel in the forest region, instead of the $\mathrm{SNR}>3$ threshold used in the baseline analysis. The cosmological constraints from the low-SNR and baseline analyses are consistent at the $0.5\,\sigma$ level, while the fit probabilities are 6 and 22\% for the first and second, respectively. The goodness of fit is likely worse for this variation because noise-related systematics are larger for low-SNR spectra \cite{Karacayli2025_p1d_dr1}. Furthermore, the value of the HCD parameters for this variation are slightly larger than for the baseline analysis, suggesting that low-SNR spectra are more strongly affected by HCD contamination than high-SNR spectra, consistent with the degraded performance of the DESI DLA finder at low SNR \cite{Brodzeller2025_desiDLA}. Taken together, these findings supports the choice of using the high-SNR sample in the baseline analysis. 

The green contours correspond to the analysis of FFT-based measurements using forests with $\mathrm{SNR}>3$ per pixel. While the resulting constraints are consistent with those of the baseline analysis at the $1\,\sigma$ level, the associated fit probability is lower by six orders of magnitude. Achieving a comparable fit probability for the FFT and QMLE analyses requires inflating the statistical uncertainties by 18\% for the FFT variation, compared to only by 5\% in the baseline. This discrepancy further motivates our choice of QMLE measurements for the baseline analysis.

\begin{figure}
    \centering
    \includegraphics[width=0.495\linewidth]{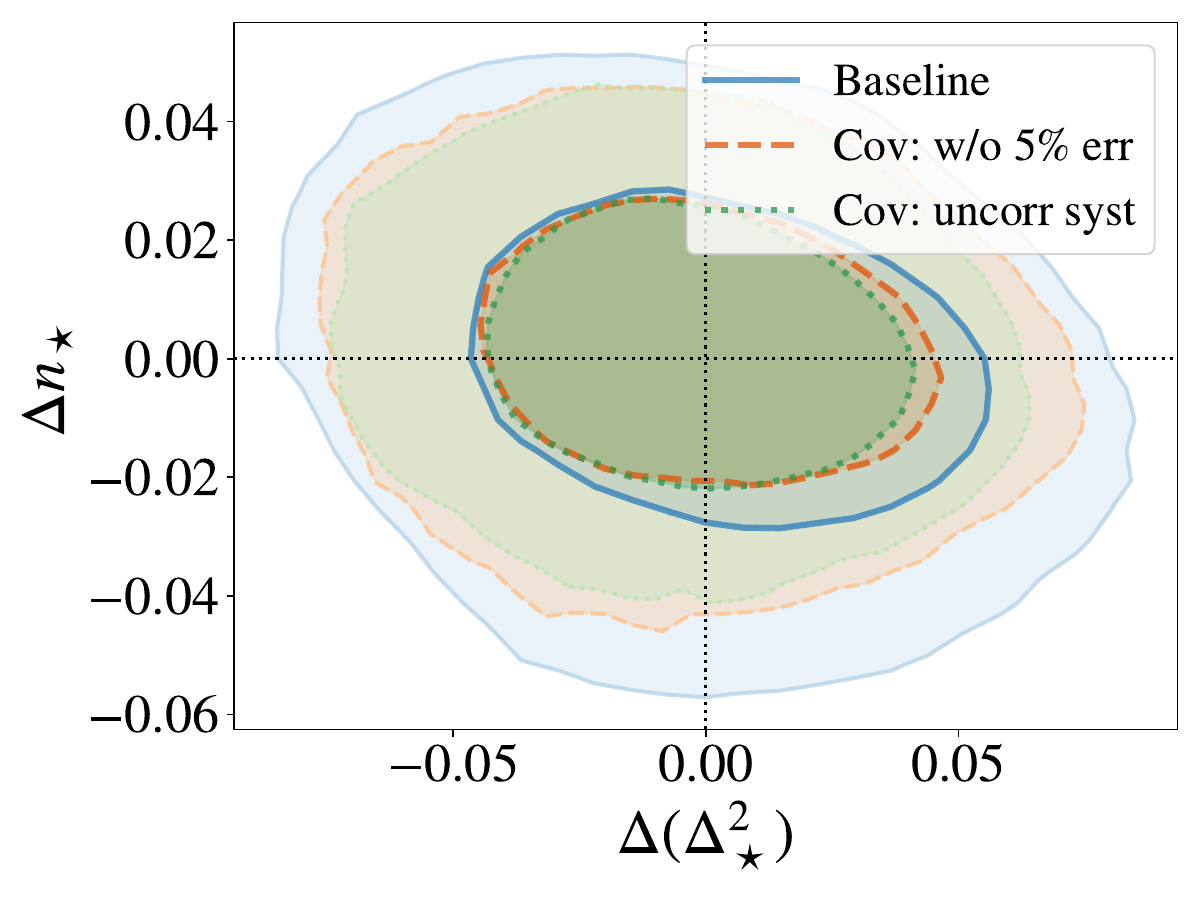}
    \includegraphics[width=0.495\linewidth]{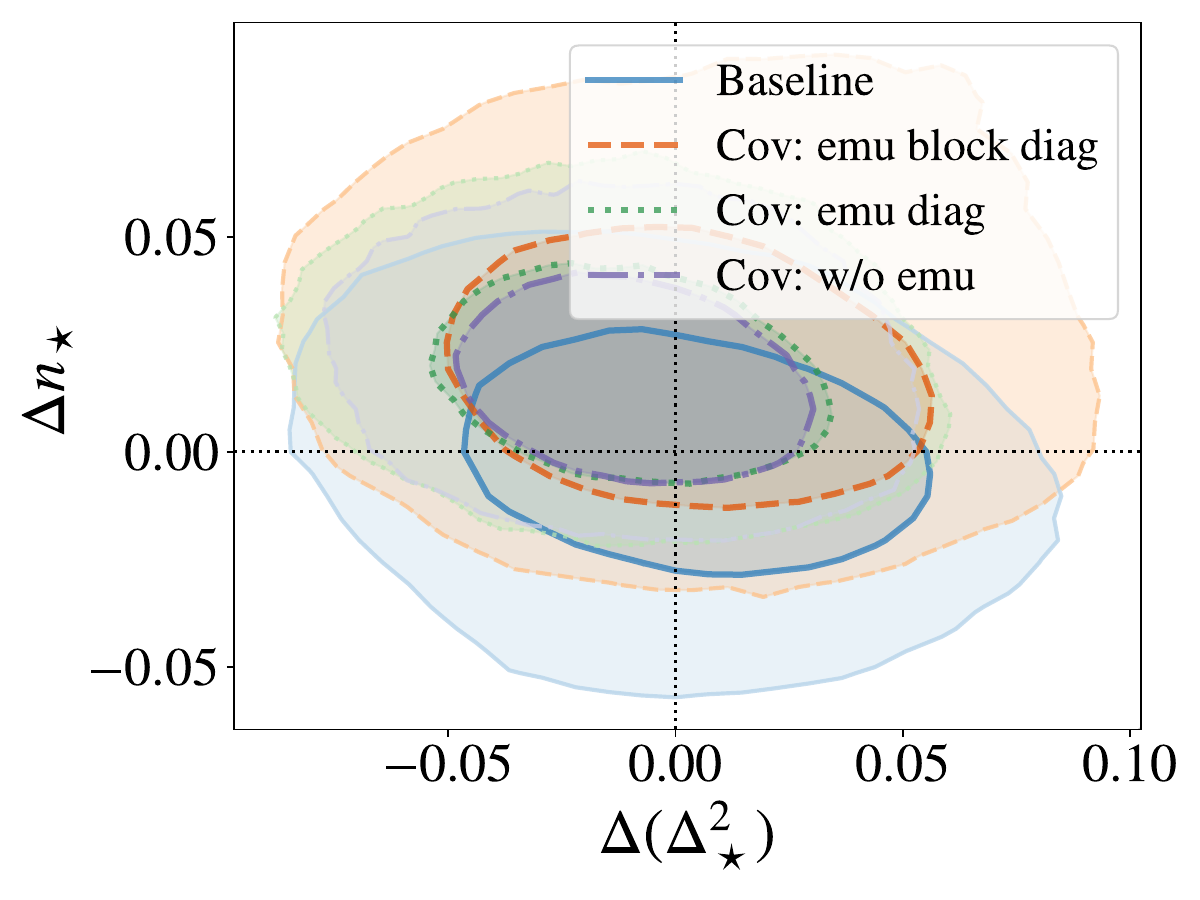}
    \caption{Same as \cref{fig:variations_data}, but for variations in the treatment of uncertainties. In the left panel, the orange contours show the results obtained without inflating the statistical errors by 5\%, while the green contours correspond to the case where systematic errors are added to the covariance matrix while assuming no correlation across scales, instead of full correlation as in the baseline setup. In the right panel, the orange, green, and purple contours show the results obtained when adding emulator errors to the covariance matrix while assuming no redshift correlation, no correlation across either redshifts or scales, and no emulator errors altogether, respectively.
}
    \label{fig:variations_cov}
\end{figure}

In the left panel of \cref{fig:variations_cov}, the orange contours show the results obtained when omitting the 5\% inflation of the statistical errors, while the green contours correspond to the case where systematic errors are added to the covariance matrix assuming no correlation across scales, instead of full correlation as in the baseline setup (see \cref{sec:data}). The cosmological constraints obtained from these variations are fully consistent with those of the baseline analysis. The fit probability decreases from 22 to 0.6\% when the statistical uncertainties are not inflated, and increases to 40\% when systematic errors are assumed to be uncorrelated. Since systematic uncertainties are unlikely to be fully correlated across scales, a more realistic treatment of these correlations would likely yield a higher fit probability. We also find that the cosmological constraints improve by only $\simeq 10\%$ when the error bars are not inflated, indicating that this pragmatic choice does not result in a significant loss of constraining power.

In the right panel, the orange, green, and purple contours show the results obtained when we include emulator errors following different approaches: uncorrelated across redshift bins, uncorrelated across both redshifts and scales, and excluded altogether, respectively. The constraints remain consistent with the baseline analysis at the $1\,\sigma$ level in all cases, with the worse agreement occurring when we ignore emulator errors and the highest when we assume no correlation across redshift bins. The fit probability rises to 96\% when we include emulator errors but neglect correlations across both redshift and scale, underscoring the importance of accurately modeling the correlation structure of emulator errors.

Finally, we perform an alternative analysis in which we decrease by 5\% the value of \pone for all redshifts and scales\footnote{For consistency, we also decrease the statistical errors by the same factor.}, mimicking the systematic offset observed between eBOSS and DESI measurements \cite{Ravoux2023, Karacayli2024_edr, Ravoux2025, Karacayli2025_p1d_dr1}. As the origin of this discrepancy remains unknown, this test assesses its potential impact on the compressed parameters. We find that both the cosmological constraints and the fit probabilities are fully consistent between this variation and the baseline analysis, with the mean flux parameters absorbing the 5\% shift. 


\subsubsection{Emulator}
\label{sec:results_robust_emulator}

\begin{figure}
    \centering
    \includegraphics[width=0.6\linewidth]{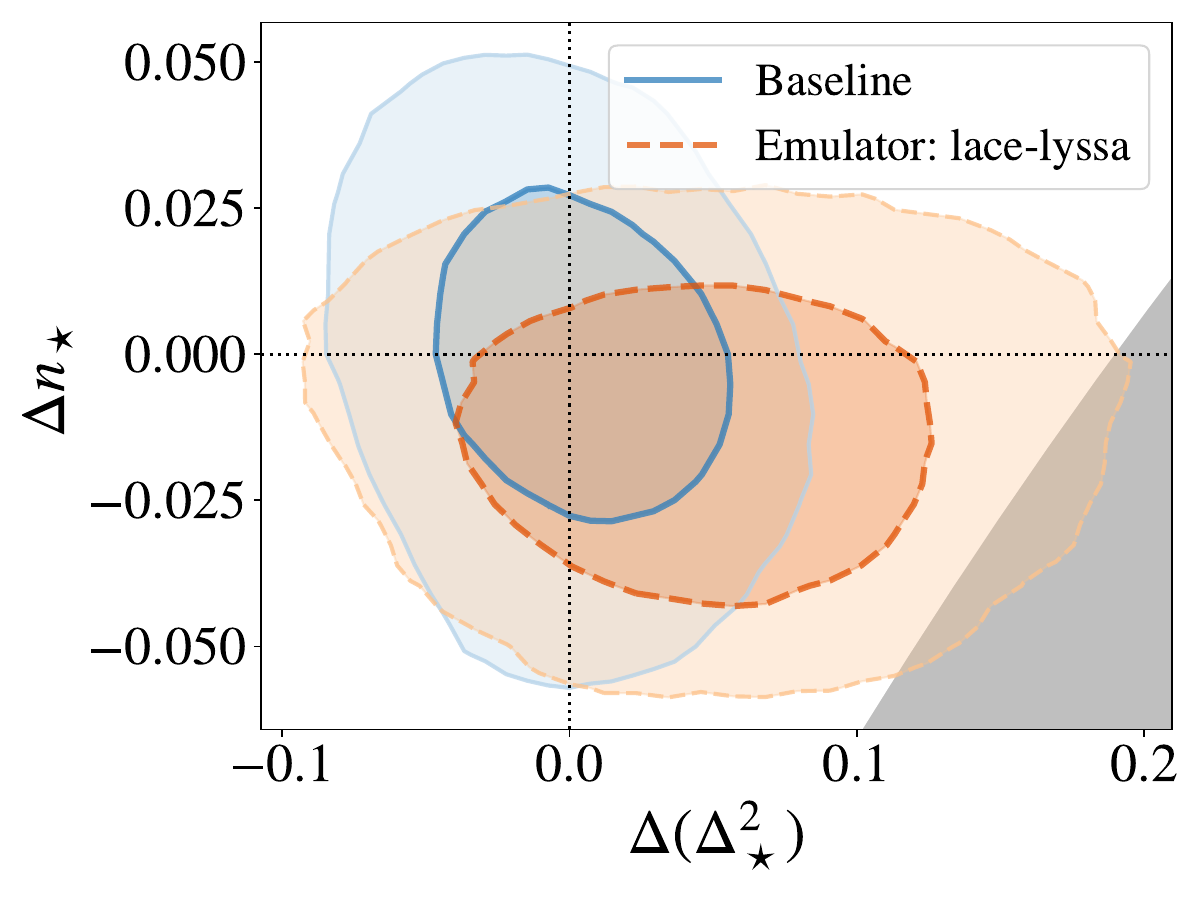}
    \caption{
    Same as \cref{fig:variations_data}, but for variations in the \pone emulator. The blue and orange contours show the results for the \lacempg and \lacelyssa emulators, respectively, trained on a suite of \mpgadget \cite{Pedersen2021} and \texttt{lyssa} simulations \cite{Walther2025_lyssa}. The gray shaded are indicates the region of the parameter space excluded by the priors for the \lacempg emulator.
}
    \label{fig:variations_emu}
\end{figure}

In \cref{fig:variations_emu}, the orange contours show the constraints obtained with an alternative emulator trained on the \texttt{lyssa} simulations (\lacelyssa; see \cref{app:lace-lyssa}). These simulations differ from the \mpgadget ones used to train \lacempg in resolution, volume, and hydrodynamical code. Despite these differences, the cosmological constraints from the \lacempg and \lacelyssa emulators agree at the $\simeq 0.5\,\sigma$ level. The main discrepancy is a 70\% larger uncertainty on \deltastar for \lacelyssa, whose possible origin we discuss in \cref{app:lace-lyssa}.

\subsubsection{Cosmological parameters fixed during inference}
\label{sec:results_robust_cosmo}

As explained in \cref{sec:model_multiple}, we carry out the cosmological analysis while holding fixed the Hubble parameter and the physical densities of cold dark matter and baryons to the best-fitting \Planck values. We follow this approach for computational efficiency. It is motivated by the fact that the Universe is effectively Einstein–de Sitter at the redshifts relevant to the \lya forest, and thus the expansion and growth histories are practically the same at these redshifts independently of the value of the fixed parameters. Furthermore, it is important to note that we set constraints on the compressed parameters defined in velocity units, and thus their value does not depend on the expansion history assumed as it would be the case if defined in comoving units. On the other hand, holding fixed the physical densities is equivalent to keeping fixed the matter transfer functions, which could also lead to a bias in the results. Nevertheless, the physical densities are measured with exquisite precision by CMB experiments. Throughout the remainder of this section, we check whether the value of the compressed parameters is sensitive to realistic variations in the value of the fixed parameters. In \cref{app:compressed}, we present a more physically motivated perspective on the sensitivity of the compressed parameters to variations in the fixed parameters.

\begin{figure}
    \centering    
    \includegraphics[width=0.495\linewidth]{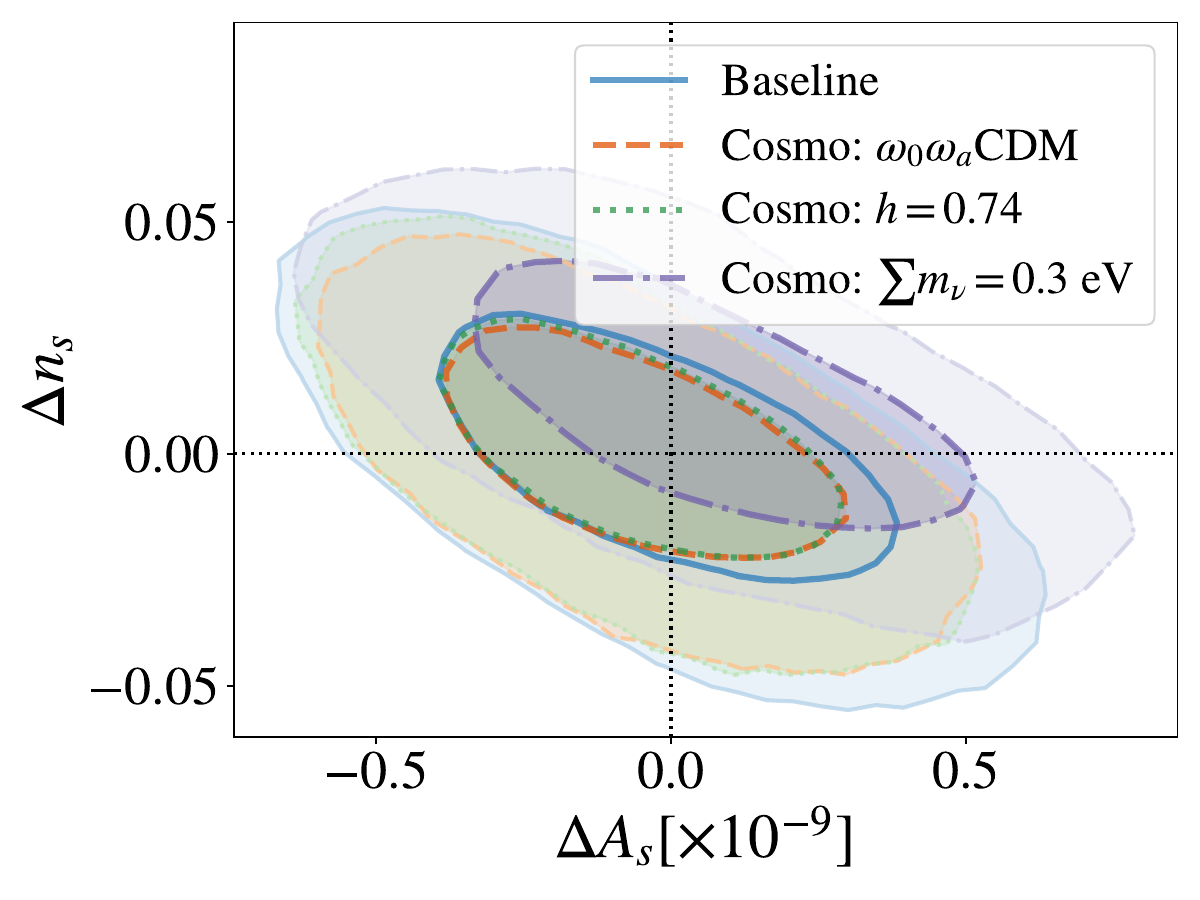}
    \includegraphics[width=0.495\linewidth]{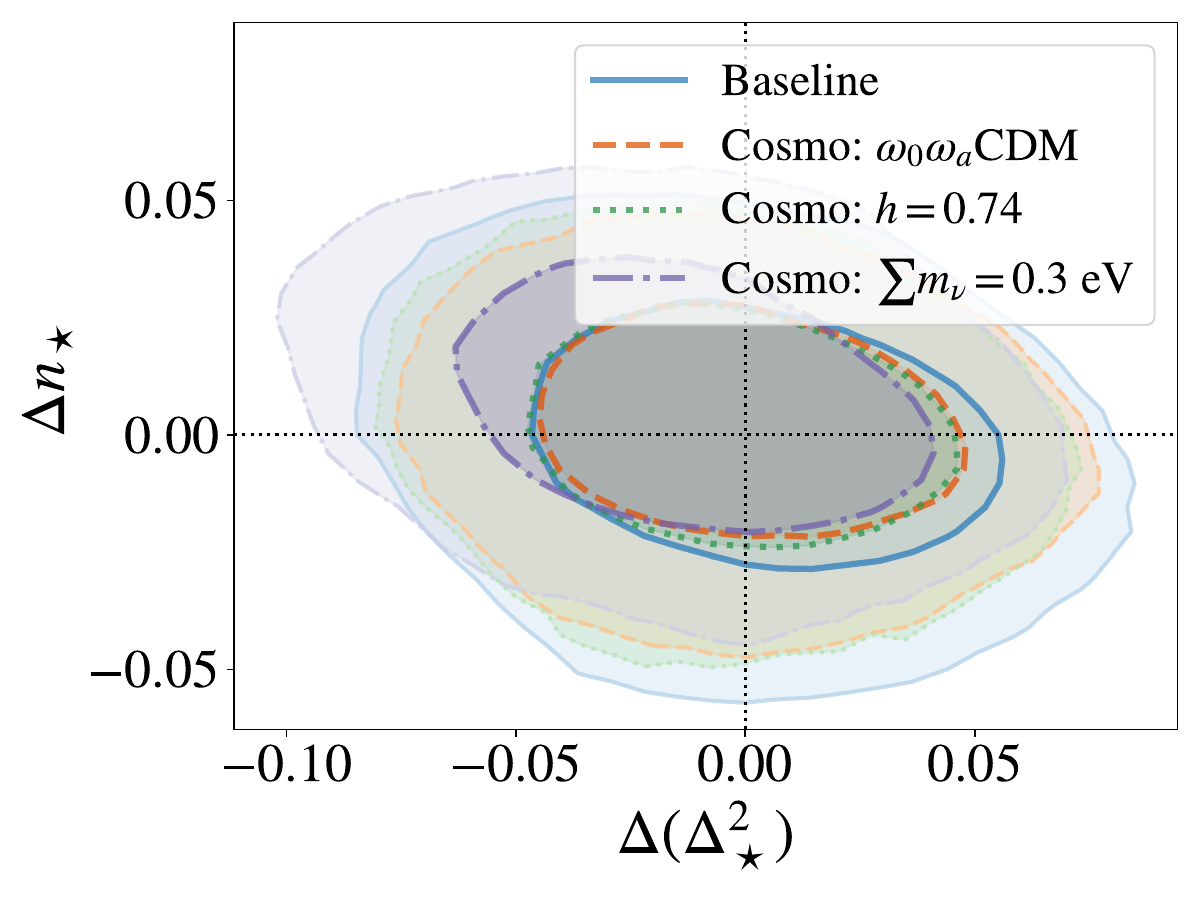}
    
    \includegraphics[width=0.495\linewidth]{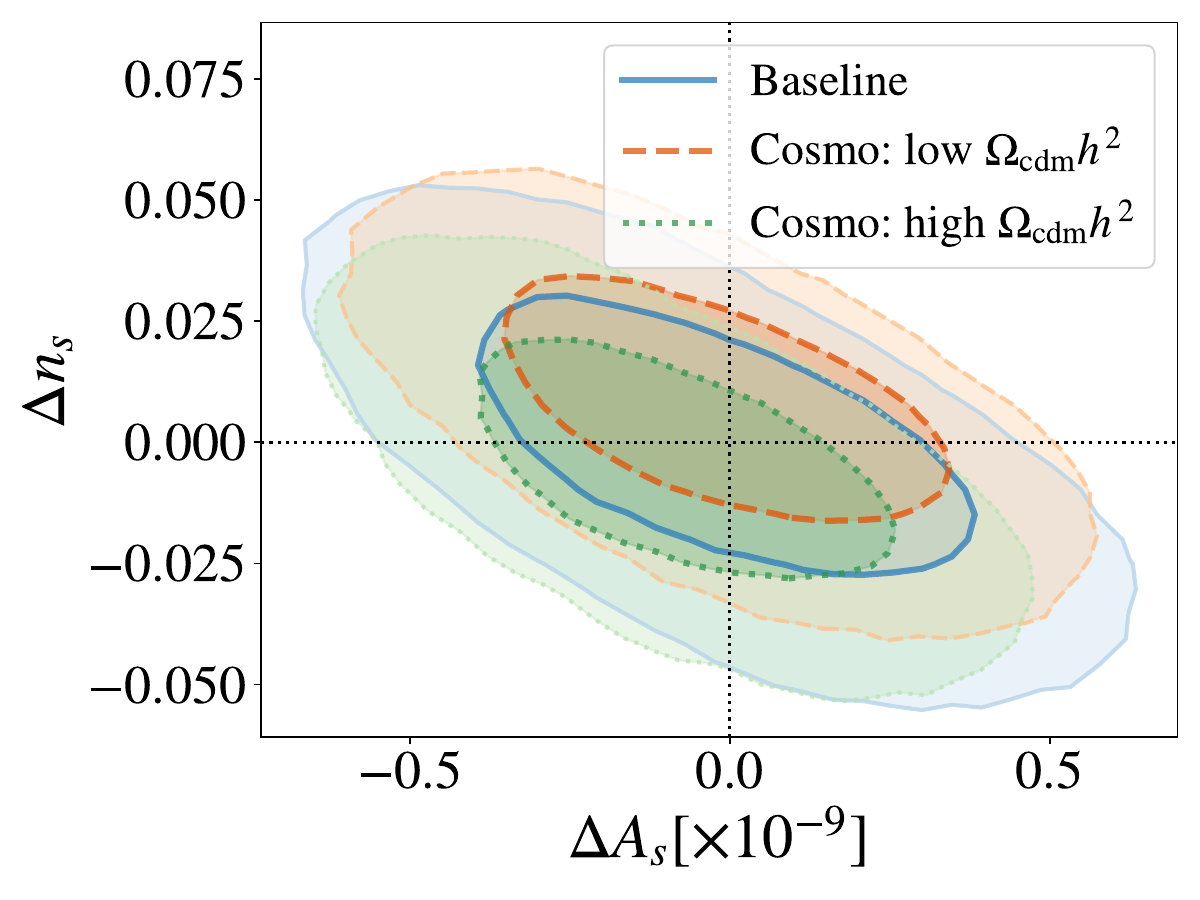}
    \includegraphics[width=0.495\linewidth]{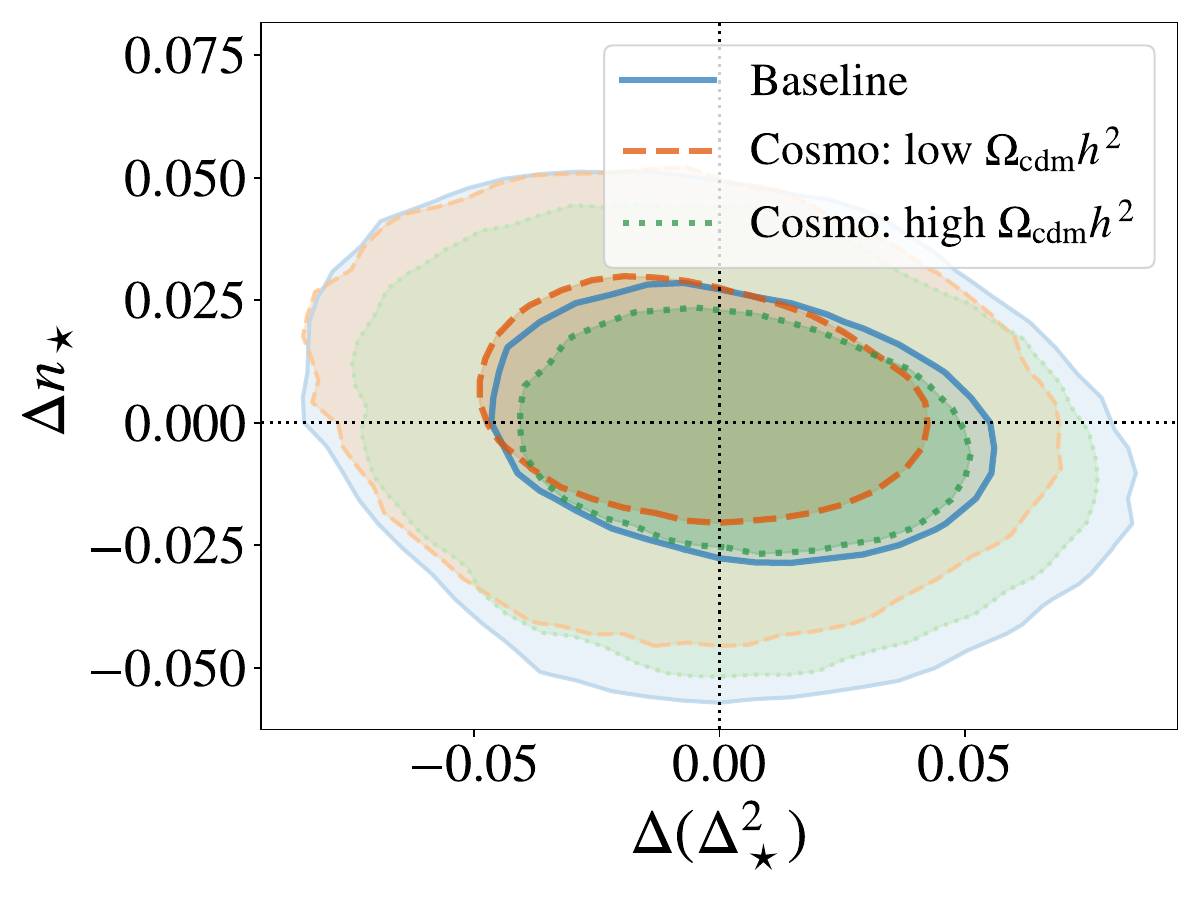}
    \caption{
    Dependence of cosmological constraints on the value of the cosmological parameters fixed during inference for computational efficiency. The left panels shows results in the $A_s – n_s$ plane, while the right panels displays those in the $\deltastar – \nstar$ plane. The blue contours show baseline results, which employ the best-fitting \Planck 2018 $\Lambda$CDM cosmology \cite{Planck2018}. In the top panels, the orange contours correspond to a variation assuming the DESI-DR2 + ACT $w_0w_\mathrm{a}$CDM cosmology \cite{Garcia_Quintero2025_desiact}, while the green and purple contours display the results for analyses assuming a \Planck~2018 $\Lambda$CDM cosmology with either a dimensionless Hubble constant of $h = 0.74$ consistent with local-universe measurements or sum of neutrino masses of $\sum m_\nu = 0.3\,\mathrm{eV}$, respectively. In the bottom figures, the orange and green contours show the results when assuming a \Planck 2018 cosmology with physical cold dark matter density decreased or increased by $3\sigma$, respectively.
}
    \label{fig:variations_cosmo}
\end{figure}

In the left and right panels of \cref{fig:variations_cosmo}, we illustrate the sensitivity of the constraints in the $A_\mathrm{s}$–$n_\mathrm{s}$ and $\deltastar$–$\nstar$ planes, respectively, to the value of the fixed parameters. The blue contours correspond to the baseline setup, which adopts the \Planck~2018 $\Lambda$CDM cosmology \cite{Planck2018}. In the top panels, the green contours display the results when using a dimensionless Hubble constant of $h=0.74$ --- a value consistent with local-universe measurements (e.g.; \cite{Casertano2025_H0tension}) --- instead of $h=0.6777$ while holding fixed the physical densities, whereas the orange contours show the results when using the best-fitting evolving dark energy cosmology from the joint analysis of DESI-DR2 BAO measurements and ACT-DR6 CMB data \cite{Garcia_Quintero2025_desiact}. Both variations effectively test the impact of changing the expansion and growth history on the value of the compressed parameters since the value of the physical densities for the second variation is compatible with the one from \Planck. As we can see, the constraints from both variations are fully consistent with the baseline analysis in both planes. This is because the differences in the expansion history, growth history, and matter transfer function of the three cosmologies at $z=2.2$ are smaller than 1\%.

The purple contours show the results when assuming a \Planck~2018 cosmology with sum of neutrino masses equal to $\sum m_\nu = 0.3\,\mathrm{eV}$ --- a value significantly larger than current large-scale structure constraints --- while varying $h$ from 0.6766 to 0.6843 in order to hold fixed $\Omega_\mathrm{m}$. In this case, the differences in the Hubble parameter and the logarithmic growth at $z=2.2$ are also around 1\%. On the other hand, while the transfer function of cold dark matter and baryons are the same in both cosmologies, the difference in the matter transfer function at $k=1\iMpc$ reaches 8\%. As a result, the posterior gets significantly shifted in the $A_\mathrm{s}$–$n_\mathrm{s}$ plane compared to the baseline analysis, while it remains fully consistent in the $\deltastar$–$\nstar$ plane, supporting the use of the compressed parameterization.

In the bottom panels, the orange and green contours represent analyses in which we decrease or increase the fixed value of the physical cold dark matter density by $3\sigma$ of \Planck while holding fixed the value of the Hubble constant. The differences in the Hubble parameter and the logarithmic growth at $z=2.2$ between the three cosmologies are below 1\%, but the differences in the matter transfer function at this redshift and $k=1\iMpc$ reach 2\%. Due to these differences, the posteriors move in the $A_\mathrm{s}$–$n_\mathrm{s}$ plane compared to the ones of the baseline analysis, while these remain largely consistent in the $\deltastar$–$\nstar$ plane.


\subsubsection{Contaminants}

As discussed in \cref{sec:model_single}, fitting the DESI data requires a substantially more detailed treatment of metal and HCD contamination than that adopted in previous \pone analyses. In this section, we examine how the cosmological constraints respond to alternative modeling choices for these contaminants. Consequently, the constraints from the following variations are not expected to necessarily agree with those from the baseline analysis.

\begin{figure}
    \centering
    \includegraphics[width=0.495\linewidth]{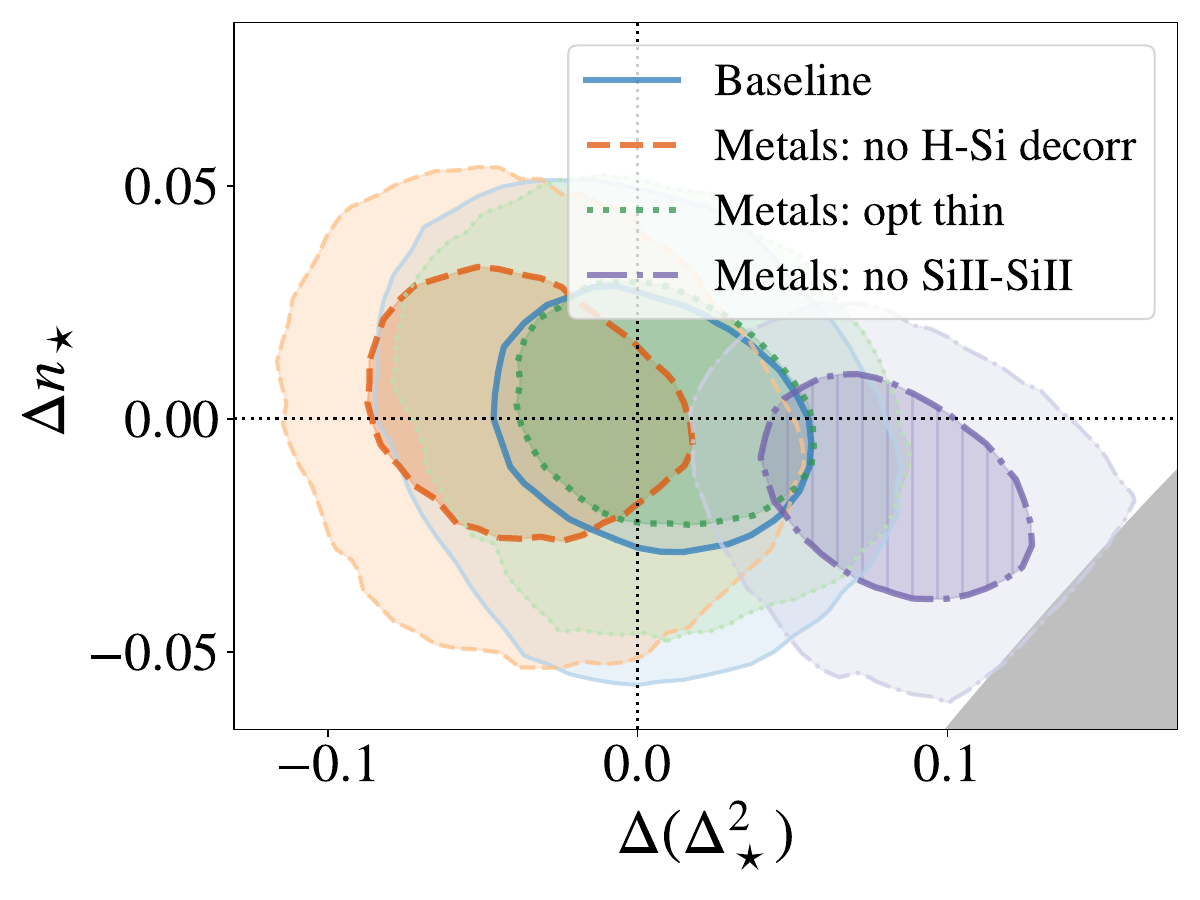} \includegraphics[width=0.495\linewidth]{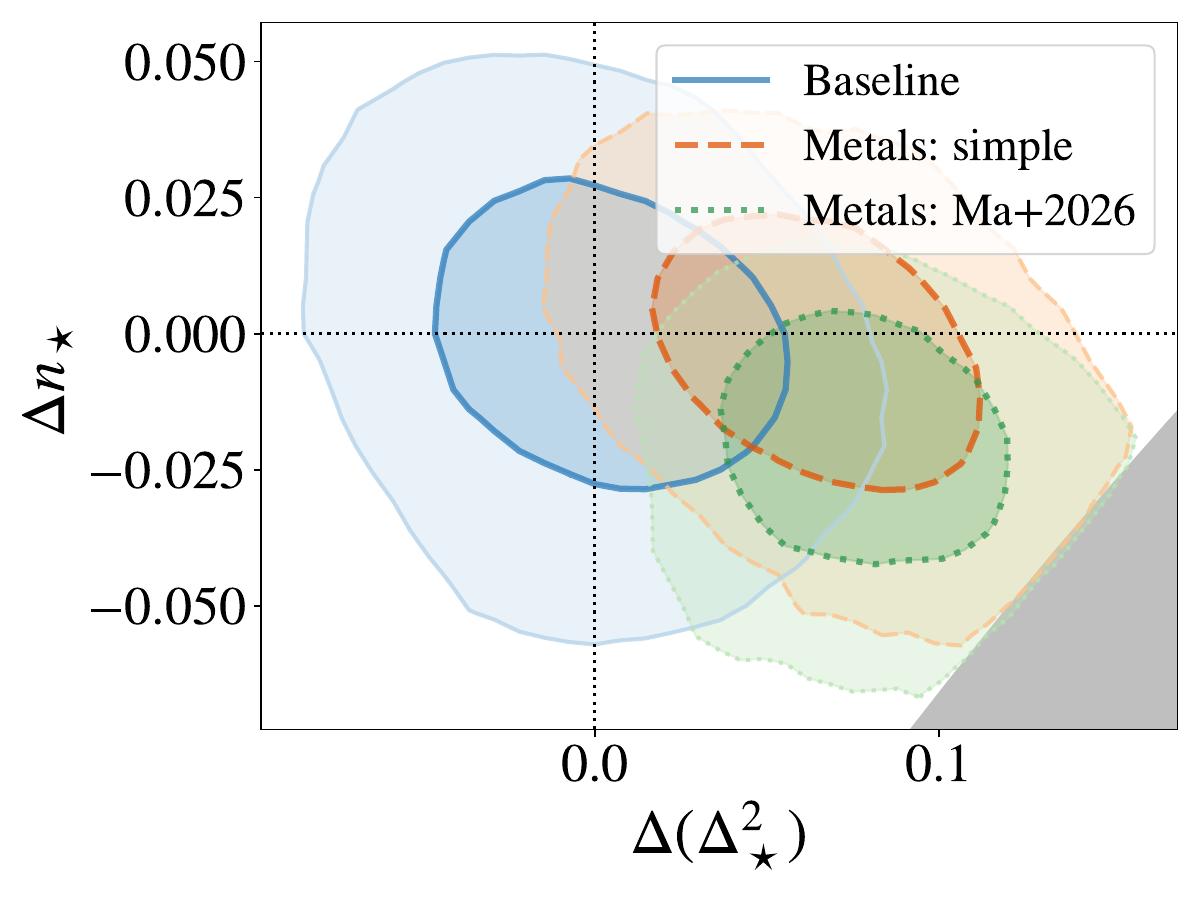}
    \caption{
    Same as \cref{fig:variations_data}, but showing the sensitivity of cosmological constraints to the treatment of the metal contamination. In the left panel, the orange, green, and purple contours show the results when omitting the small-scale decorrelation between H and Si, deviations from the optically-thin limit, or \siisii contamination, respectively. In the right panel, the orange contours show the results for the simplistic metal model employed in previous analyses of SDSS data \cite{McDonald2006, palanque-delabrouille2015ConstraintNeutrinoMasses, p1d_Chabanier2019, Fernandez2024_priya, Walther2025_lyssa}, while the blue contours displays the results for the metal model recently presented in \cite{Ma2025_metals}. 
}
    \label{fig:variations_metals}
\end{figure}

In the left panel of \cref{fig:variations_metals}, we study the sensitivity of the constraints to the omission of individual components of the metal model. The orange contours display the results for a variation in which the small-scale damping of metal contamination --- arising from the decorrelation between the hydrogen and silicon distributions on small scales --- is neglected. The green contours show the results obtained when adopting the optically thin approximation, rather than allowing for deviations from it through two free parameters as in the baseline analysis. The purple contours correspond to the case where \siisii contamination is excluded. Omitting the first two components does not shift the cosmological constraints more than $1\,\sigma$, but neglecting the \siisii term alters the cosmological constraints by more than $2\,\sigma$. Moreover, the fit probability decreases from 0.22 for the baseline analysis to $1.2\times10^{-58}$, $5.7\times10^{-3}$, and $1.0\times10^{-3}$ for the first, second, and third variation, respectively, signaling that all components of the metal model are essential for an accurate description of the data.

In the right panel, the orange contours correspond to a simplified metal model adopted in multiple analyses of SDSS data \cite{McDonald2006, palanque-delabrouille2015ConstraintNeutrinoMasses, p1d_Chabanier2019, Fernandez2024_priya, Walther2025_lyssa}, which includes a single parameter to control the amplitude of \lyasiii contamination and another for \lyasii. The green contours show the results obtained using a more recent model that employs four parameters to describe \lyasiii contamination \cite{Ma2025_metals}. For a fairer comparison with our baseline setup, we allow the parameters of these variations to evolve with redshift as a power law. In both cases, the inferred values of the cosmological parameters are shifted significantly relative to the baseline result, likely due to the omission of \siisii contamination for the first model and both the omission of both \siisii and \lyasii for the second. As a result, these metal models are strongly disfavored by the data, with fit probabilities of $7.1\times10^{-76}$ and $5.1\times10^{-18}$ for the first and second, respectively.

\begin{figure}
    \centering
    \includegraphics[width=0.495\linewidth]{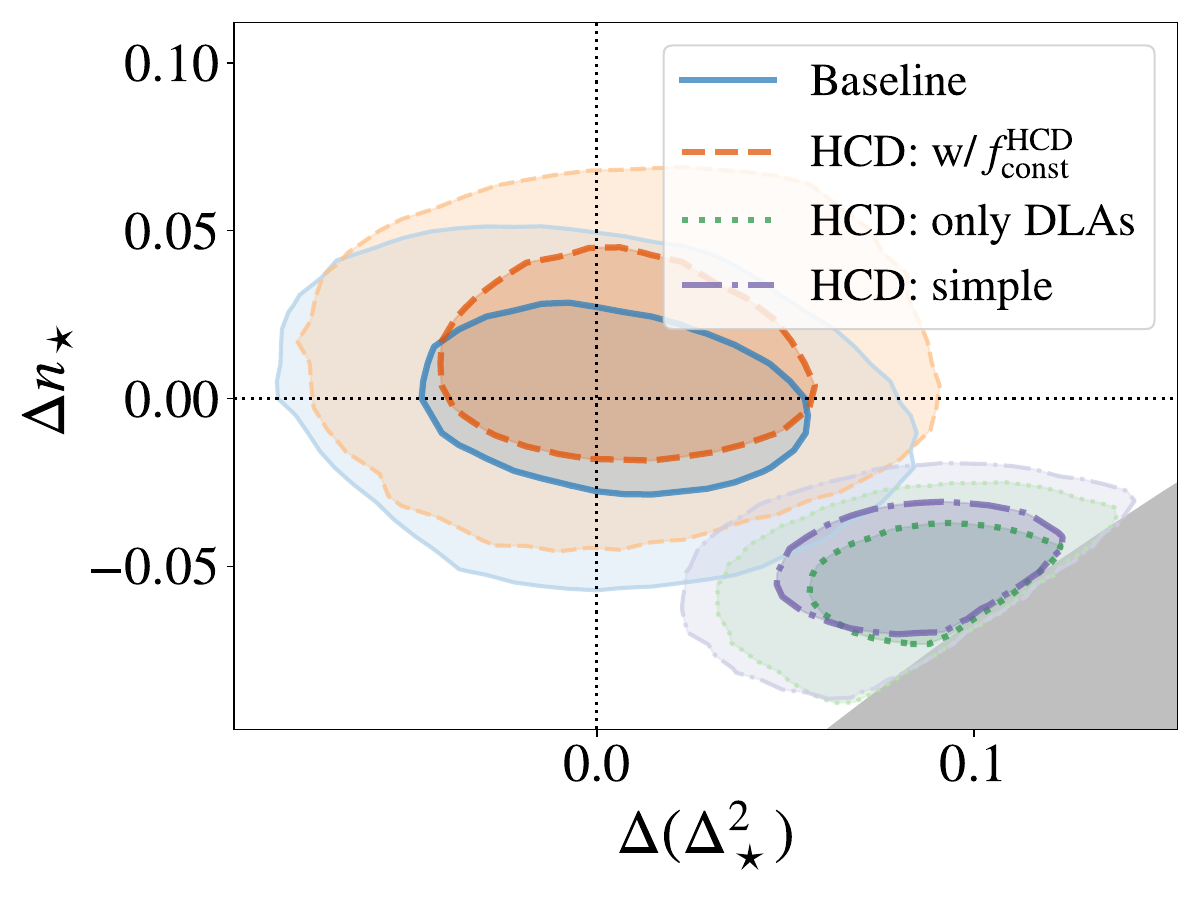} 
    \caption{
    Same as \cref{fig:variations_data}, but showing the sensitivity of cosmological constraints to variations in the modeling of HCD contamination. The orange contours show the results obtained when allowing the normalization parameter $f_\mathrm{norm}^\mathrm{HCD}$ to vary instead of fixing it to zero, the green contours correspond to the case where the modeling of LLS and sub-DLA contamination is omitted, and the purple contours show the results obtained using a simplistic HCD model adopted in some of the previous analyses of SDSS data \cite{palanque-delabrouille2015ConstraintNeutrinoMasses, p1d_Chabanier2019, Walther2025_lyssa}.
    }
    \label{fig:variations_hcd}
\end{figure}

In \cref{fig:variations_hcd}, we examine the sensitivity of cosmological constraints to the modeling of HCD contamination. The orange contours correspond to a variation in which the $k$-independent term $f_\mathrm{norm}^\mathrm{HCD}$ is allowed to vary with redshift instead of being fixed to zero. The green contours show the results obtained when omitting the modeling of LLS and sub-DLA contamination, while the red contours correspond to a simplified HCD model with a single free parameter adopted by multiple analyses of SDSS data \cite{palanque-delabrouille2015ConstraintNeutrinoMasses, p1d_Chabanier2019, Walther2025_lyssa}. For a fairer comparison with our baseline setup, we allow the parameters of these variations to vary with redshift as a power law. The cosmological constraints and fit probability for the $f_\mathrm{norm}^\mathrm{HCD}$ variation are perfectly compatible with those from the baseline analysis, whereas omitting the LLS and sub-DLA contributions or employing the simplified HCD model result in $>3\,\sigma$ shifts in the inferred cosmological parameters. Furthermore, the fit probability decreases modestly from 22 to 18\% when LLS and sub-DLA contributions are omitted, and more significantly to 4\% when using the simplified HCD contamination model. In these two cases, the posteriors are truncated in the lower right corner as they reach the edges of the likelihood priors. After unblinding, we noticed that this region of the parameter space is not well covered by the \mpgadget simulations.

\begin{figure}
    \centering
    \includegraphics[width=0.6\linewidth]{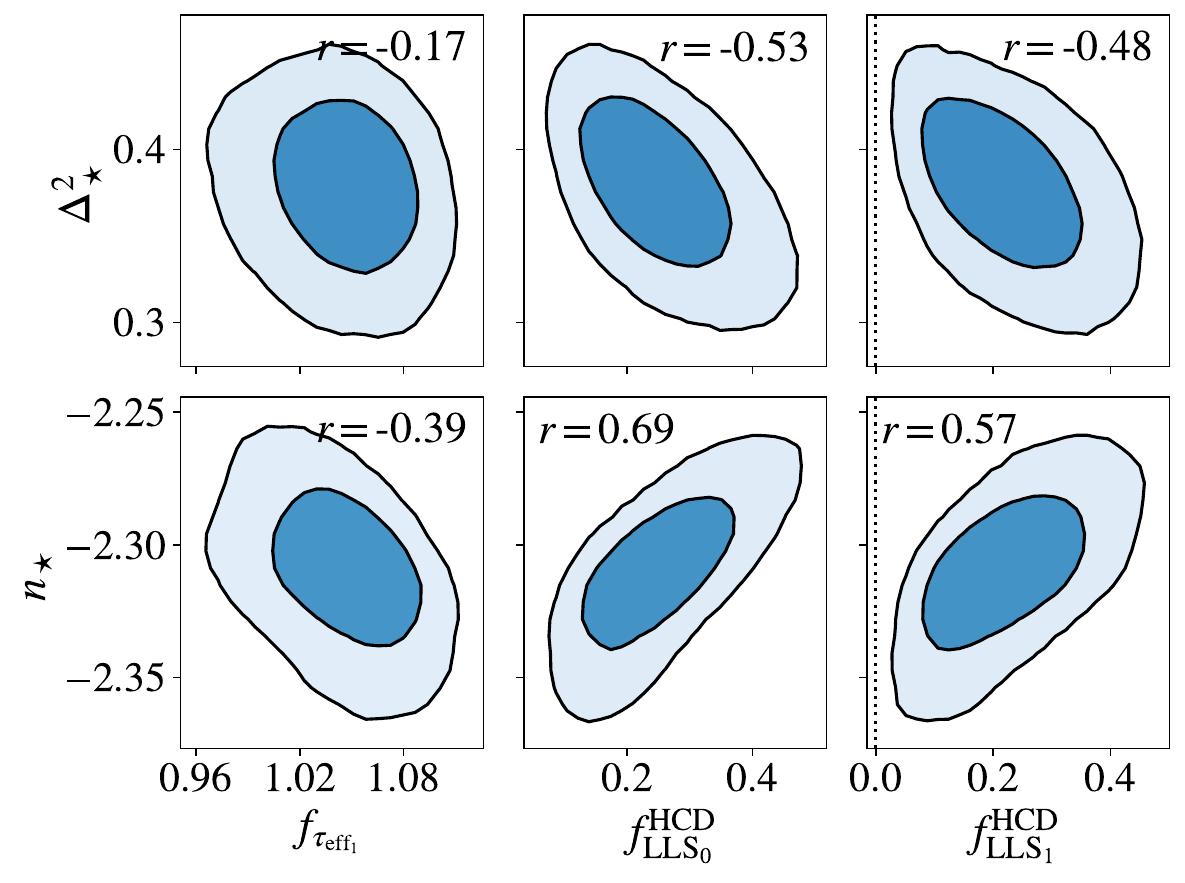}
    \caption{
    Posterior distributions of the compressed parameters and the nuisance parameters most strongly correlated with them. The left panel illustrates the correlation between the compressed parameters and the ones scaling the mean flux, while the central and right panels show the correlation between the compressed parameters and LLS contamination. The Pearson correlation coefficient is indicated at the top of each panel.
}
    \label{fig:corr_comp_mix}
\end{figure}

To better understand these trends, we compute the correlation between all cosmological and nuisance parameters for the baseline analysis. We find that only the mean flux and LLS parameters are strongly correlated with the cosmological parameters (see \cref{app:nuisance}). In \cref{fig:corr_comp_mix}, we show the two-dimensional projection of posterior for the compressed parameters, a representative mean flux parameter, and the LLS parameters at the low and high redshift nodes. Both \deltastar and \nstar are anticorrelated with the mean flux, whereas the LLS parameters are strongly correlated with \nstar and anticorrelated with \deltastar. The correlation between the LSS and compressed parameters explain the sensitivity of cosmological constraints to the modeling of LLS discussed above (see also \cite{Fernandez2024_priya, Ho2025_priya}), while the correlation between the mean flux and the compressed parameters explain the increased constraining power of the mock analyses when IGM parameters are not marginalized over (see \cref{sec:model_validation}). 

The correlation between the IGM, LLS, and compressed parameters motivate three alternative analyses expected to agree with the baseline. In the first two, we model the redshift evolution of the IGM (LLS) parameters using six (four) nodes instead of the four (two) adopted in the baseline analysis, aiming to test whether an insufficiently flexible modeling of the redshift evolution could bias the cosmological constraints. In the third analysis, we enlarge by 10\% the priors on the IGM parameters beyond their original range, which is set by the values covered by the \mpgadget simulations. We find that the resulting cosmological constraints and fit probabilities are fully consistent between these three variations and the baseline analysis.


\subsection{Cosmological constraints}
\label{sec:results_cosmo}

In the previous section, we presented a broad set of blinded analysis variations that demonstrate that our cosmological constraints are robust. We therefore proceeded to unblind the cosmological measurements, obtaining  
\begin{equation}
    \deltastar = 0.379^{+0.032}_{-0.033}, \qquad
    \nstar = -2.309^{+0.019}_{-0.019},
\end{equation}
where the central values and the uncertainties correspond to the medians and 16 and 84th percentiles, respectively. The Pearson correlation coefficient between these parameters is $r = -0.1738^{+0.0006}_{-0.0006}$, obtained by computing the correlation coefficient over 100 random subsamples of the chain, each containing half of the posterior samples. Note that minor differences may exist in the definition of the compressed parameters compared to other analyses in the literature; for reproducibility, our code to compute the value of the compressed parameters is publicly available\footnote{\url{https://github.com/igmhub/cobaya_lya_p1d}}. In \cref{app:pivot}, we also report the constraints at the optimal pivot scale for DESI DR1.

\begin{figure}
    \centering
    \includegraphics[width=0.75\linewidth]{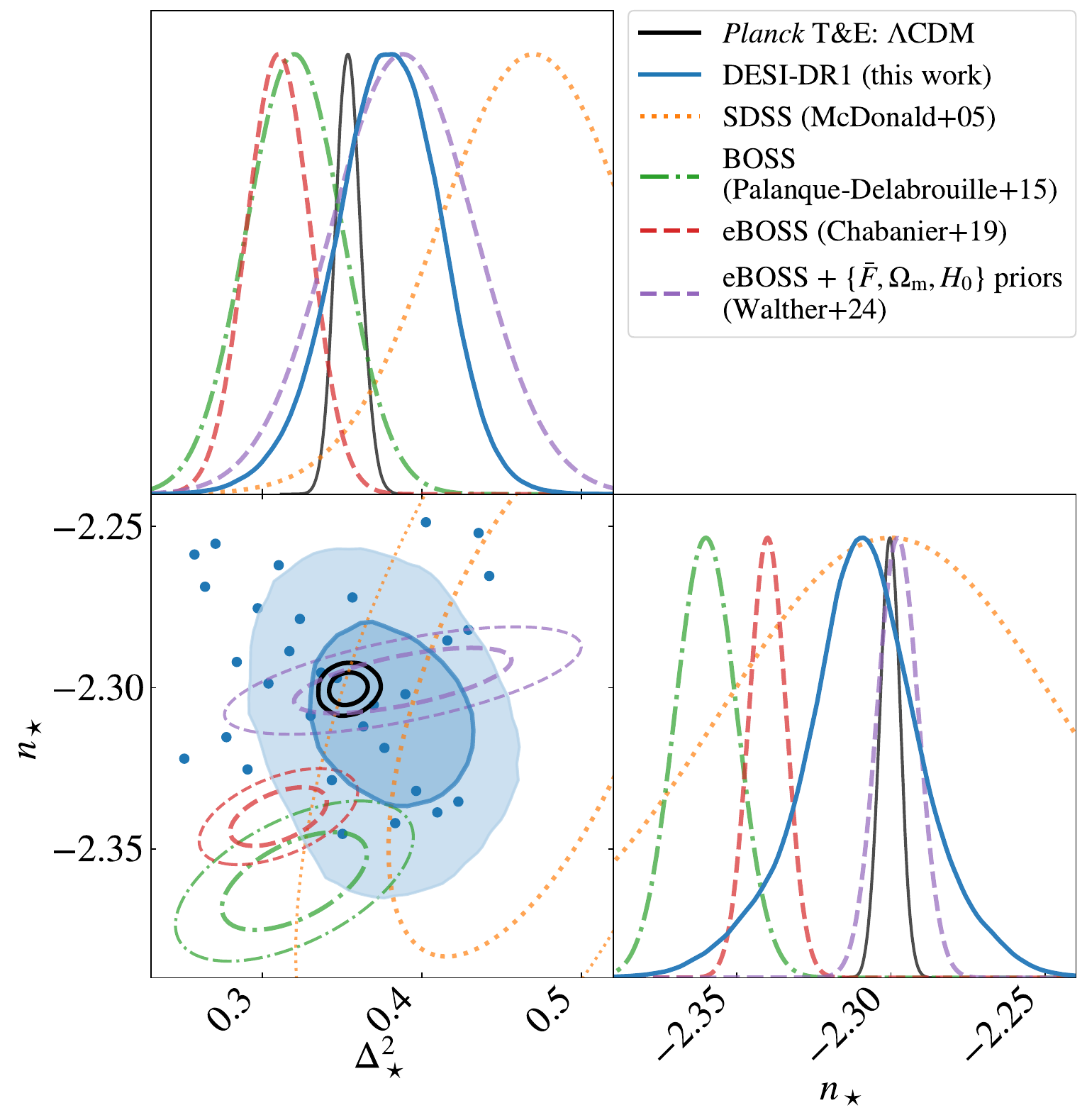}
    \caption{Constraints on the amplitude and logarithmic slope of the linear matter power spectrum at \kstarval and $z_\star = 3$. The blue contours show the results from our analysis, while the black contours show \Planck T\&E constraints when assuming a $\Lambda$CDM cosmology with $\sum m_\nu=0.06$ eV. For comparison, the orange, green, red, and purple contours correspond to the analyses of SDSS DR2 \cite{McDonald2005a, McDonald2006}, SDSS DR9 (BOSS) \cite{palanque-delabrouille2015ConstraintNeutrinoMasses}, and SDSS DR14 (eBOSS) from \cite{p1d_Chabanier2019} and \cite{Walther2025_lyssa}, respectively. The inner and outer contours denote the 68 and 95\% credible intervals, respectively. Blue dots show the position of the 30 fixed-and-paired training simulations across the parameter space.}
    \label{fig:constraints}
\end{figure}

In \cref{fig:constraints}, the blue contours show the constraints from our baseline analysis, while the orange, green, red, and purple contours correspond to the results from the analysis of SDSS DR2 \cite{McDonald2005a, McDonald2006}, SDSS DR9 (BOSS) \cite{palanque-delabrouille2015ConstraintNeutrinoMasses}, and SDSS DR14 (eBOSS) from \cite{p1d_Chabanier2019} and \cite{Walther2025_lyssa}, respectively. The black contours indicate constraints from the analysis of temperature and polarization measurements from \Planck~2018 (\Planck T\&E \cite{planck2018_like, Planck2018}) under the $\Lambda$CDM model with $\sum m_\nu=0.06$ eV. Blue dots indicate the locations of the 30 fixed-and-paired \mpgadget simulations, which sample the $1\,\sigma$ credible interval of the posterior precisely. Consequently, our emulator error estimates (see \cref{sec:emulator_performance}) are reliable within this region, though they may be underestimated beyond it.

Despite being derived from 62\,807 high-SNR spectra, our \nstar constraints are weaker than those from the aforementioned analyses of BOSS and eBOSS data, extracted from 13\,821 and 43\,751 high-SNR spectra, respectively. This is likely caused by several methodological differences between the analyses. The most significant ones are that we marginalize over both LLS and \siisii contamination and account for emulator errors, all neglected in the previous analyses. Each of these variations increases the uncertainties in the compressed parameters by approximately 50, 20, and 25\%, respectively; taken together, these account for most of the difference in constraining power between these analyses. Furthermore, we find that the uncertainty on \nstar increases more than on \deltastar, which partially explains the different degeneracy direction between our and previous analyses.

It is also worth noting that the main distinction between the \cite{p1d_Chabanier2019} and \cite{Walther2025_lyssa} analyses of eBOSS data lie in the emulator construction --- the former employs a Taylor expansion around a central simulation, whereas the latter uses as training data the more accurate \texttt{lyssa} suite --- and in the inclusion of mean-flux priors in the latter. Nevertheless, the \deltastar constraints from \cite{Walther2025_lyssa} are noticeably weaker than those of \cite{p1d_Chabanier2019}, even though the inclusion of informative mean flux priors should have tightened the posteriors. These discrepancies suggest that not accounting for emulator errors in earlier analyses may have significantly underestimated the uncertainty on the compressed parameters.

The location of the posterior also differs noticeably across previous \lya analyses, a discrepancy likely exacerbated by the possibly underestimation of uncertainties in earlier work. In contrast, our constraints are consistent with all previous analyses at the $2\,\sigma$ level, mostly because our \nstar uncertainties are significantly larger. The lack of agreement at the $1\,\sigma$ level is plausibly explained by the use of simplified models for metal and HCD contamination in earlier studies, which can induce significant shifts in the inferred posteriors (\cref{tab:variations_model}). In \cref{tab:literature}, we summarize the main characteristics of the aforementioned analyses. On the other hand, our results are fully consistent with those from \Planck~2018.

\begin{table}[]
    \centering
    \begin{tabular}{c|cccccccc}
         Sample & $n_\mathrm{forest}$ & Simulations, emu & $\sigma_\mathrm{emu}$ & $n_\mathrm{metal}$ & $n_\mathrm{HCD}$ & $\sigma_{\deltastar}$ & $\sigma_{\nstar}$ \\ \hline
         
         SDSS \cite{McDonald2005a, McDonald2006} & 3\,035 & HPM, grid & No & 1 & 1 & 0.06 & 0.055\\
         
         BOSS \cite{palanque-delabrouille2015ConstraintNeutrinoMasses} & 13\,821 & \texttt{Gadget-3}, Taylor & No & 1 & 0 & 0.03 & 0.01\\
         
         eBOSS \cite{p1d_Chabanier2019} & 43\,751 & \texttt{Gadget-3}, Taylor & No & 2 & 0 & 0.02 & 0.006\\
         
         \makecell{eBOSS \cite{Walther2025_lyssa} \\ (incl. $\bar{F}$ prior)} & 43\,751 & \texttt{Nyx}, GP & No & 2 & 1 & 0.045 & 0.0067\\
         
         \makecell{DESI DR1 \\ (this work)} & 62\,807 & \mpgadget, GP & Yes & 16 & 8 & 0.032 & 0.019\\
    \end{tabular}
    \caption{Summary of the main characteristics of the \pone analyses reporting constraints on \deltastar and \nstar.}
    \label{tab:literature}
\end{table}


\subsection{Constraints on the IGM, contaminants, and resolution systematics}
\label{sec:results_nuisance}

In previous sections, we performed a series of mock and alternative data analyses to test the robustness of the constraints on the compressed parameters. We did not, however, carry out an analogous validation for the IGM, metal, HCD, and resolution parameters. As a result, these parameters may absorb inaccuracies arising from the limited resolution of the simulations used to train our emulator, unmodeled systematic or physical effects, or other unknown sources of error. In this section, we present constraints on the nuisance parameters with the caveat that their physical interpretation may be affected by these uncertainties, and gather their numerical values in \cref{app:nuisance}.

\begin{figure}
    \centering
    \includegraphics[width=0.6\linewidth]{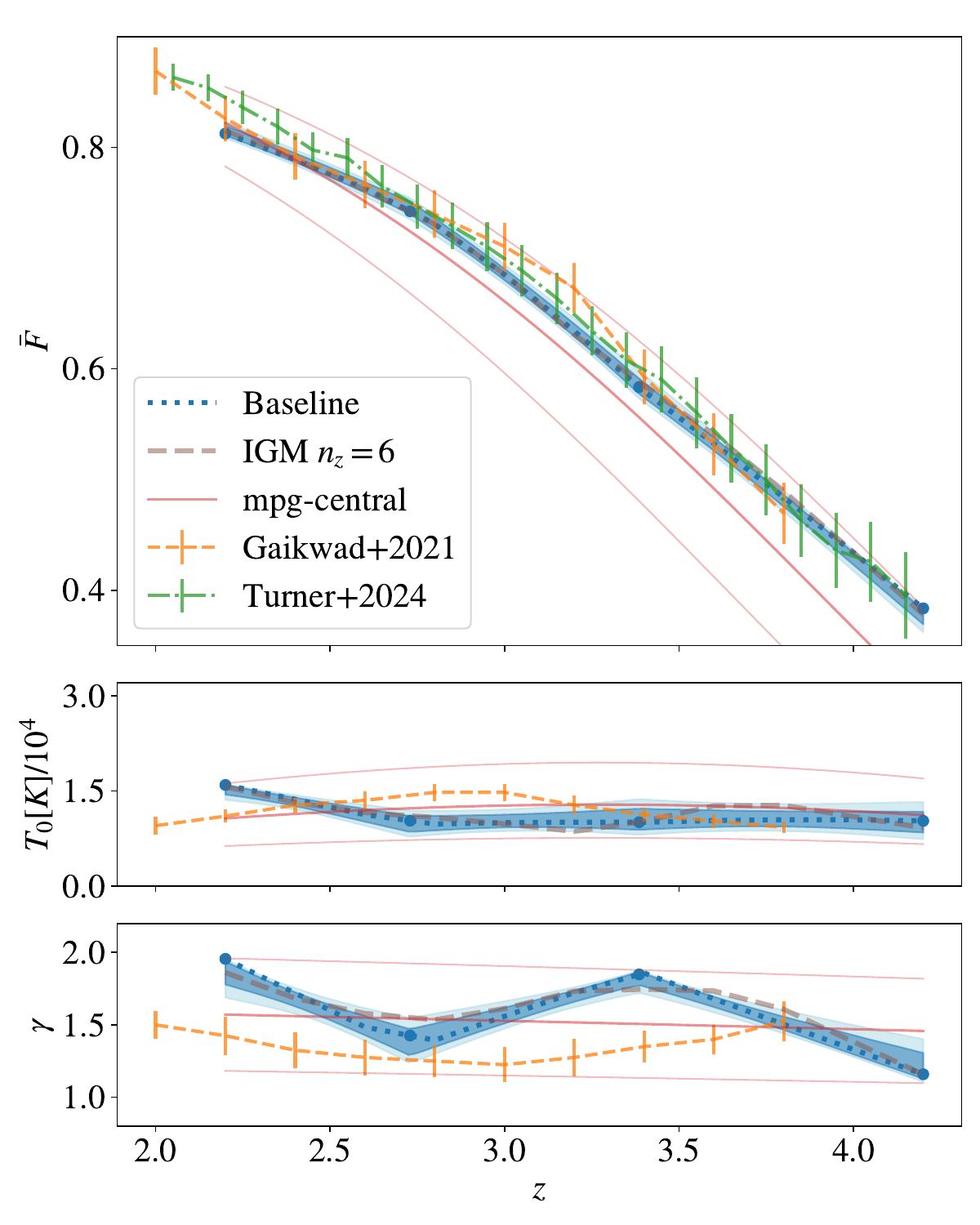}
    \caption{
    Best-fitting constraints on the mean flux (top panel), and the amplitude and slope of the temperature–density relation (middle and bottom panels). The blue lines show the best-fitting results from this work, while the orange and green lines correspond to the measurements from \cite{Gaikwad2021} and \cite{Turner2024}, respectively. Thick red lines display predictions from the \texttt{mpg-central} simulation, while the thin red lines indicate the range of values covered by the \mpgadget simulations. Blue dots denote the best-fitting constraints at the four redshift nodes, while the light and dark blue shaded regions represent the 1 and $2\sigma$ credible intervals, respectively. Brown dashed lines show the results for the variation using six nodes to capture the redshift evolution of the IGM parameters rather than four as in the baseline analysis.
    }
    \label{fig:IGM_best}
\end{figure}

In \cref{fig:IGM_best}, we display the best-fitting redshift evolution of the mean flux (top panel) and of the amplitude and slope of the temperature–density relation (middle and bottom panels) from the baseline analysis, and we compare these with results from high-resolution KODIAQ quasar spectra \cite{Gaikwad2021} and DESI quasar spectra analyzed with LyCAN \cite{Turner2024}. In \cref{app:nuisance}, we also show the results for the alternative analysis using the \lacelyssa emulator. The light blue and dark blue shaded regions show the 1 and $2\,\sigma$ credible intervals from our analysis, respectively, obtained by computing the redshift evolution of the IGM parameters for 20\,000 randomly selected points from the chains. As explained in \cref{sec:model_multiple}, the free parameters of the model capture deviations from the IGM histories predicted by the \texttt{mpg-central} simulation, shown by the thick red lines.

As we can see, the mean flux constraints from the baseline analysis are in good agreement with the KODIAQ and LyCAN measurements, as well as with those from the variation using six nodes to capture the redshift evolution of the IGM parameters rather than four as in the baseline analysis. Note that the best-fitting measurement at $z = 4.2$ approaches the upper prior boundary, highlighting the need for future simulations that cover a wider range of mean flux values at high redshift. The constraints on the amplitude and slope of the temperature–density relation between the baseline and and the variant using six nodes are also broadly consistent. Nonetheless, the constraints from both analyses are in tension with KODIAQ measurements and reach the prior boundaries at some redshifts. This behavior likely reflects the similar absolute impact of all IGM parameters on \pone measurements at large scales and the comparable influence of the temperature parameters and those describing resolution biases at the smallest DESI scales (see \cref{app:nuisance}). Because the sensitivity of \pone to the IGM state and resolution systematics is different for scales smaller than those probed by DESI, a joint analysis of DESI data with high-resolution \pone measurements could break these degeneracies and result in more meaningful IGM constraints. We plan to carry out such an analysis in the future.

\begin{figure}
    \centering
    \includegraphics[width=0.495\linewidth]{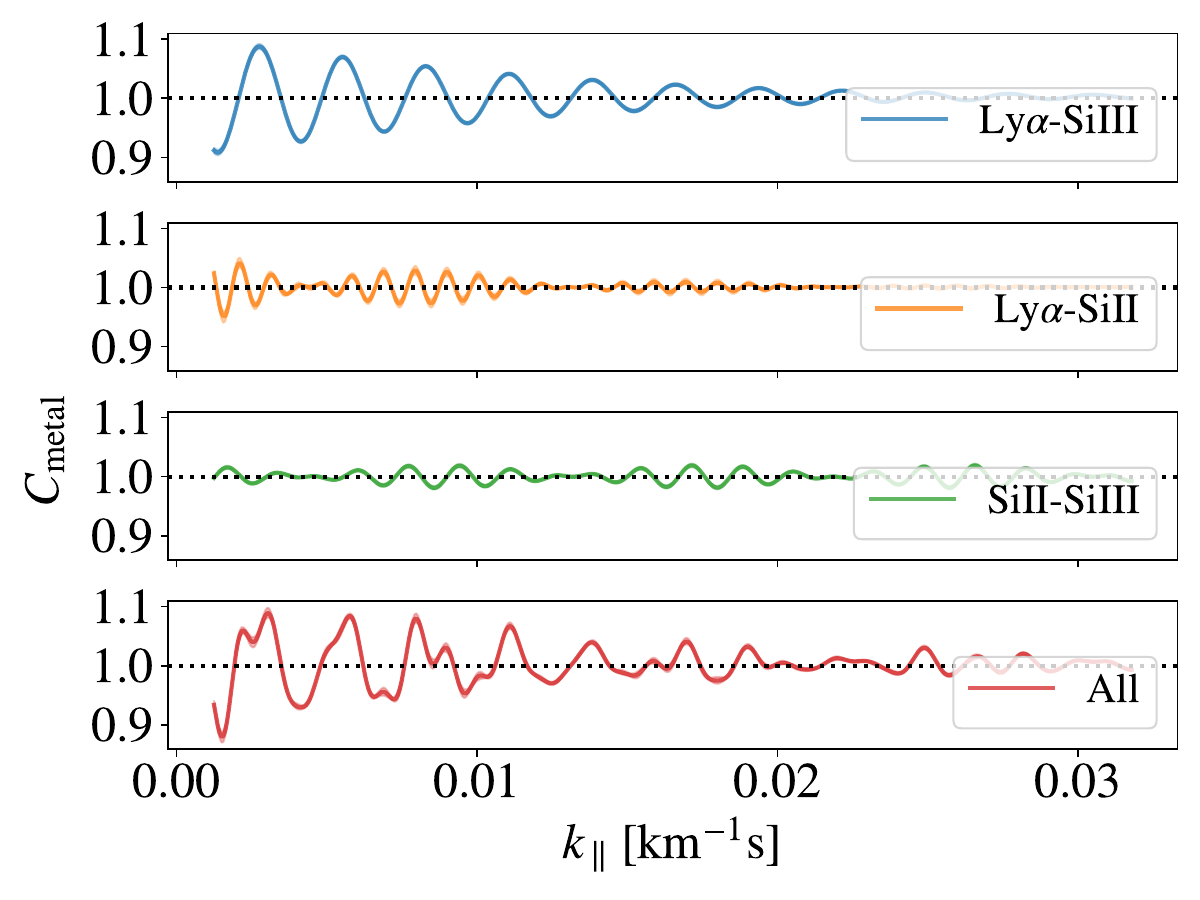}
    \includegraphics[width=0.495\linewidth]{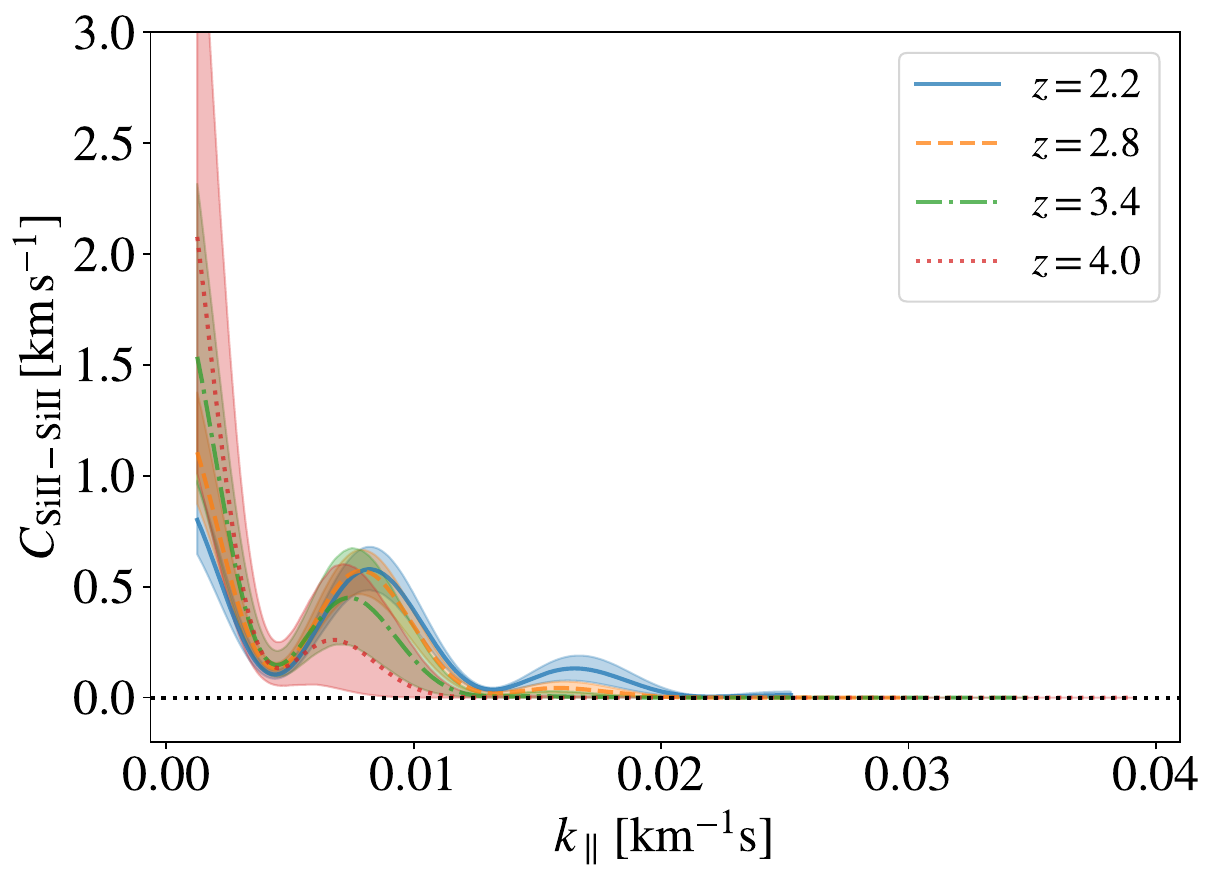}
    
    \includegraphics[width=0.495\linewidth]{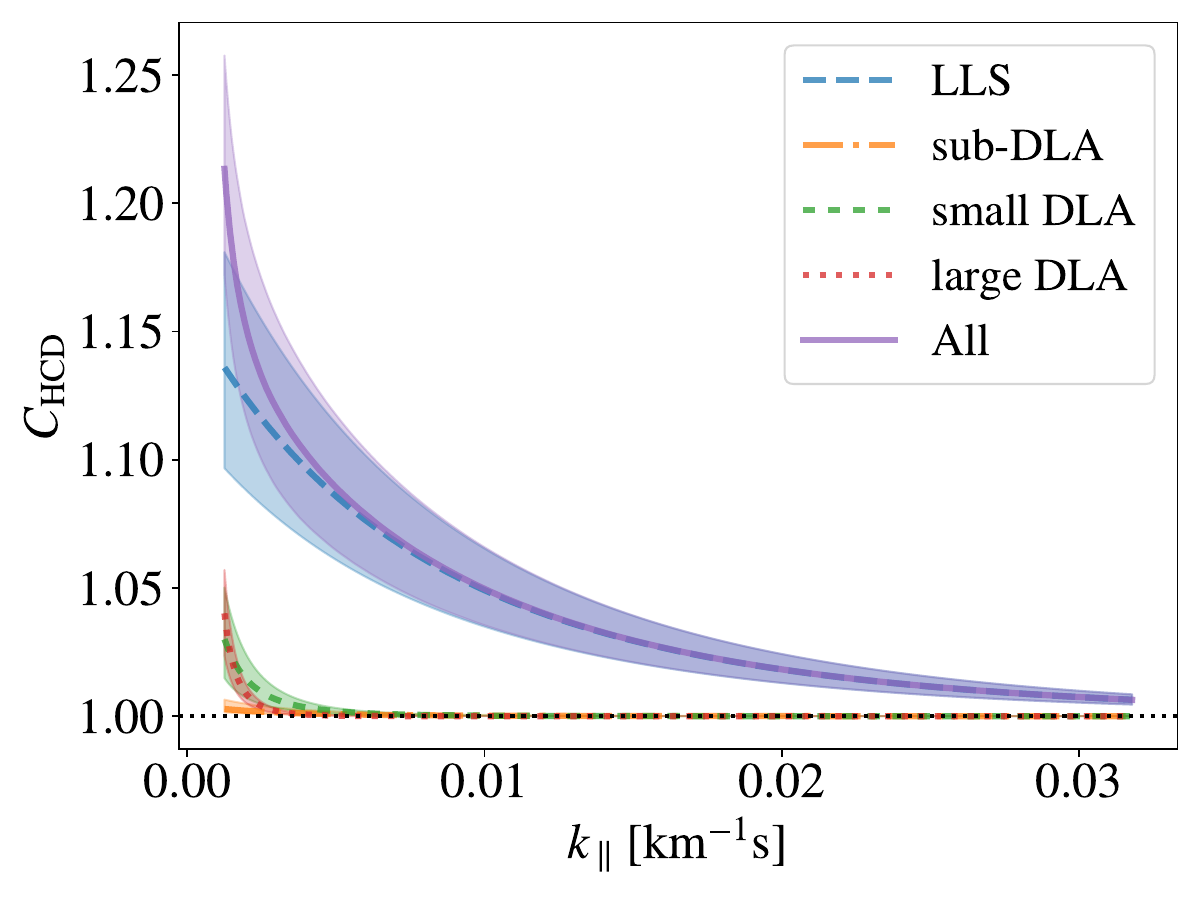}
    \includegraphics[width=0.495\linewidth]{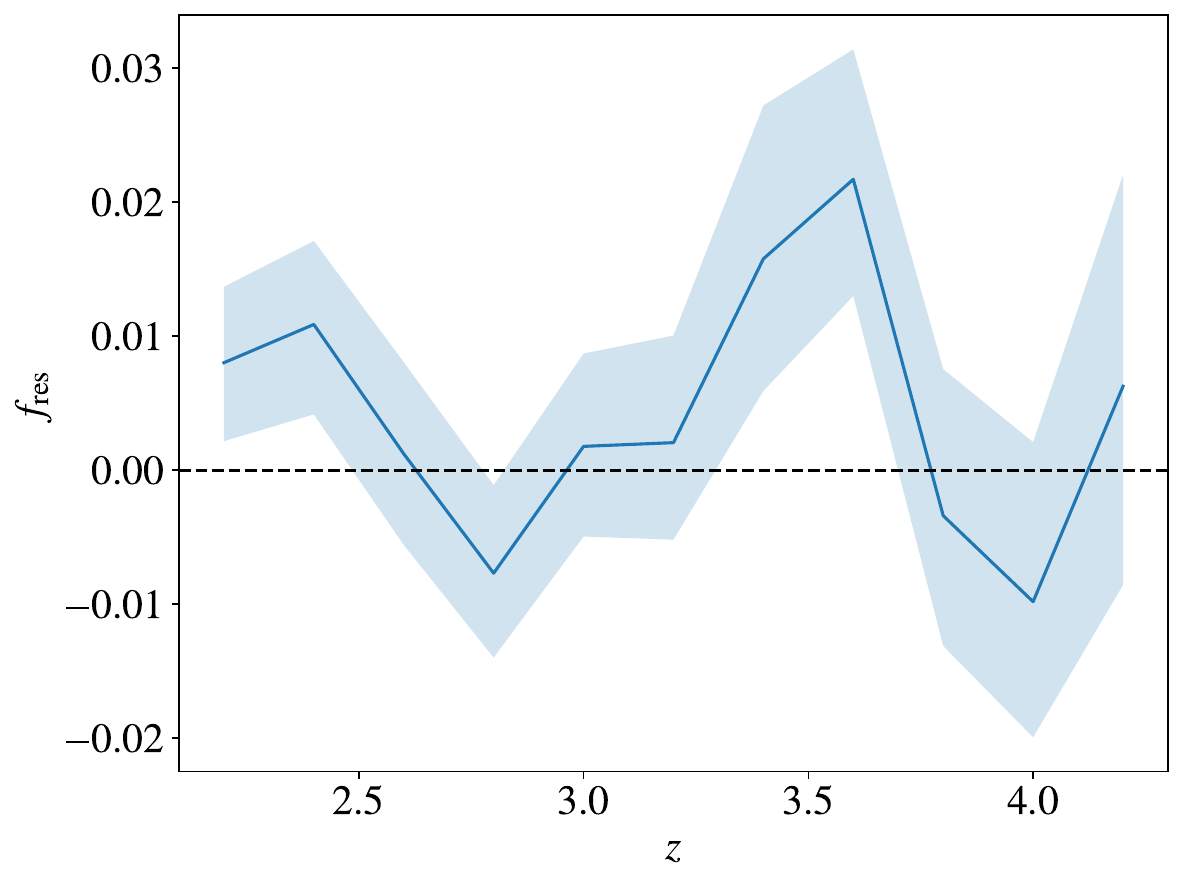}
    \caption{
        Best-fitting results for metal contamination, HCD contamination, and resolution biases. The top-left panel shows the multiplicative metal correction at $z=3$ decomposed into the contributions of \lyasiii, \lyasii, and \siisiii, while the top-right panel displays the additive metal correction from \siisii as a function of redshift. In the bottom-left panel, the blue, orange, green, and red colors show the impact of LLS, sub-DLA, small DLA, and large DLA contamination on \pone, respectively, while the purple color displays their combined effect. The bottom-right panel displays the best-fitting resolution bias as a function of redshift. Shaded areas indicate the $1\,\sigma$ credible interval.
    }
    \label{fig:nuisance_cont}
\end{figure}

In the top panels of \cref{fig:nuisance_cont}, we present the best-fitting results for metal contamination. The top-left panel shows the multiplicative contamination from different ion pairs at $z = 3$, decomposed into the contributions of \lyasiii, \lyasii, and \siisiii. The \lyasiii term dominates across all wavenumbers, while the \lyasii and \siisiii terms become relevant only at large and small scales, respectively. Shaded areas indicate the $1\,\sigma$ region, computed following the same approach as to estimate credible intervals for the IGM parameters. We detect the amplitude of the \lyasiii (\lyasii) contamination at the low and high redshift nodes with 17.9 and $5.5\,\sigma$ ($6.1$ and $2.0\,\sigma$) significance, respectively, and its scale-dependent damping with 14.2 and $4.2\,\sigma$ ($4.7$ and $1.2\,\sigma$). As shown in \cref{app:nuisance}, the best-fitting results indicate a redshift dependence in the amplitudes, while the scale-dependent dampings are consistent with no redshift evolution.

We also detect deviations from the optically-thin limit for the \siia and \siib lines with $2.4\,\sigma$ evidence. Reordering the free parameters of the metal contamination model, the best-fitting effective strength ratio for these lines is $r_\mathrm{eff} \equiv r_{\siia} r = 1.10^{+0.29}_{-0.24}$ instead of 0.48 as expected in the optically-thin limit. As a result, the strengths of the two lines are consistent with being equal, which seems to be supported by measurements of the one-dimensional correlation function \cite{Karacayli2025_p1d_dr1}. We also find that the \siisiii contamination at $z = 2.2$ is $80\%$ larger than the expectation from \cref{eq:sii-siii} at the $2.0\,\sigma$ level.

The top-right panel shows the additive \siisii contamination as a function of redshift, for which we find evidence at the $9.3\,\sigma$ and $2.5\,\sigma$ levels for the low- and high-redshift nodes, respectively. The amplitude of this contaminant varies slowly with redshift, and because the \lya-only \pone increases rapidly with redshift, the relative impact of \siisii decreases sharply toward higher redshift. In the bottom-left panel, we display the best-fitting multiplicative correction factors accounting for HCD contamination, decomposed into the contributions from LLSs, sub-DLAs, small DLAs, and large DLAs. As shown, the LLS contribution dominates the total HCD contamination across all scales, and the other components become mildly relevant only on the largest scales. In fact, we detect LLS contamination at the 3.1 and $2.4\,\sigma$ level for the low- and high-redshift nodes, respectively, while we detect the other components with at most $2\,\sigma$ significance.

In the bottom-right panel, we display the best-fitting values of the resolution parameters. Their redshift evolution is generally smooth, with two notable exceptions around $z = 2.6$ and 3.7. Interestingly, the first corresponds to observed-frame wavelengths of approximately $4\,400\,\mathrm{\AA}$, where the spectrograph collimator mirror exhibits a dip in reflectivity, while the second occurs near $5\,800\,\mathrm{\AA}$, coinciding with the transition of the \lya line between the blue and red channels of the DESI spectrograph \cite{Spectro.Pipeline.Guy.2023}.

%% file: external_journal.tex
\section{Constraints on $\Lambda$CDM extensions}
\label{sec:extensions}

\begin{figure}
    \centering
    \includegraphics[width=0.8\linewidth]{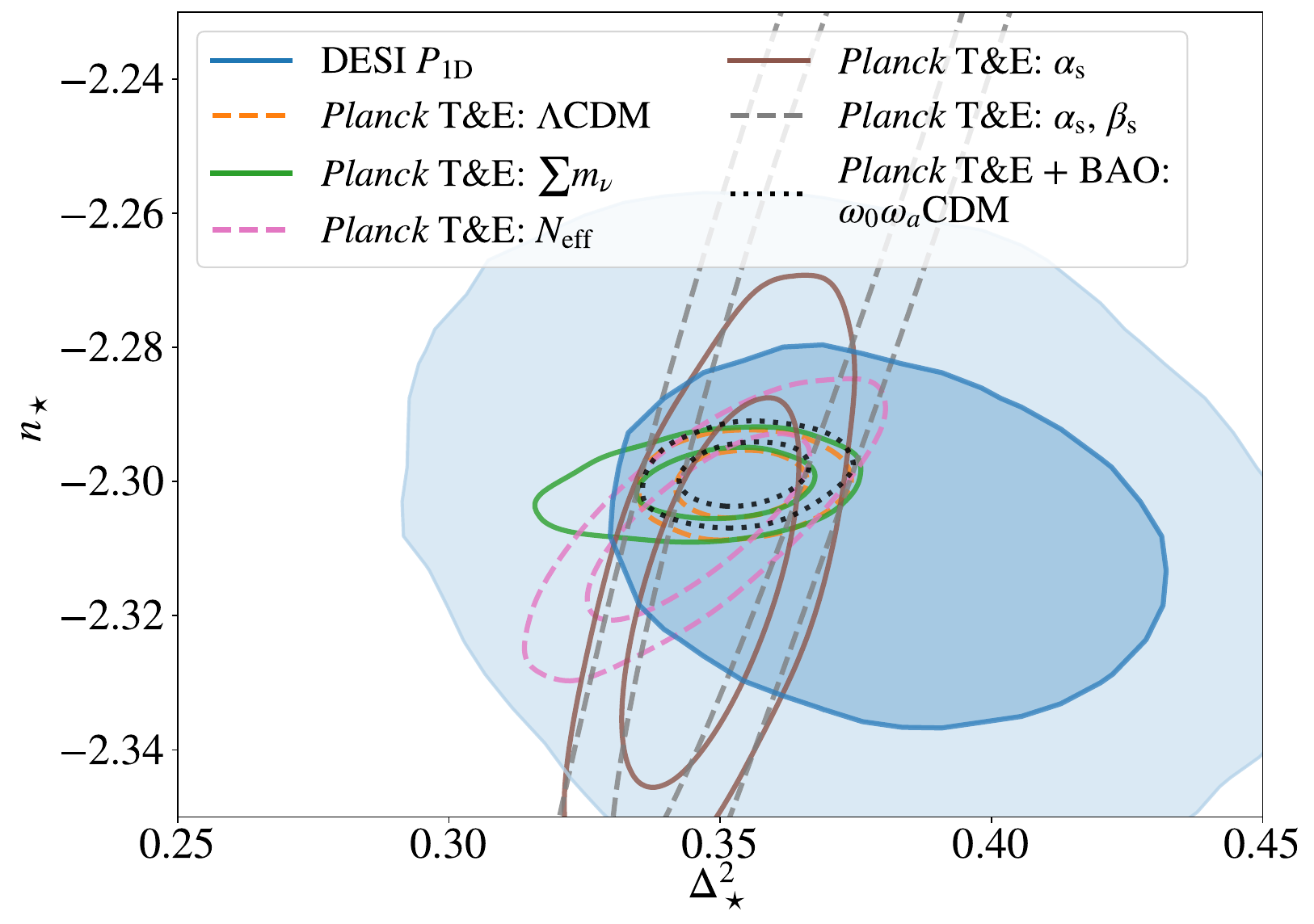}
    \caption{Constraints on the compressed parameters from this work and from \Planck T\&E when assuming different cosmological models. The blue contours show the results of our analysis, while the orange, green, pink, brown, and gray contours correspond to \Planck T\&E constraints within the $\Lambda$CDM framework with $\sum m_\nu=0.06$ eV and when additionally varying the sum of neutrino masses, the effective number of relativistic species, the running of the spectral index, and the running-of-the-running of the spectral index, respectively. The black contours show the results from the combination of \Planck T\&E constraints and and BAO measurements from the 6dF Galaxy Survey \cite{beutler2011_6df}, SDSS DR7 \cite{Ross2015_sdssdr7}, and SDSS DR12 \cite{Alam2017_sdssdr12} when assuming the $\omega_0\omega_a$CDM model. The inner and outer contours denote the 68 and 95\% credible intervals, respectively.
    }
    \label{fig:const_star}
\end{figure}

In \cref{fig:const_star}, we present our constraints on the compressed parameters together with those from \Planck T\&E assuming $\Lambda$CDM with $\sum m_\nu=0.06\,\mathrm{eV}$ and four extensions: variations in the sum of neutrino masses, the effective number of relativistic species, the running of the spectral index, and the running of the running. We also present constraints on the compressed parameters from the combination of \Planck T\&E and BAO measurements from the 6dF Galaxy Survey \cite{beutler2011_6df}, SDSS DR7 \cite{Ross2015_sdssdr7}, and SDSS DR12 \cite{Alam2017_sdssdr12} when assuming the $\omega_0\omega_a$CDM model. Our posterior is significantly broader than those assuming the plain $\Lambda$CDM model and the $\omega_0\omega_a$CDM model, and thus our measurements are not precise enough to tighten the constraints on $\Lambda$CDM parameters or measure the dark energy equation of state \cite{Garza2026_p1d_DE}. On the other hand, our \pone measurements provide competitive constraining power for the remaining extensions, specially for the running of the spectral index and the running of the running. The \lyaf is also sensitive to the nature of dark matter (e.g.; \cite{Rogers2021_ULADM, Rogers2025_running, GarciaGallego2025_wdm}) and primordial magnetic fields \cite{pavicevic2025_primordial_magnetic}, but constraining these effects is beyond the scope of this work.

\begin{figure}
    \centering
    \includegraphics[width=0.75\linewidth]{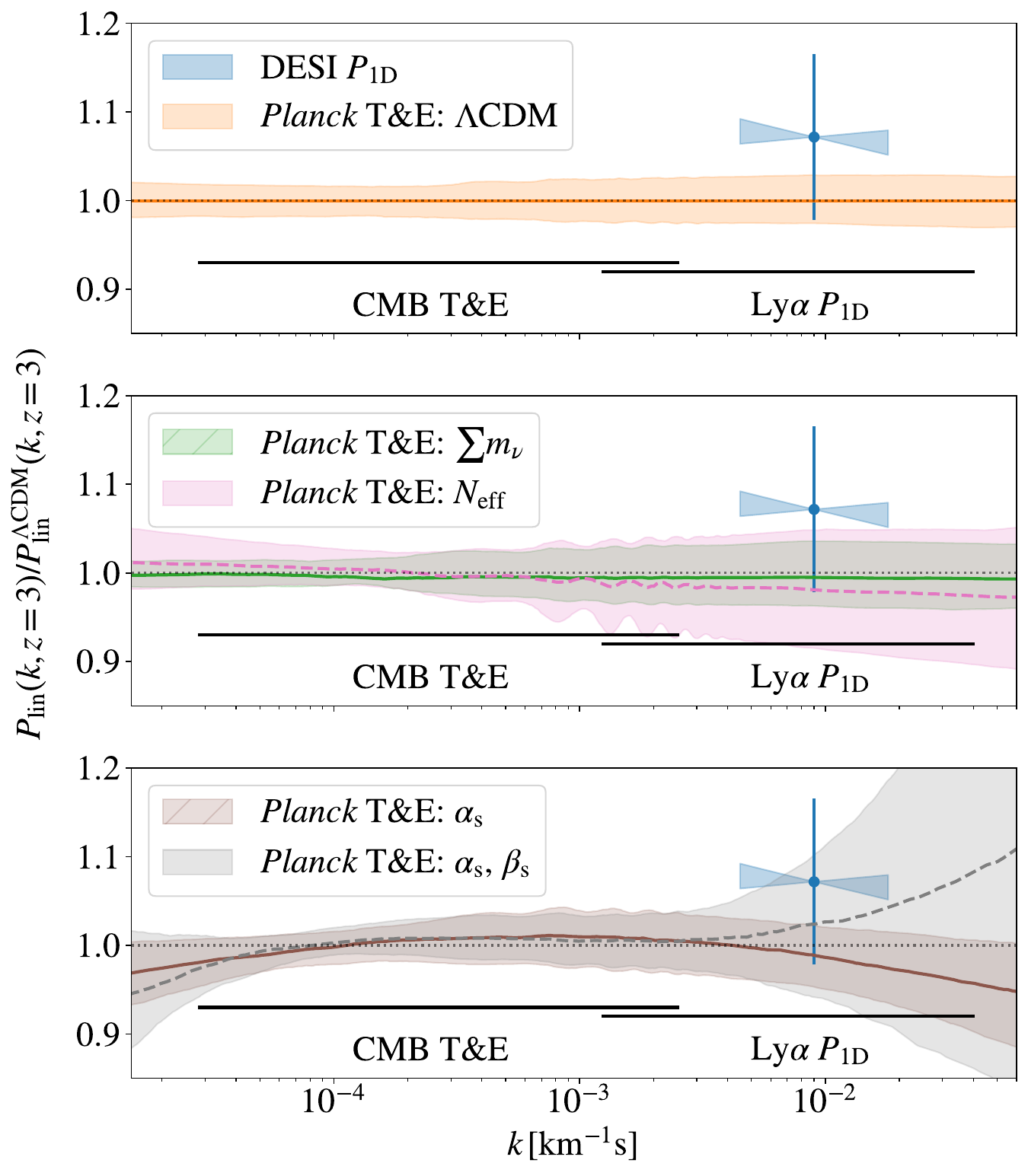}
    \caption{
    Constraints on the shape of the linear matter power spectrum at $z=3$. The blue datapoints display the ratio of our constraints on \deltastar and \nstar and those from \Planck T\&E for $\Lambda$CDM, with the error bar and shaded area indicating the $1\,\sigma$ region for the first and second, respectively. Orange, green, pink, brown, and gray regions display the ratio of the \Planck T\&E linear matter power spectrum for plain $\Lambda$CDM and the extensions varying the sum of neutrino masses, the effective number of relativistic species, the running of the spectral index, and the running of the running, respectively, relative to the $\Lambda$CDM results. Shaded areas denote 68\% credible intervals. The left and right horizontal black lines indicate the approximate range of scales probed by \Planck T\&E and DESI \lya measurements, respectively.
    }
    \label{fig:Pk_star}
\end{figure}

To illustrate why \lya measurements are valuable for constraining the previous $\Lambda$CDM extensions, \cref{fig:Pk_star} shows the linear matter power spectrum constraints at $z=3$ from this work and \Planck T\&E under different cosmologies. The blue datapoints display the ratio between our constraints on \deltastar and \nstar and those from \Planck T\&E for $\Lambda$CDM with $\sum m_\nu = 0.06\,\mathrm{eV}$, with the error bar and shaded area indicating the $1\,\sigma$ region for the first and second, respectively. The width of the blue region reflects the range of scales used to compute the compressed parameters (from half to twice the value of the pivot scale; see \cite{Pedersen2021}). The orange region shows the 68\% credible interval from the analysis of \Planck T\&E when assuming the $\Lambda$CDM model with $\sum m_\nu = 0.06\,\mathrm{eV}$, which we estimated using 500 randomly selected samples from the chain. As expected, the small-scale shape of the linear power spectrum is tightly determined by its large-scale behavior. Combined with the much broader extent of our constraints relative to those from \Planck T\&E, this explains why our measurements add little additional constraining power for the $\Lambda$CDM and $\omega_0\omega_a$CDM models. Nevertheless, it is important to note that our measurements are compatible with \Planck constraints within the $\Lambda$CDM model, while this is not the case for some of the previous eBOSS analyses \cite{Rogers2025_running} as explained in \cref{sec:results_cosmo}.

The green, pink, brown, and gray lines and colored regions show the median and 68\% credible intervals for the ratio for extensions varying the sum of neutrino masses, the effective number of relativistic species, the running of the spectral index, and the running of the running, respectively. As we can see, the constraints on the linear power spectrum are nearly identical over the range of scales probed by the CMB when using the plain $\Lambda$CDM model or the previous extensions. However, we can see progressively larger differences in the constraints as we move towards smaller scales, especially in the small-scale slope when varying the running and the running of the running of the spectral index. As a result, our constraints become quite significant for these extensions, which is explained because \pone measurements probe much smaller scales than those accessed by \Planck.

We derive constraints on $\Lambda$CDM extensions by combining our measurements with three different datasets: a) \Planck T\&E measurements; b) the combination of DESI-DR2 BAO measurements (DESI BAO; \cite{DESI.DR2.BAO.lya, DESI.DR2.BAO.cosmo}) and temperature, polarization, and lensing data from \Planck, ACT-DR6, and SPT-3G (CMB-SPA; \cite{Camphuis2025_spt3gd1}); and c) the combination BAO \cite{DESI2024.III.KP4, DESI2024.IV.KP6, DESI2024.VI.KP7A} and full-shape measurements \cite{DESI2024.V.KP5} from DESI DR1 (DESI FS; \cite{DESI2024.VII.KP7B}) and Big Bang nucleosynthesis constraints (BBN; \cite{schoneberg2024_bbn}). We perform the combination using importance sampling \cite{lewis_bridle2002}, a technique that updates existing Markov chains according to the likelihood of new measurements, thereby avoiding the computational cost of recomputing the full joint posterior. For efficiency, we model our constraints as a Gaussian posterior centered on the values reported in \cref{sec:results_cosmo}, with widths $\sigma_{\deltastar}=0.032$ and $\sigma_{\nstar}=0.019$ estimated from the standard deviations of the chains and correlation $r=-0.1738$\footnote{\url{https://github.com/igmhub/cobaya_lya_p1d}}. We discuss the resulting constraints throughout the remainder of this section and summarize these in \Cref{tab:lcdm_extensions}.

\begin{table}[]
    \centering
    \begin{tabular}{ccccc}
         & \Planck T\&E$^*$ & \makecell{\Planck T\&E$^*$ \\+ DESI \pone$^{\dagger}$} & \makecell{CMB-SPA$^{\ddagger}$ \\+ DESI BAO$^{\mathsection}$} & \makecell{CMB-SPA$^{\ddagger}$ \\+ DESI BAO$^{\mathsection}$ \\+ DESI \pone$^{\dagger}$}\\
        \hline
        $\sum m_\nu$
        & $<_{95\%} 257$ meV   
        & $<_{95\%} 205$ meV 
        & $<_{95\%} 55.1$ meV 
        & $<_{95\%} 53.7$ meV\\
        $N_\mathrm{eff}$    
        & $2.92\pm 0.19$        
        & $2.96\pm 0.16$ 
        & $3.03\pm 0.12$ 
        & $3.02\pm 0.10$\\
        $\alpha_\mathrm{s}$ 
        & $-0.0055\pm 0.0067\;\;\,$ 
        & $-0.0033\pm 0.0046\;\;\,$ 
        & $0.0042\pm 0.0052$ 
        & $0.0014\pm 0.0041$\\
        \hline
        \multicolumn{5}{c}{$\alpha_\mathrm{s}$ and $\beta_\mathrm{s}$}\\
        \hline
        $\alpha_\mathrm{s}$ 
        & $0.001\pm 0.010$ 
        & $-0.0047\pm 0.0049\;\;\,$ 
        & $0.0075\pm 0.0058$ 
        & $0.0016\pm 0.0045$\\
        $\beta_\mathrm{s}$  
        & $0.012\pm 0.013$ 
        & $0.0041\pm 0.0049$       
        & $0.0110\pm 0.0091$ 
        & $-0.0006\pm 0.0048\;\;\,$\\
    \end{tabular}
    \caption[]{
    Constraints on $\Lambda$CDM extensions from different combinations of the following datasets:\\
    $^*$ \Planck T\&E \cite{planck2018_like, Planck2018}: \Planck~2018 \texttt{Plik} high-$\ell$ T\&E, \texttt{Commander} low-$\ell$ TT, and \texttt{SimAll} low-$\ell$ EE.\\
    $^\dagger$ DESI \pone (this work): DESI DR1 Ly$\alpha$ \pone.\\
    $^\ddagger$ CMB-SPA \cite{Camphuis2025_spt3gd1}: P-ACT T\&E \cite{Louis2025_actdr6like} (\Planck T\&E \cite{planck2018_like, Planck2018} and ACT DR6 T\&E \cite{Louis2025_actdr6like, Calabrese2025_actdr6cosmo}), SPT-3G D1 T\&E \cite{Camphuis2025_spt3gd1}, and CMB-SPA $\phi\phi$ \cite{Qu2025_cpalensing} (\Planck \texttt{NPIPE} PR4 \cite{Carron2022_planck_lensing}, ACT-DR6 \cite{Qu2024_act_lensing, Madhavacheril2024_act_lensing}, and SPT-3G D1 \cite{Ge2025_sptlensing}).\\
    $^\mathsection$ DESI BAO \cite{DESI.DR2.BAO.lya, DESI.DR2.BAO.cosmo}: DESI-DR2 baryon acoustic oscillation measurements from galaxy clustering, quasar clustering, and the \lyaf.
    }
    \label{tab:lcdm_extensions}
\end{table}


\subsection{Shape of the primordial power spectrum}
\label{sec:ext_shape}

The standard $\Lambda$CDM model predicts that the initial conditions for structure formation are given by purely adiabatic, scalar perturbations characterized by a power-law curvature power spectrum, $\Delta^2_{\mathcal{R}, \Lambda\mathrm{CDM}} = A_\mathrm{s}(k/k_\mathrm{s})^{n_\mathrm{s}-1}$, with $k_\mathrm{s}=0.05\,\iMpc$. However, inflationary models generally predict small deviations from a power law \cite{Kosowsky1995_running, Planck2014_inflation, Planck2020_inflation}, which can be captured by expanding the primordial power spectrum,
\begin{equation}
\log \Delta_\mathcal{R}^2(k) =
\log A_\mathrm{s} + (n_\mathrm{s} - 1) \log(k/k_\mathrm{s})
+ \frac{\alpha_\mathrm{s}}{2} \log(k/k_\mathrm{s})^2
+ \frac{\beta_\mathrm{s}}{6} \log(k/k_\mathrm{s})^3
+ \mathcal{O}[\log(k/k_\mathrm{s})^4],
\end{equation}
where $\alpha_\mathrm{s}$ and $\beta_\mathrm{s}$ denote the running and the running of the running of the spectral index, respectively.

CMB observations tightly constrain the shape of the primordial power spectrum near $k_\mathrm{s}$ but, as shown in \cref{fig:Pk_star}, these are less sensitive to smaller scales, limiting the ability of CMB studies to detect deviations from a pure power law. In contrast, \pone measurements probe the linear power spectrum at much smaller scales than CMB studies, making the combination of both ideal for constraining the shape of the primordial power spectrum. Previous studies found that the combination of \pone measurements from eBOSS and \Planck T\&E constraints preferred a value of the running of the spectral index different from zero at the $5\,\sigma$ level \cite{Rogers2025_running}, providing motivation to measure both $\alpha_\mathrm{s}$ and $\beta_\mathrm{s}$. More recently, \cite{Fairbairn2025_running} found a $5\,\sigma$ tension between \pone measurements from eBOSS and CMB-SPA data under the $\Lambda$CDM model, and a $\simeq3\,\sigma$ tension when varying the running of the spectral index or both the running and running of the running. However, as shown in \cref{fig:const_star,fig:Pk_star}, our measurements (hereafter DESI \pone) agree well with \Planck T\&E constraints.

\begin{figure}
    \centering
    \includegraphics[width=0.495\linewidth]{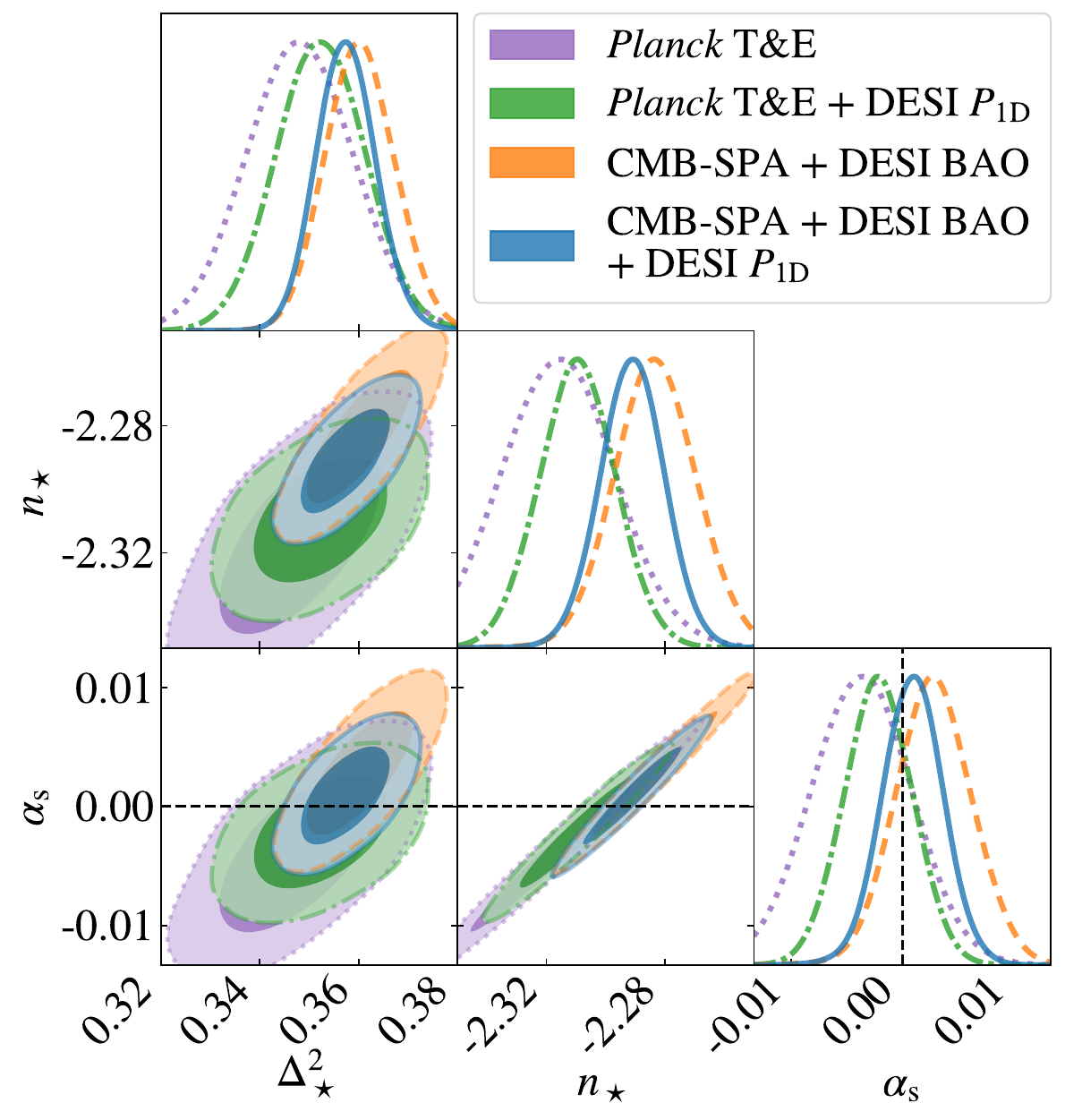}
    \includegraphics[width=0.495\linewidth]{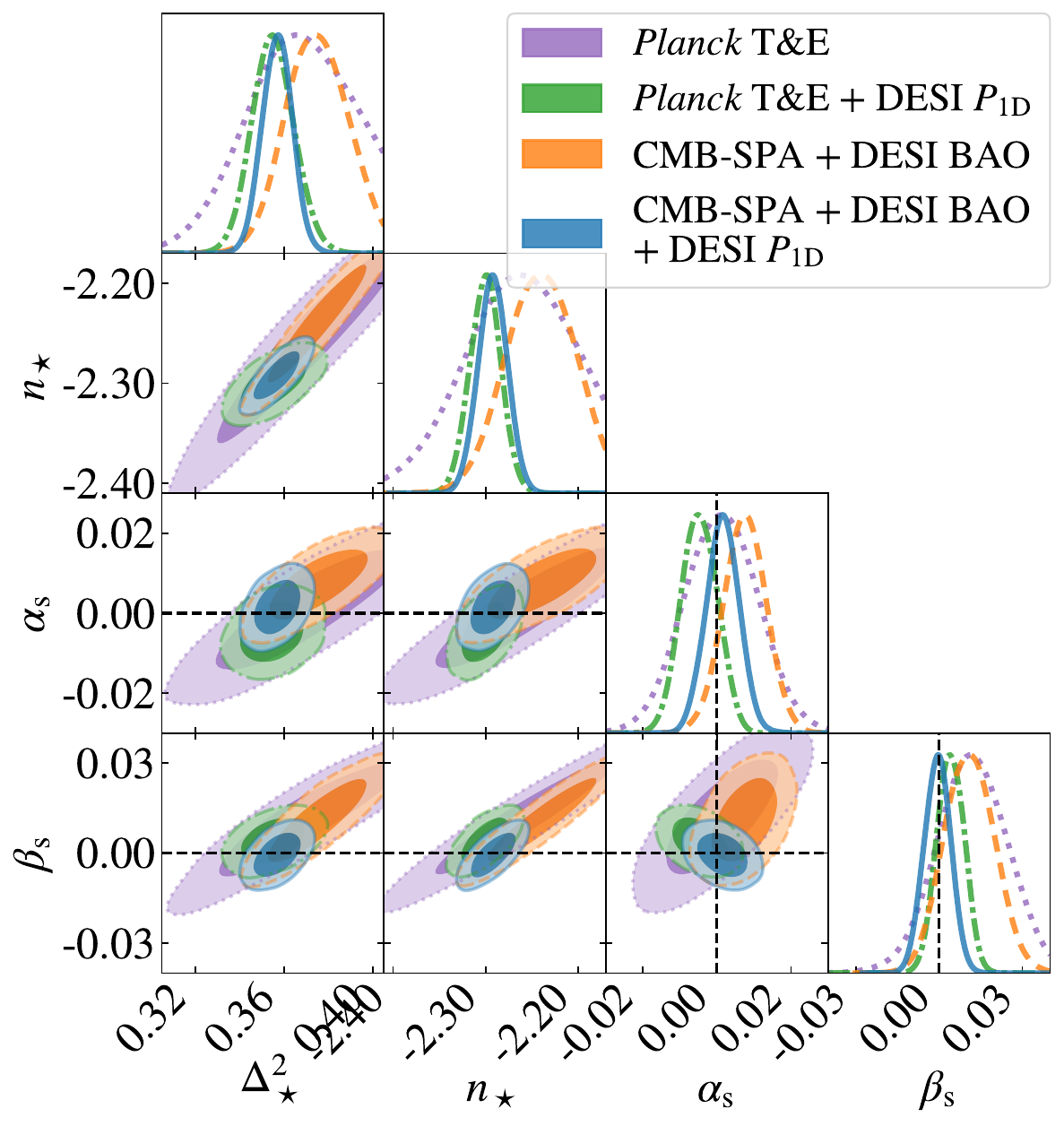}
    \caption{Constraints on the running of the spectral index (left panel) and both the running and the running-of-the-running (right panel). The purple, green, orange, and blue contours display the results from the analysis of \Planck T\&E; the combination \Planck T\&E and DESI \pone; the combination of CMB-SPA and DESI BAO; and the combination of CMB-SPA, DESI BAO, and DESI \pone; respectively. The inner and outer contours denote the 68 and 95\% credible intervals, respectively. Dashed lines indicate $\alpha_\mathrm{s}=0$ and $\beta_\mathrm{s}=0$.}
    \label{fig:import_nrun}
\end{figure}

In the left panel of \cref{fig:import_nrun}, we show constraints on the compressed parameters and the running of the spectral index from the analysis of \Planck T\&E; the combination of \Planck T\&E and DESI \pone; the combination of CMB-SPA and DESI BAO; and the combination of CMB-SPA, DESI BAO, and DESI \pone. The combination of \Planck T\&E and DESI \pone yields $\alpha_\mathrm{s} = -0.0034 \pm 0.0046$, improving the precision of \Planck T\&E constraints by a factor of 1.46. In contrast, the combination of CMB-SPA and DESI BAO yields $\alpha_\mathrm{s}=0.0042\pm0.0052$, which is a factor of 1.13 worse than the combination of \Planck T\&E and our constraints. This result highlights the importance of combining probes sensitive to the linear power spectrum over widely separated scales when testing deviations from a power law. The tightest constraints arise from the combination of CMB-SPA, DESI BAO, and DESI \pone, $\alpha_\mathrm{s} = 0.0014 \pm 0.0041$, which improve those from combination of CMB-SPA and DESI BAO by a factor of 1.27. These constraints are fully consistent with the running of the spectral index being zero, and neither data combination shows any preference for departures from this value. 

To further explore possible deviations of the primordial power spectrum from a power law, the right panel displays constraints on both the running and the running of the running. The combination of \Planck T\&E and DESI \pone yields $\alpha_\mathrm{s} = -0.0047 \pm 0.0049$ and $\beta_\mathrm{s} = 0.0041 \pm 0.0049$, reducing \Planck T\&E uncertainties by a factor of 2.04 and 2.65, respectively. Consequently, the relative constraining power of \pone measurements becomes even more significant when allowing for broader departures from a power-law primordial spectrum, consistent with the trends shown in \cref{fig:Pk_star}. The combination of CMB-SPA, DESI BAO, and DESI \pone yields $\alpha_\mathrm{s} = 0.0016 \pm 0.0045$ and $\beta_\mathrm{s} = -0.0006 \pm 0.0048$, which increases the precision of the constraints from the combination of CMB-SPA and DESI BAO by a factor of 1.29 and 1.90, respectively. The constraints on both $\alpha_\mathrm{s}$ and $\beta_\mathrm{s}$ are fully consistent with a pure power-law primordial spectrum.

In \cref{app:primodial_pk}, we present constraints on the parameters defining the primordial power spectrum: $A_\mathrm{s}$, $n_\mathrm{s}$, $\alpha_\mathrm{s}$, and $\beta_\mathrm{s}$. We find that \pone measurements do not tighten the constraints on $A_\mathrm{s}$; this is expected since \pone measurements probe scales around $k\simeq0.7\,\iMpc$, which is much smaller than $k_\mathrm{s}=0.05\,\iMpc$, the pivot scale at which $A_\mathrm{s}$ is defined. On the other hand, the combination of CMB-SPA, DESI BAO, and DESI \pone yields $n_\mathrm{s} = 0.9738 \pm 0.0036$, increasing the precision of the constraints from the combination of CMB-SPA and DESI BAO by a factor of 1.11. The improvement in the spectral index reflects the fact that this parameter influences a much larger range of scales than $A_\mathrm{s}$.


\subsection{Effective number of relativistic species}
\label{sec:ext_neff}

In the early Universe, neutrinos remained in thermal equilibrium with the primordial plasma through weak interactions until the temperature dropped to $T \sim 2$~MeV, when these interactions became inefficient and started to decouple. The neutrino-to-photon density ratio after the electron–positron annihilation is given by
\begin{equation}
    \frac{\rho_\nu}{\rho_\gamma} = 1 + \frac{7}{8} \left(\frac{4}{11}\right)^{4/3} N_\nu,
\end{equation}
where $\rho_\gamma$ and $\rho_\nu$ are the photon and neutrino energy densities, respectively, and $N_\nu\simeq 3.046$ stands for the effective number of neutrino species, slightly exceeding three because the three Standard Model neutrinos were not fully decoupled during electron–positron annihilation \cite{Mangano2005_neff}. Any additional relativistic relic would increase this value such that $N_\mathrm{eff} \equiv N_\nu + \Delta N_\mathrm{eff}$, where $N_\mathrm{eff}$ denotes the effective number of relativistic species. A measurement of $\Delta N_\mathrm{eff} > 0$ would therefore constitute evidence for physics beyond the Standard Model \cite{Antel2023_reportneff}.

\begin{figure}
    \centering
    \includegraphics[width=0.495\linewidth]{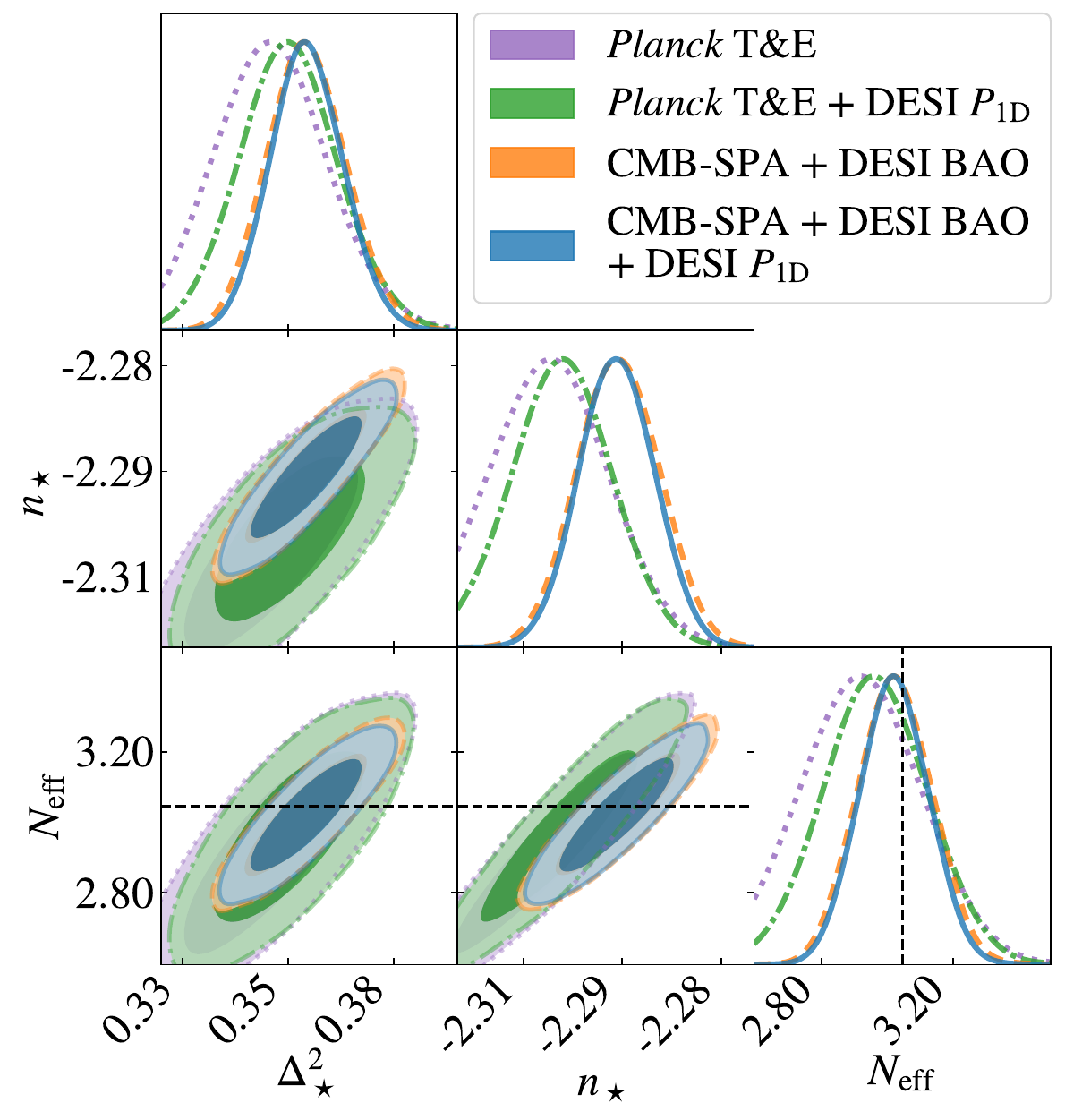}
    \includegraphics[width=0.495\linewidth]{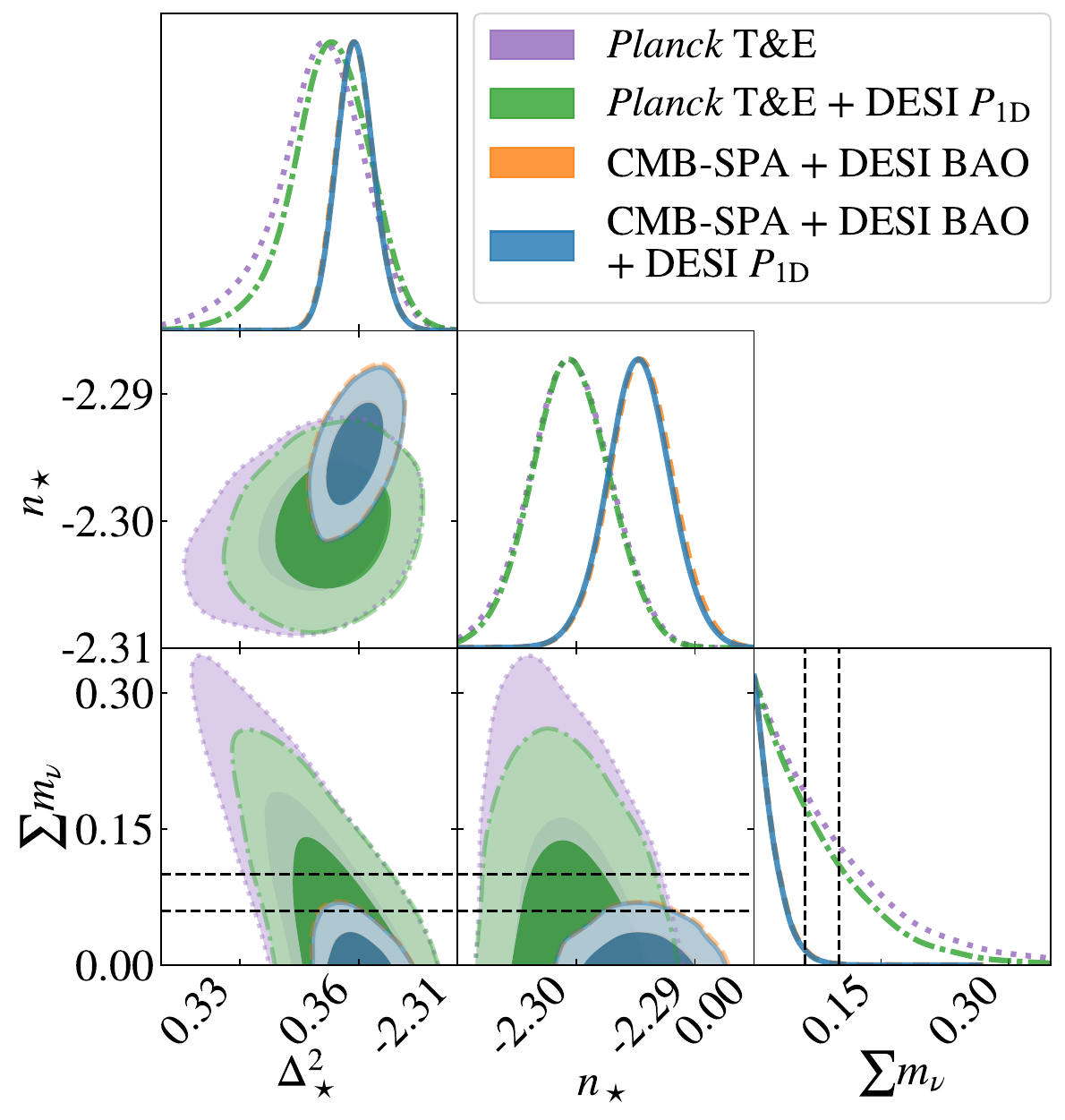}
    \caption{Same as \cref{fig:import_nrun}, but for constraints on the effective number of relativistic species (left panel) and the sum of neutrino masses (right panel). The dashed lines in the left panel correspond to $N_\mathrm{eff} = 3.046$, the value predicted by the Standard Model of particle physics, while the dashed lines in the right panel indicate the minimum sum of neutrino masses allowed for the normal ($\sum m_\nu \gtrsim 0.06$ eV) and inverted ($\sum m_\nu\gtrsim0.10$ eV) mass orderings.}
    \label{fig:import_mnu_nnu}
\end{figure}

In the left panel of \cref{fig:import_mnu_nnu}, we display constraints on $N_\mathrm{eff}$ from the same data combinations considered in \cref{fig:import_nrun}. The combination of \Planck T\&E and DESI \pone measurements yields $N_\mathrm{eff} = 2.96 \pm 0.16$, reducing \Planck T\&E uncertainties by a factor of 1.17. This improvement comes from the fact that letting $N_\mathrm{eff}$ vary degrades the precision of CMB constraints on $n_\mathrm{s}$, and adding \pone constraints help to break this degeneracy. The combination of CMB-SPA, DESI BAO, and DESI \pone gives $N_\mathrm{eff} = 3.02 \pm 0.10$, with DESI \pone measurements increasing the precision of the constraints from the combination of CMB-SPA and DESI BAO by a factor of 1.18. This increase is similar to the one found when adding DESI BAO constraints to CMB-SPA measurements \cite{Camphuis2025_spt3gd1}. For all data combinations, the constraints on $N_\mathrm{eff}$ are fully compatible with the Standard Model expectation.

\subsection{Sum of neutrino masses}
\label{sec:ext_neutrinos}

The detection of solar \cite{Davis1968_solar_neutrinos, Bahcall1976_solar_neutrinos, Anselmann1995_solar_neutrinos, Fukuda1998_solar_neutrinos} and atmospheric \cite{Fukuda1998_atmos_neutrinos} neutrino oscillations demonstrates that neutrinos are massive, with at least two species being non-relativistic today, which provides direct evidence for physics beyond the Standard Model. Oscillation experiments are sensitive to the difference between the squared masses of the neutrinos, with two of these masses close to each other and the other either larger --- normal ordering --- or smaller --- inverted ordering. By setting the lightest of the three masses to zero, we can determine the smallest possible sum of neutrino masses allowed by normal ordering, $\sum m_\nu > 58.980 \pm 0.304$ meV, and inverted ordering, $\sum m_\nu > 99.824 \pm 0.581$ meV \cite{Chebat2025_neutrinos}. Cosmological observations are sensitive to the sum of neutrino masses; therefore, when combined with constraints from neutrino oscillation experiments, a precise measurement of this sum would allow the individual neutrino masses to be inferred.

In the right panel of \cref{fig:import_mnu_nnu}, we show that the joint analysis of \Planck T\&E and DESI \pone yields 95\% upper limits on the sum of neutrino masses of $\sum m_\nu <_{95\%} 205$~meV, which is are a factor of 1.25 tighter than the \Planck T\&E upper limits. This gain is driven by the sensitivity of \pone measurements to the small-scale suppression of clustering induced by massive neutrinos. The combination of CMB-SPA, DESI BAO, and DESI Ly$\alpha$ data yields $\sum m\nu <_{95\%} 53.7$~meV\footnote{We quote limits allowing $\sum m_\nu < 60$ meV because we intend this as a diagnostic of whether the $\Lambda$CDM cosmological model is consistent with Standard Model neutrinos, not as a constraint on the latter.}, improving constraints from the the combination of CMB-SPA and DESI BAO by a factor of 1.03. Thus, \pone measurements add little constraining power beyond that provided by the combination of CMB and BAO data.

Nevertheless, it is important to emphasize that the combinations of CMB and \pone or BAO measurements are sensitive to neutrino masses through different physical mechanisms. While the combination of CMB and \pone is sensitive to the neutrino-induced suppression of the small-scale matter clustering, the combination of CMB and BAO primarily provides geometric constraints on $\sum m_\nu$ \cite{Loverde2024_neutrinos}. In particular, BAO measurements tightly constrain the matter density and the product of the Hubble parameter and the sound horizon at the drag epoch, thereby enhancing the sensitivity of CMB constraints through the neutrino-induced shift in the angular diameter distance to last scattering \cite{Garcia_Quintero2025_desiact}.

We also set constraints on the sum of neutrino masses using a frequentist profile likelihood analysis \cite{Herold2025_profile_like}. We briefly discuss this approach in what follows, and refer the reader to \cite{Chebat2025_neutrinos} for more details. We construct the profile likelihood by maximizing the log-likelihood over all cosmological and nuisance parameters but $\sum m_\nu$, which we fix at a set of equally spaced positive values starting from $\min(\sum m_\nu)=5~\mathrm{meV}$. The resulting curve corresponds to a $\chi^2$ distribution for a Gaussian likelihood, from which we set 95\% confidence level limits accounting for the physical boundary $\sum m_\nu>0$ following the Feldman-Cousins prescription \cite{Feldman1998_profile_like}.

\begin{figure}
    \centering
    \includegraphics[width=0.6\linewidth]{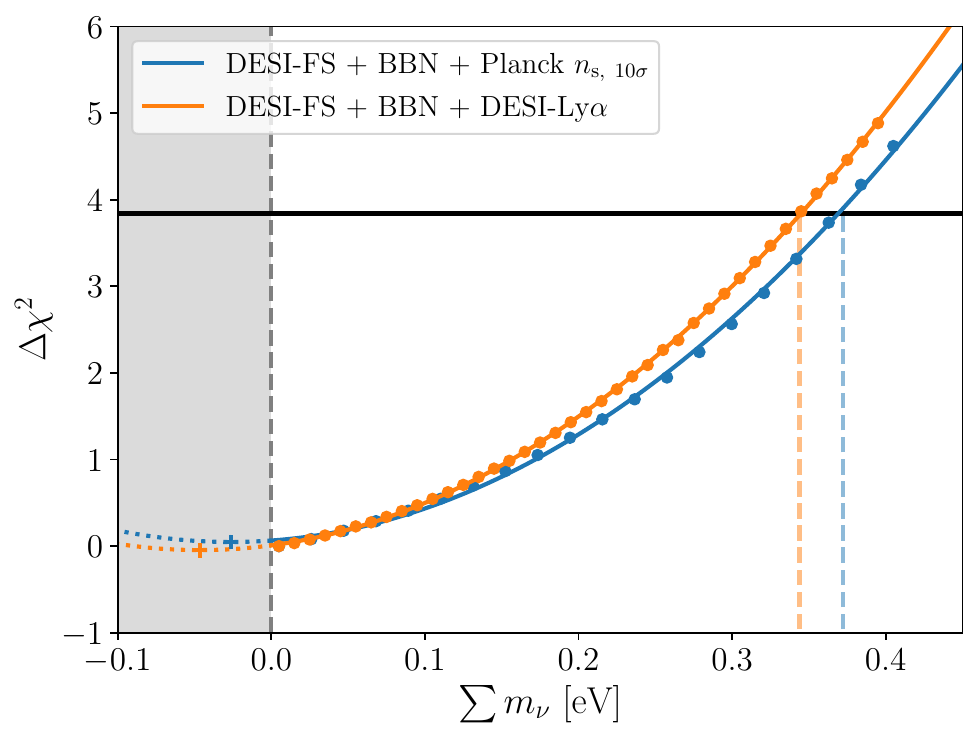}
    \caption{Profile likelihood constraints on the sum of neutrino masses. The blue color shows constraints from the combination of DESI FS, BBN and a $10\,\sigma$ prior on $n_\mathrm{s}$ from \Planck T\&E constraints, while the orange color does so for the combination of DESI-FS, BBN, and DESI \pone measurements. Dots mark the profile likelihood evaluations, while lines denote the corresponding best-fitting parabolas. The curves are extended to negative values of $\sum m_\nu$ to illustrate that no significant pull toward the unphysical regime is present. Crosses indicate the minimum of the parabolas.
    }
    \label{fig:mnu_chebat}
\end{figure}

In \cref{fig:mnu_chebat}, we display constraints from the combination of DESI FS, BBN, and a $10\,\sigma$ prior on $n_\mathrm{s}$ from \Planck T\&E constraints, as well as from the combination of DESI FS, BBN, and our measurements. Lines depict the best-fitting parabolas to the profile likelihoods, which are extended to negative values to verify the absence of any significant pull toward the unphysical regime. We find that the combination of DESI FS, BBN, and DESI \pone measurements yields $\sum m_\nu <_{95\%} 344$~meV fully independent of CMB data and with no significant pull towards negative neutrino masses. Furthermore, this upper limit is slightly tighter than that obtained from the combination of DESI FS, BBN, and the $10\,\sigma$ \Planck prior, which is commonly adopted in DESI FS analyses when deriving $\sum m_\nu$ constraints without explicitly including CMB data \cite{DESI2024.VII.KP7B, Chebat2025_neutrinos, Y3.cpe-s2.Elbers.2025}. The addition of DESI \pone measurements to the DESI FS and BBN combination therefore demonstrates that meaningful constraints on the sum of neutrino masses can be achieved without relying on any CMB information.

%% file: discussion_journal.tex
\section{Discussion}
\label{sec:discussion}

In this work, we derive cosmological constraints from the analysis of DESI DR1 \pone measurements using a novel implementation of \texttt{cup1d} that incorporates several improvements over previous analyses. Nonetheless, multiple avenues remain for further enhancing the robustness of these constraints or to tighten them further. We discuss several of these directions throughout the remainder of this section.

Our baseline analysis uses the \lacempg emulator, trained on a suite of 30 fixed-and-paired \mpgadget simulations \cite{Pedersen2021, cabayol-garcia2023NeuralNetworkEmulator}. Because of the limited number of simulations, emulator uncertainties are comparable to the DESI DR1 statistical errors up to $z=2.8$ (see \cref{sec:data_obs}), degrading the precision of the \deltastar and \nstar constraints by 30 and 19\%, respectively. Since statistical errors will substantially shrink in future DESI releases, expanding the simulation suite is essential to prevent emulator uncertainties from becoming the dominant source of error in future analyses. 

The limited number of simulations also forces us to rely on leave-one-out tests to estimate emulator errors, which tend to overestimate uncertainties because they often involve extrapolation beyond the training domain. Ideally, we would instead use multiple test simulations evenly distributed across the parameter space. Furthermore, even though the \lacempg emulator recovers the true cosmology when analyzing the \pone mocks based on the high-resolution \texttt{lyssa-central} and \texttt{sherwood} simulations, we need a better understanding of the impact of resolution, box size, and initial conditions on cosmological constraints.

It would also be valuable to refine the treatment of known systematics affecting the \pone measurements. Following \cite{Karacayli2025_p1d_dr1, Ravoux2025}, our baseline analysis assumes that different systematic uncertainties are fully correlated across scales and uncorrelated with one another. This is a simplification that warrants further investigation because, as shown in \cref{sec:results_robust_data}, adopting the alternative assumption --- used in multiple previous analyses of SDSS data \cite{palanque-delabrouille2015ConstraintNeutrinoMasses, p1d_Chabanier2019, Walther2025_lyssa} --- that systematic uncertainties are uncorrelated across scales increases the fit probability by a factor of two compared to the baseline analysis.

We also need to perform additional tests to identify unknown systematics. A standard approach is to search for internal inconsistencies across data subsamples. In this work, we only split the sample by redshift; future analyses should explore a broader range of splits --- such as those considered in \cite{Karacayli2025_p1d_dr1, Ravoux2025} --- to better identify potential correlations between data selection choices and the inferred cosmological constraints.

Another key improvement would be to perform a joint analysis of DESI measurements and high-resolution data. While the mean flux, the amplitude and slope of the temperature–density relation, and pressure smoothing have similar effects on the absolute value of \pone at large scales, their impact differs substantially on small scales. Therefore, high resolution measurements would help to break these degeneracies and, as shown in \cref{sec:model_validation}, significantly tighten cosmological constraints. However, such a study would require extending the range of our emulator and a more careful assessment of the impact of simulation resolution on the results. Another possible approach is to impose informative priors on these parameters based on external analyses; for example, a recent reanalysis of eBOSS data required a prior on the mean flux to obtain meaningful cosmological constraints \cite{Walther2025_lyssa}. We do not pursue such an analysis here, as we have not validated that \texttt{cup1d} recovers IGM parameters in an unbiased manner (see \cref{sec:model_validation}). In the absence of this validation, informative IGM priors could introduce biases in the inferred cosmological constraints.

The two most uncertain aspects of our analysis concern the modeling of metal and HCD contamination. The metal and HCD models employed in previous analyses of SDSS data \cite{palanque-delabrouille2015ConstraintNeutrinoMasses, p1d_Chabanier2019, Walther2025_lyssa} are strongly disfavored by DESI data, and neglecting the modeling of LLS or \siisii contamination leads to significant shifts in the inferred cosmological parameters. The \siisii line pair is robustly detected in the one-dimensional correlation function of DESI DR1 data \cite{Karacayli2025_p1d_dr1} and in our analysis, as it produces a distinctive oscillatory feature that cannot be reproduced by variations in cosmological, IGM, or HCD parameters. Nevertheless, the parameters describing the amplitude and damping of the \siisii contamination show little correlation with the cosmological parameters. Conversely, while omitting LLS contamination only slightly reduces the fit probability, the parameters describing this component are strongly correlated with the compressed parameters. It is therefore concerning that possible small deviations from the assumed functional form for LLS contamination could induce noticeable variations in the inferred parameters. Future analyses should aim to improve the modeling of this contaminant. Note that external priors on the abundance of LLSs would also increase the robustness of the results.

Finally, future analyses should assess the impact on cosmological constraints of additional physical effects not considered in this work, including large-scale fluctuations in the ionizing background \cite{Fan2006_ionizingfluct, pontzen2014ScaledependentBiasBaryonicacousticoscillationscale, gontchoagontcho2014EffectIonizingBackground, Becker2015_ionizingfluct, Bosman2018_ionizingfluct}, temperature fluctuations due to inhomogeneous reionization \cite{Aloisio2015_patchyreio, Montero_Camacho2019_patchyreio}, inhomogeneous helium reionization \cite{Miralda1994_reioheii, Hui1997_reioheii, Hui2003_reioheii, Upton_Sanderbeck2020_heii}, and feedback processes associated with galaxy formation \cite{Viel2013_agn, chabanier2020ImpactAGNFeedback, Tillman2025_agn}.

%% file: conclusions_journal.tex
\section{Summary and conclusions}
\label{sec:conclusions}

In this work, we analyzed the one-dimensional \lyaf flux power spectrum (\pone) from the first data release of the DESI survey \cite{Karacayli2025_p1d_dr1, Ravoux2025} using a new version of the publicly available likelihood code \texttt{cup1d}\footnote{\url{https://github.com/igmhub/cup1d}}. We summarize our main findings below.

\begin{itemize}
    \item We analyze \pone measurements from $z=2.2$ to 4.2 in bins of $\Delta z=0.2$ obtained with the optimal quadratic maximum likelihood estimator (QMLE) from forests with an average signal-to-noise ratio (SNR) greater than 3 per pixel in the \lya region \cite{Karacayli2025_p1d_dr1}, a sample selected to minimize systematic uncertainties.
    
    \item We train an emulator on a suite of hydrodynamical simulations to capture the dependence of \pone on cosmology and intergalactic medium (IGM) physics. The emulator attains sub-percent accuracy over the scales relevant to the DESI analysis, and its uncertainties are incorporated into the total error budget for cosmological inference.
    
    \item We model the impact of metal lines and high column density systems on emulator predictions using analytical expressions inspired by \cite{McDonald2006, Karacayli2025_p1d_dr1, Karacayli2023_doublet} and \cite{Rogers2018a}, respectively. In total, our model includes 53 free parameters: 2 cosmological parameters, 16 describing the dependence of the signal on IGM physics, 24 accounting for metal and HCD contamination, and 11 to marginalize over small biases in the characterization of the DESI spectrograph resolution.
    
    \item We set constraints on the amplitude, $\deltastar = 0.379^{+0.032}_{-0.033}$, and logarithmic slope, $\nstar = -2.309^{+0.019}_{-0.019}$, of the linear matter power spectrum at the pivot scale \kstarval and redshift $z_\star = 3$, which are fully consistent with temperature and polarization measurements from \Planck \cite{Planck2018} and can be found here\footnote{\url{https://github.com/igmhub/cobaya_lya_p1d}}. We validate these constraints through the analysis of \pone mocks based on distinct hydrodynamical simulations and an extensive suite of alternative data analyses in which we vary the measurements, covariance matrix, emulator, and the modeling of contaminants. We kept the constraints blinded during the model validation and optimization to avoid experimenter bias. 

    \item We combine our measurements with DESI DR2 BAO measurements (DESI BAO \cite{DESI.DR2.BAO.lya, DESI.DR2.BAO.cosmo}) and temperature, polarization, and lensing data from \Planck, ACT, and SPT-3G (CMB-SPA \cite{Camphuis2025_spt3gd1}) to constrain the effective number of relativistic species, $N_\mathrm{eff}=3.02\pm{}0.10$, the running of the spectral index, $\alpha_\mathrm{s}=0.0014\pm{}0.0041$, and the running of the running, $\beta_\mathrm{s}=-0.0006\pm{}0.0048$, improving the precision of constraints from the analysis of CMB-SPA and DESI BAO measurements by a factor of 1.18, 1.27, and 1.90, respectively.
    
    \item We also combine our measurements with \Planck T\&E constraints to set upper limits on the sum of neutrino masses of $\sum m_\nu <_{95\%} 205$~meV, which increase the precision of \Planck T\&E upper limits by a factor of 1.25 due to the sensitivity of \pone to the small-scale suppression of the clustering caused by massive neutrinos. On the other hand, the combination of CMB-SPA, DESI BAO, and our measurements yields $\sum m_\nu <_{95\%} 53.7$~meV, only improving the precision of constraints from the combination of CMB-SPA and DESI BAO by a factor of 1.03. Finally, we combine our measurements with DESI full-shape \cite{DESI2024.VII.KP7B} and BBN \cite{schoneberg2024_bbn} constraints to set $\sum m_\nu <_{95\%} 344$~meV upper limits fully independent of CMB data.
\end{itemize}

In \cref{sec:discussion}, we outline several avenues to improve future \pone analyses. We note that increasing the number of training simulations will be essential for preventing emulator uncertainties from dominating the error budget of future DESI analyses. We also discuss possible refinements to the cosmological parameterization to enhance robustness, and emphasize the need for a more thorough characterization of systematic uncertainties. Moreover, we discuss that a joint analysis combining DESI and high-resolution \pone measurements would break key parameter degeneracies and substantially tighten cosmological constraints. Finally, we mention that future work should assess the sensitivity of the results to small variations in the metal and HCD contamination models, and explore additional physical effects not captured in our framework, including large-scale ionizing-background fluctuations, inhomogeneous helium reionization, and galaxy formation feedback.

%% file: app_covariance_journal.tex
\section{Emulator covariance}
\label{app:emu_covariance}

In \cref{sec:emulator_performance}, we use leave-one-out tests in order to estimate the average emulator accuracy. In this section, we employ these tests to estimate whether emulator errors are correlated among scales of redshifts and compute an emulator covariance matrix, which we employ when analyzing \pone data in \cref{sec:data_obs}.

\begin{figure}
    \centering
    \includegraphics[width=0.9\linewidth]{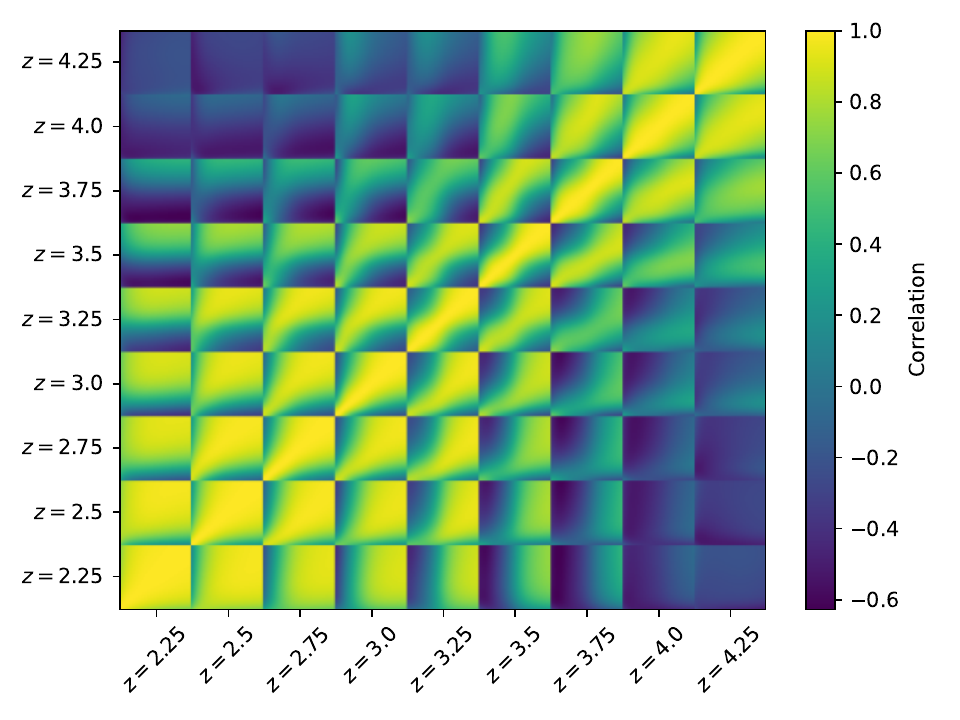}
    \caption{Correlation of emulator errors as a function of scale and redshift. Each small square in the diagonal displays the correlation as a function of scale for a particular redshift, while the off-diagonal squares depict the correlation between different redshifts.}
    \label{fig:emu_corr_matrix}
\end{figure}

We compute the emulator covariance matrix by comparing simulation measurements and predictions from the 30 leave-one-out emulators for each of the wavenumbers of the 11 snapshots of the \mpgadget simulations. Therefore, the resulting covariance matrix captures errors coming from the sparsity of the training set. In \cref{fig:emu_corr_matrix}, we display the corresponding correlation matrix, which shows that emulator errors are strongly correlated as a function of scale and redshift. It is important to note that we smooth the simulation measurements according to the process described in \cref{sec:emulator_implementation} to mitigate the impact of cosmic variance resulting from the limited size of the simulations. We do not incorporate to the covariance matrix errors coming from inaccuracies in the smoothing procedure since we estimate such errors to be approximately a factor of 5 smaller than those caused by the sparsity of the training set.

To check the robustness of the emulator covariance matrix, we computed 30 new covariance matrices while using only 29 out of the 30 leave-one-out tests at a time. Then, we computed the standard deviation of these covariance matrices, which provides an estimate of the error affecting the original covariance matrix. We found that the error is of the order of a few percent, and thus we do not expect of our estimation of the covariance matrix to be too noisy due to the limited number of the leave-one-out tests. Furthermore, we checked that there is no systematic bias between simulation measurements and emulator predictions across the parameter space.

It is important to note that we compute the emulator covariance for the native redshifts and wavenumbers of the \mpgadget simulations. To add this source of uncertainty to the statistical and systematic covariance matrices, we interpolate the emulator covariance matrix to the wavenumbers and redshifts of DESI measurements. We verify that the resulting covariance matrix can be inverted without numerical instability. We do not incorporate errors resulting from cosmic variance since, as shown in \cref{sec:emulator}, their size is approximately one order of magnitude smaller than that of emulator errors. Furthermore, we do not consider errors due to the limited resolution of the \mpgadget simulations, but we validate emulator predictions using the high-resolution \lyssa and \texttt{sherwood} simulations in \cref{sec:model_validation}. 

%% file: app_compressed_params_journal.tex
\section{Compressed parameters}
\label{app:compressed}

Throughout this work, we compress the cosmological information contained in the \pone measurements into the small-scale amplitude and logarithmic slope of the linear power spectrum. During inference, for computational efficiency, we hold fixed the value of the Hubble parameter and the physical density of cold dark matter and baryons. In \cref{sec:results_robust}, we show that constraints in the $\deltastar$–$\nstar$ plane are substantially less sensitive to variations in the fixed parameters than in the $A_\mathrm{s}$–$n_\mathrm{s}$ plane. In this section, we explore whether this sensitivity can be further reduced by adopting an alternative compressed parameterization.

We perform the following exercise to assess how efficiently the compressed parameters capture cosmological information. First, we compute the linear power spectrum for the \Planck~2018 $\Lambda$CDM cosmology and integrate it to obtain the corresponding linear prediction for \pone
\begin{equation}
    \label{eq:p1d}
    \pone(k_\parallel)=(2\pi)^{-1}\int_0^\infty \mathrm{d} k_\perp\, k_\perp\, P_\mathrm{lin}(k_\parallel,\, k_\perp) \exp(-k^2/k^2_\mathrm{pressure}),
\end{equation}
where $k_\parallel$ and $k_\perp$ denote the parallel and perpendicular components of the wavevector, respectively, $k^2 = k_\parallel^2 + k_\perp^2$, and $k_\mathrm{pressure} = 0.4\,\ikms$ is the pressure smoothing scale, a value we choose to reduce the dependency of the linear \pone on the value of the linear power spectrum on very small scales.

Then, we select an alternative cosmology, and compute its corresponding compressed parameters. After that, we rescale the value of $A_\mathrm{s}$ and $n_\mathrm{s}$ of the alternative cosmology so the value of the compressed parameters for this new cosmology matches the ones for the initial cosmology \cite{Pedersen2023}. Finally, we evaluate \cref{eq:p1d} in order to obtain the linear \pone for the rescaled cosmology. Any difference between the linear \pone for the initial and rescaled cosmology would indicate that the compressed parameters do not completely capture the cosmological dependence of this observable.

\begin{figure}
    \centering
    \includegraphics[width=0.475\linewidth]{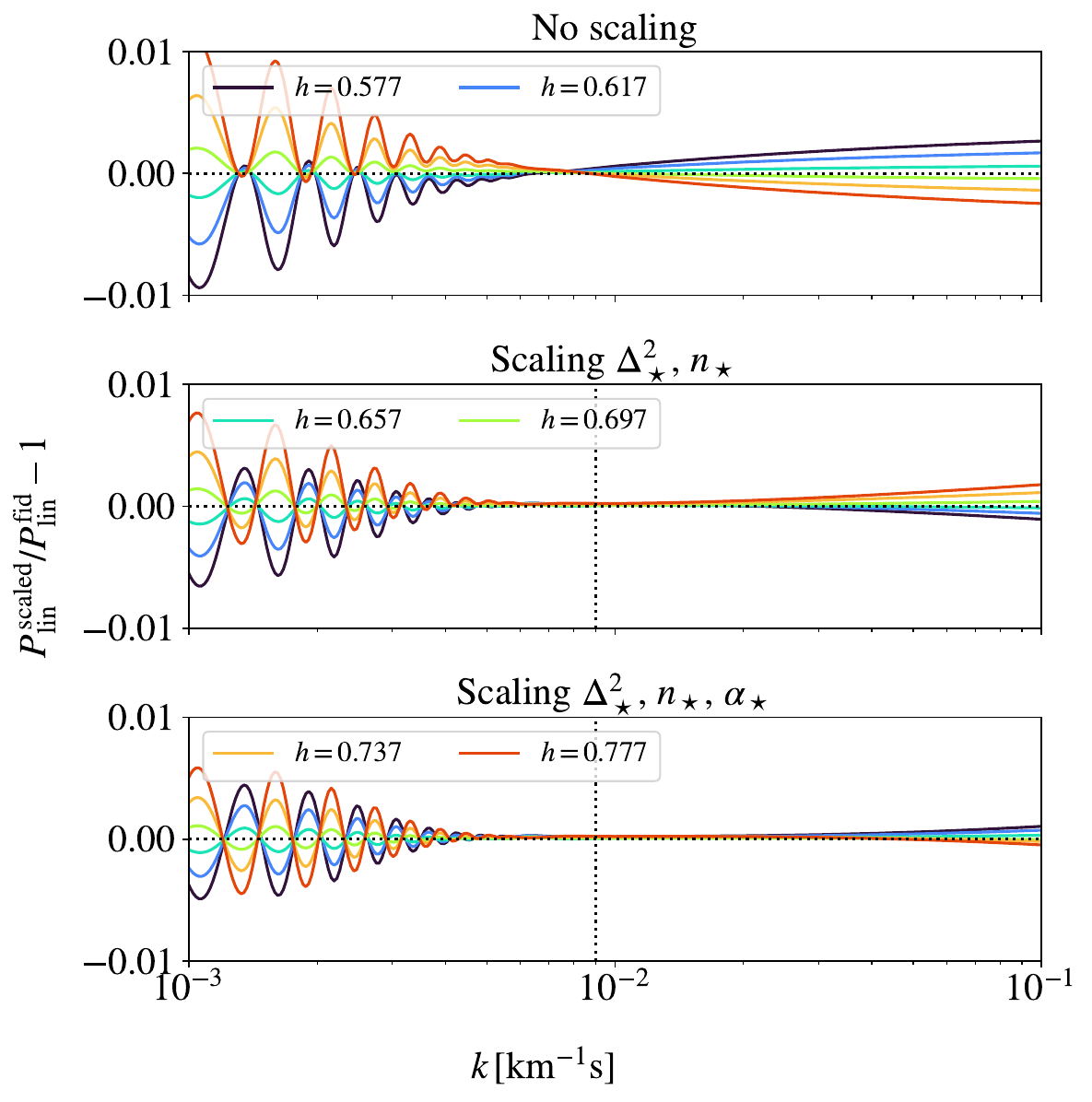}
    \includegraphics[width=0.475\linewidth]{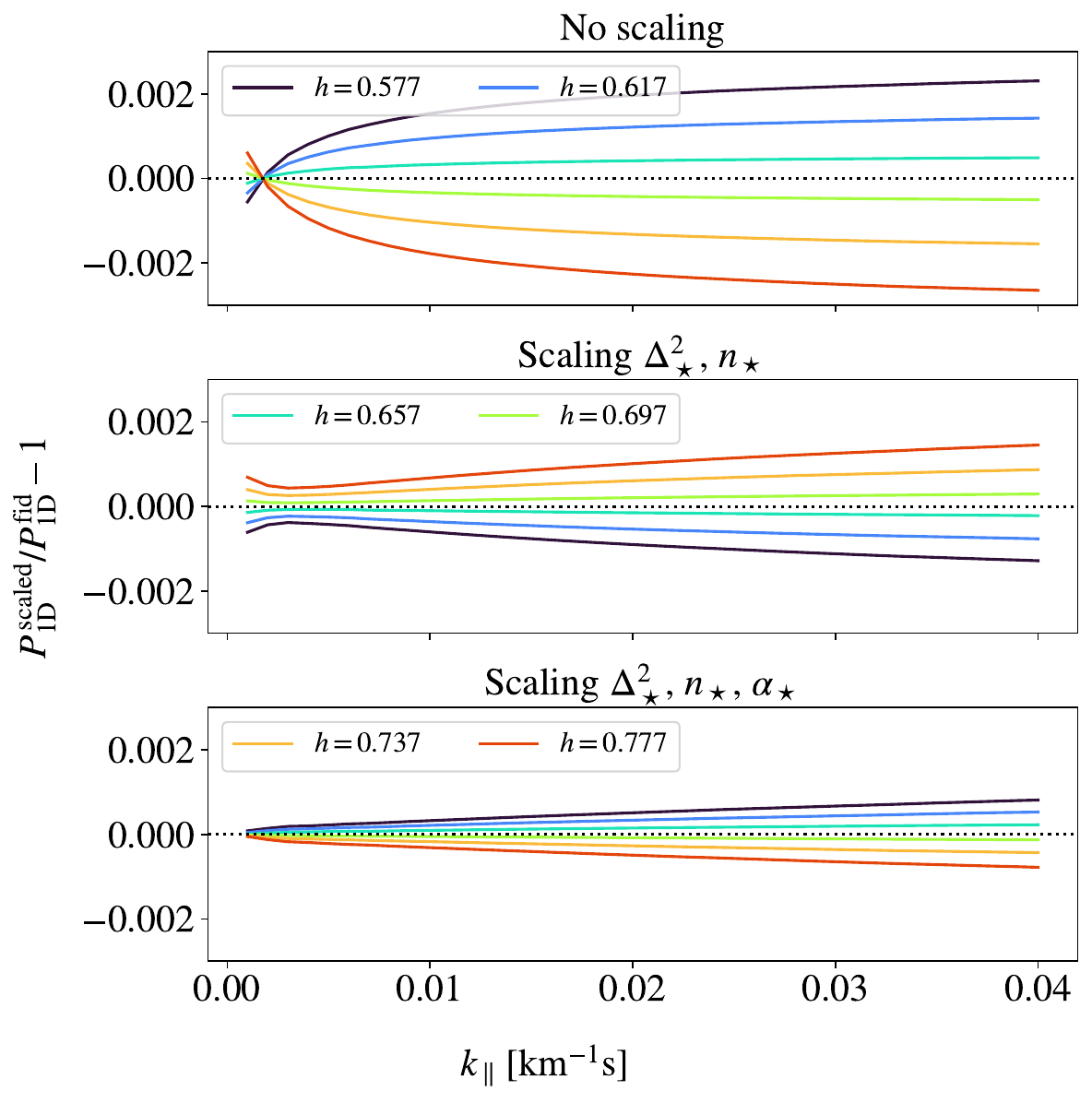}
    \caption{
    Effect of information compression on the linear power spectrum (left) and linear \pone (right) at $z=3$ when varying the Hubble parameter while holding the physical densities fixed. The top, middle, and bottom panels show the residuals relative to the fiducial cosmology, after rescaling the varied cosmology to match \deltastar and \nstar, and after additionally matching $\alpha_\star$, respectively. In the middle- and bottom-left panels, the vertical dashed lines indicate the value of $k_\star$.
    }
    \label{fig:scaling_h}
\end{figure}

In the top-left (top-right) panels of \cref{fig:scaling_h}, we show the residuals between the linear power spectrum (linear \pone) at $z=3$ for cosmologies with different values of the Hubble parameter and for the fiducial cosmology. These variations are taken symmetrically around the value of the Hubble parameter for the fiducial cosmology while keeping the physical baryon and matter densities fixed. The residuals remain below 1.0\% for the linear power spectrum and 0.3\% for the linear \pone, reflecting the near–Einstein–de Sitter behavior of the universe at $z=3$. 

The middle panels show the residuals after rescaling $A_\mathrm{s}$ and $n_\mathrm{s}$ in the varied cosmologies to match the value of the compressed parameters for the fiducial cosmology, which reduces the residuals by a factor of two for the linear \pone. This result explains the consistency between the posteriors for the baseline analysis and the variation adopting a \Planck cosmology with $h=0.74$ (see \cref{sec:results_robust_cosmo}). The bottom panels repeat the exercise after additionally rescaling $\alpha_\mathrm{s}$ in the varied cosmologies to match the small-scale curvature of the linear power spectrum for the fiducial cosmology
\begin{equation}
    \label{eq:alpha_star}
    \alpha_\star = \frac{\mathrm{d}^2\log P_\mathrm{lin}(k, z)}{\mathrm{d}\log k^2}\bigg|_{k_\star, \, z_\star}.
\end{equation}
As shown, matching the small-scale curvature further reduces the residuals.

\begin{figure}
    \centering
    \includegraphics[width=0.475\linewidth]{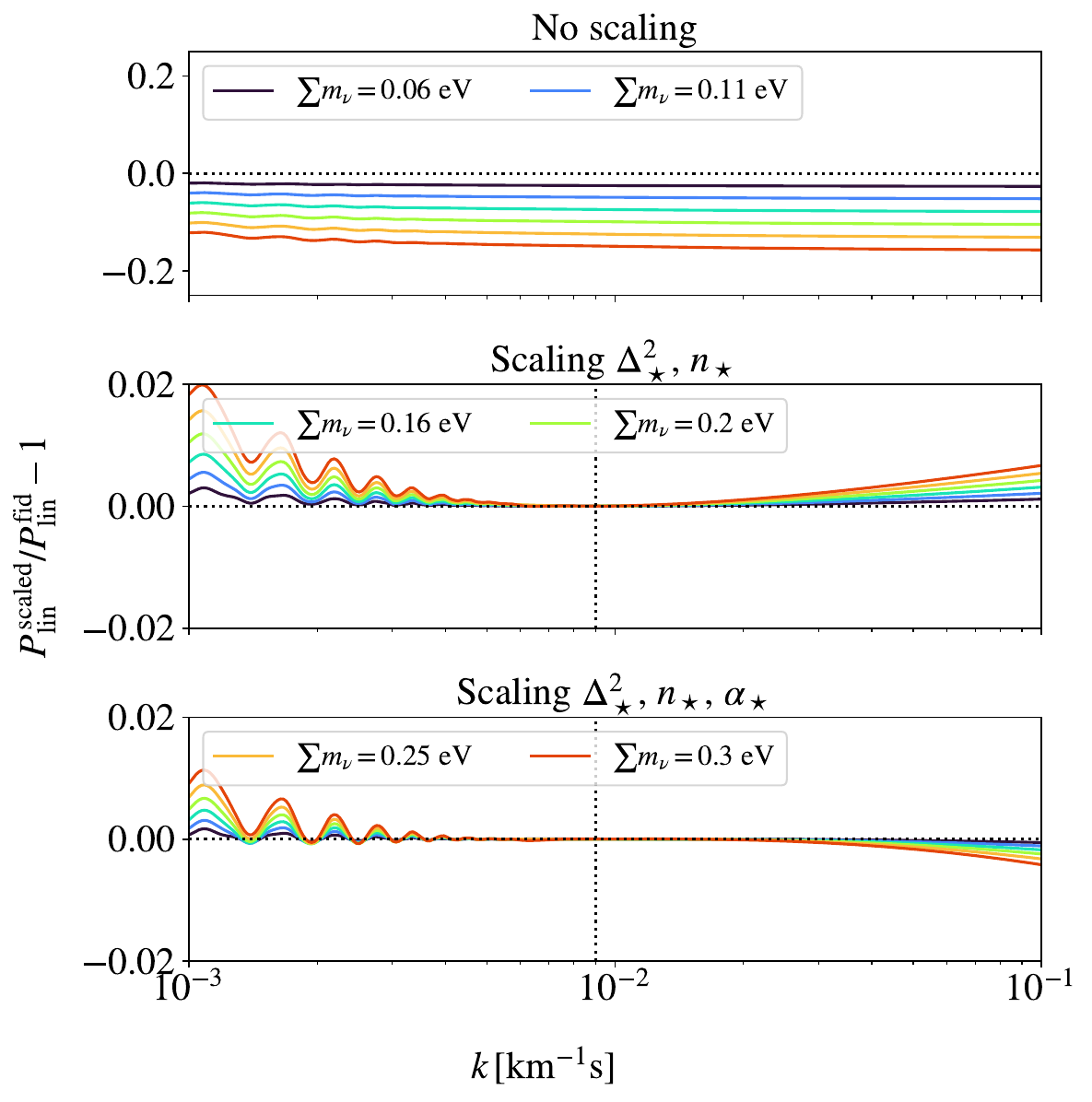}
    \includegraphics[width=0.475\linewidth]{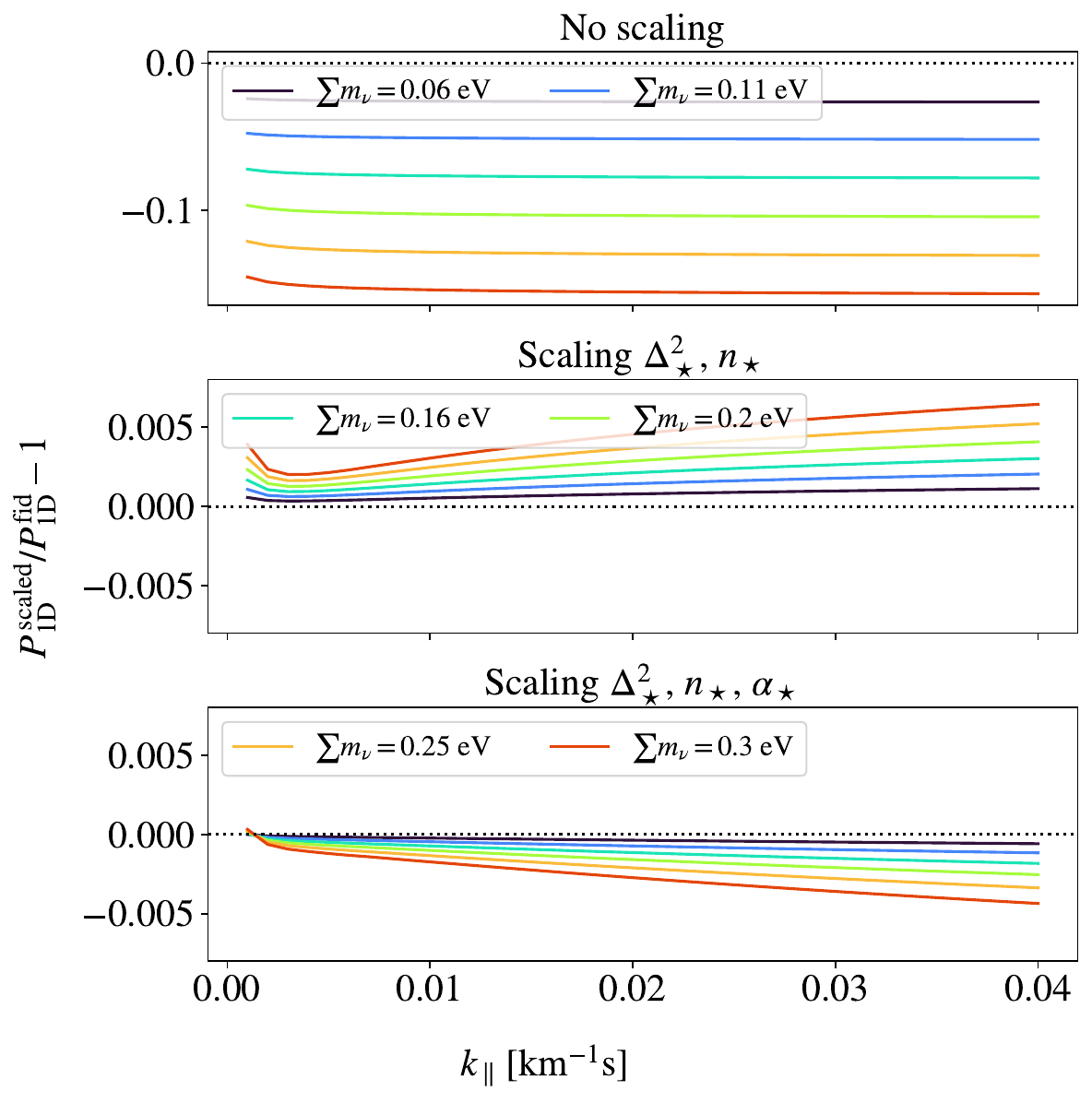}
    \caption{Same as \cref{fig:scaling_h} but for variations in the sum of neutrino masses at fixed physical densities.}
    \label{fig:scaling_mnu}
\end{figure}

In \cref{fig:scaling_mnu}, we repeat the same exercise as before but for variations in the sum of neutrino masses at fixed physical matter density. As expected, increasing the sum of neutrino masses decreases the value of the linear power spectrum on small scales. However, rescaling the cosmologies to match the value of the compressed parameters reduces these differences significantly. The residual for the linear \pone approaches 0.7\% for the most extreme case of $\sum m_\nu=0.3$ eV, corresponding to the cosmology adopted in one of the alternative analyses considered in \cref{sec:results_robust_cosmo}, which explains the small shift observed in the posterior for this variation.

\begin{figure}
    \centering
    \includegraphics[width=0.475\linewidth]{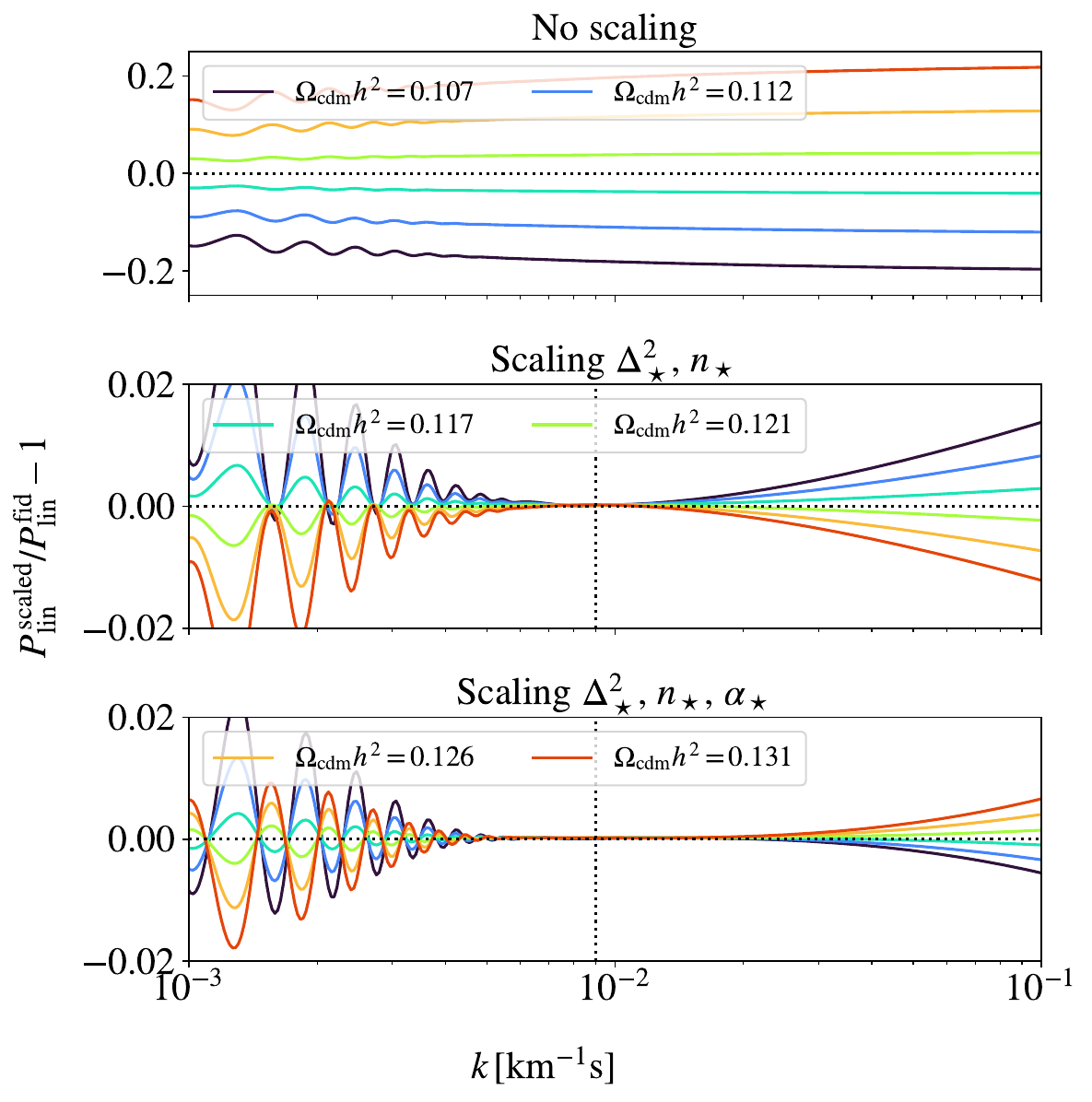}
    \includegraphics[width=0.475\linewidth]{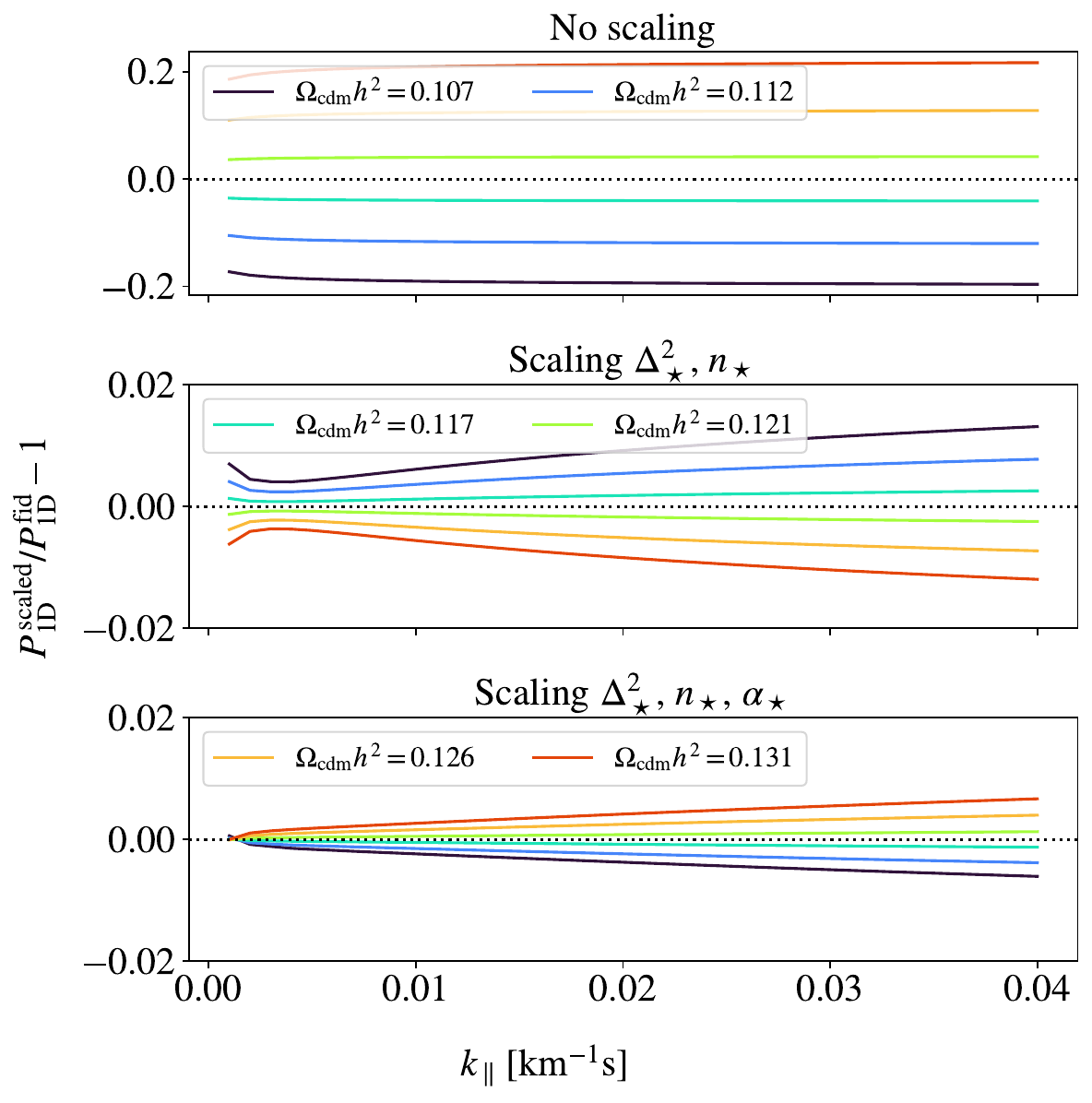}
    \caption{Same as \cref{fig:scaling_h} but for variations in the physical cold dark matter density at fixed value of the Hubble parameter. The cosmologies analyzed span $10\,\sigma$ around the best-fitting \Planck 2018 $\Lambda$CDM cosmology.}
    \label{fig:scaling_omh2}
\end{figure}

In \cref{fig:scaling_omh2}, we perform a similar exercise for up to $10\,\sigma$ variations in the physical cold dark matter density around \Planck constraints at fixed value of the Hubble parameter. As shown, the variations in the linear power spectrum are stronger for the most extreme cases than when varying the Hubble parameter or the sum of neutrino masses. Applying the rescaling of the small-scale amplitude and slope reduces the residual for the linear \pone by more than an order of magnitude. However, the residual for the extreme positive and negative cases are above 1\%. As we can see in the bottom panels, further rescaling the small-scale curvature significantly reduces the residuals. However, we do not include $\alpha_\star$ in our analysis because the \lacempg emulator was trained on a suite of hydrodynamical simulations in which only $A_\mathrm{s}$ and $n_\mathrm{s}$ were varied (see \cref{sec:data_sim}), and thus the value of $\alpha_\star$ is the same for all the simulations in the training set.

%% file: app_lyssa_emulator_journal.tex
\section{The \lacelyssa emulator}
\label{app:lace-lyssa}

Throughout most of this work, we employed an emulator trained on a suite of \mpgadget simulations, \lacempg. In this section, we introduce \lacelyssa, an alternative emulator trained on the suite of \texttt{lyssa} simulations described in \cref{sec:data_sim_lyssa}.

We train this emulator following the same methodology as the one adopted to train \lacempg (see \cref{sec:emulator}), with the sole difference being the number of input parameters. To enhance the accuracy of the information compression technique (see \cref{app:compressed}), we additionally include the curvature of the linear power spectrum at a pivot scale of $k_\mathrm{p}=0.7\,\iMpc$ and redshift $z$, denoted $\alpha_\mathrm{p}(z)$. This parameter can be incorporated because the \lyssa simulations span different values of the physical cold dark matter density. Note that we do not include the simulation \texttt{lyssa-14} in the training set due to known inconsistencies with other simulations.

\begin{figure}
    \centering
    \includegraphics[width=0.495\linewidth]{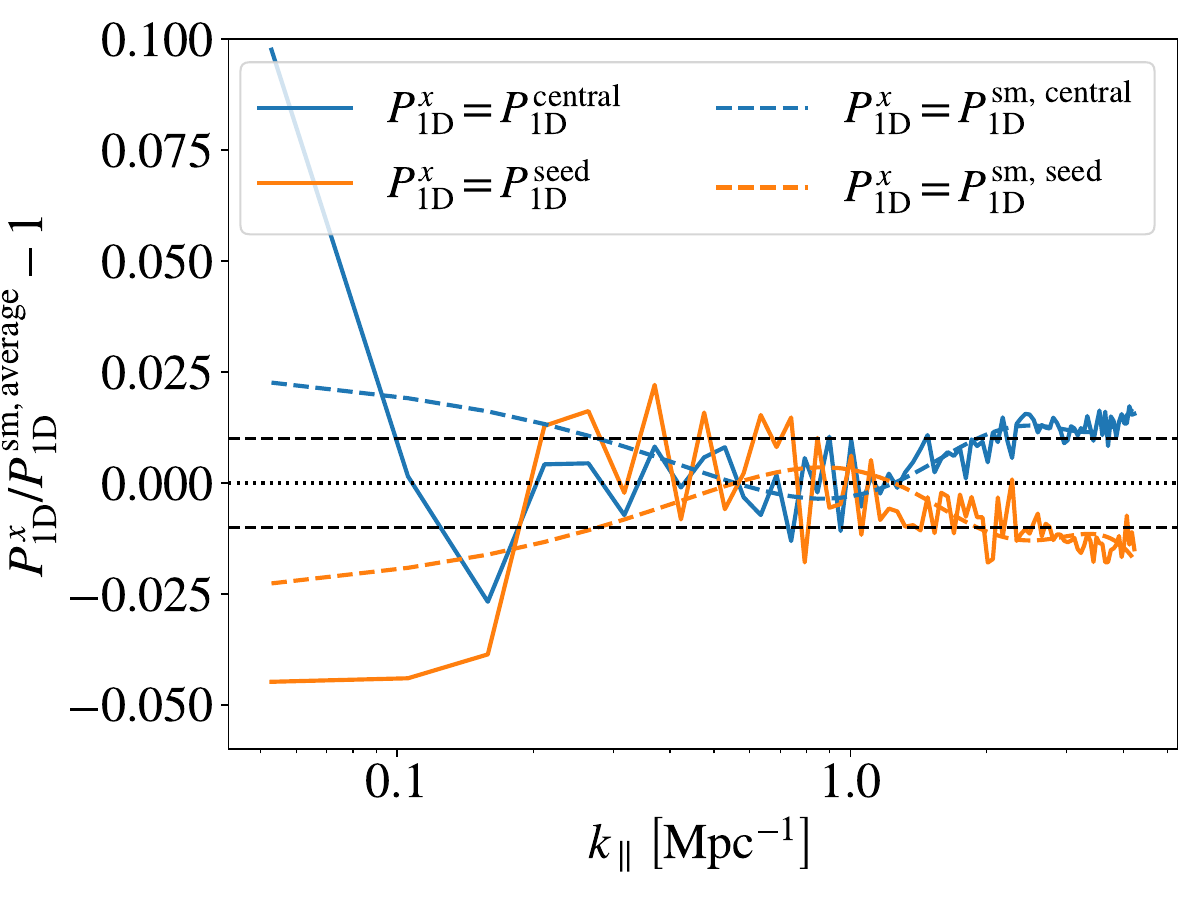}
    \includegraphics[width=0.495\linewidth]{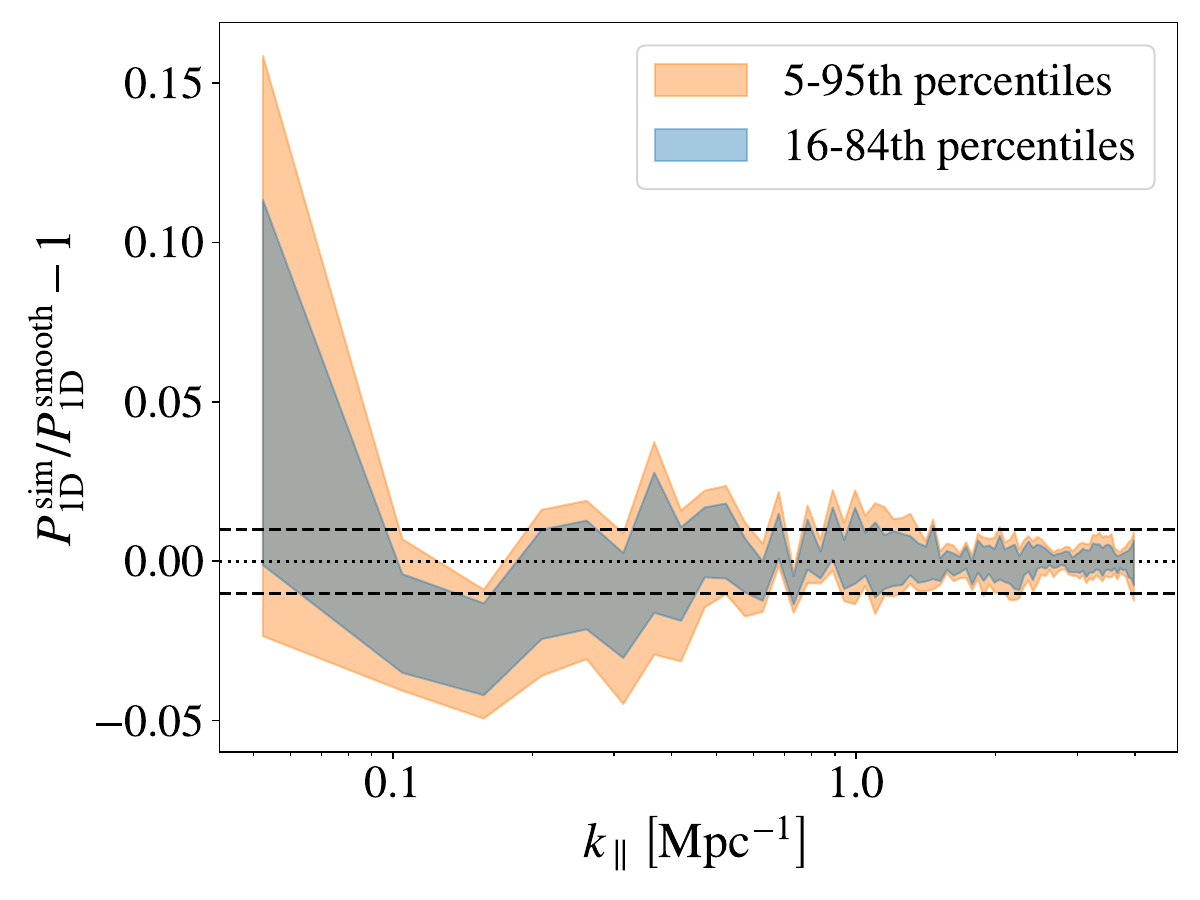}
    \caption{Same as \cref{fig:emulator_smooth} for the \lacelyssa emulator. In the left panel, we show the ratio between different samples and the mean of the best-fitting smooth models to the \texttt{lyssa-central} and \texttt{lyssa-seed} simulations at $z = 3$. The blue and orange solid lines correspond to ratios using \texttt{lyssa-central} and \texttt{lyssa-seed} predictions in the numerator, respectively, while the blue and orange dashed lines show the ratios obtained from their respective best-fitting smooth models. In the right panel, the blue and orange shaded areas show the 16 to 84th and the 5 to 95th percentile regions, respectively, of the relative difference between \pone predictions from all the \lyssa simulations in the training set and the best-fitting smooth model to each of these.}
    \label{fig:lyssa_emulator_smooth}
\end{figure}

In the left panel of \cref{fig:lyssa_emulator_smooth}, we evaluate the accuracy of the smooth \pone model (\cref{eq:psmooth}) in reproducing predictions from \texttt{lyssa-central} and \texttt{lyssa-seed}, an additional simulation run with the same configuration as \texttt{lyssa-central} but a different initial Fourier phase distribution. We show the ratio between each simulation sample and the mean of the best-fitting smooth models to each of the two simulations. The differences between the simulation predictions and the average smooth model reach up to 10\% due to cosmic variance, while the largest differences for the \mpgadget simulations were smaller than 2\%. Consequently, even though the \lyssa simulations have a box size roughly 75\% larger than that of the \mpgadget simulations, the impact of cosmic variance on the latter is much smaller due to the fixed-and-paired technique used to generate their initial conditions \cite{angulo2016CosmologicalNbodySimulations, anderson2019CosmologicalHydrodynamicSimulations, fixedpaired_Villaescusa}.

\begin{sloppypar}
On the other hand, the difference between the best-fitting smooth model to the \texttt{lyssa-central} and \texttt{lyssa-seed} simulations and their average remains below 2.5\%, indicating that applying a smooth model to pre-process the emulator input reduces the impact of cosmic variance by roughly a factor of four. However, the impact of this residual cosmic variance on the smooth models is still nearly an order of magnitude larger than the one affecting the smooth models to the \mpgadget simulations, underscoring again the advantage of the fixed-and-paired technique.
\end{sloppypar}

In the right panel of \cref{fig:lyssa_emulator_smooth}, we show the accuracy of the smooth model in reproducing \pone predictions from the all the \lyssa simulations in the training data. The largest deviations occur at the same wavenumbers as for the \texttt{lyssa-central} simulation, reflecting the fact that the performance of the smooth model is similar for different redshifts and cosmologies and that all training simulations share the same initial Fourier phase distribution. It is important to note that the largest deviations reach up to 15\% at the $2\,\sigma$ level, nearly an order of magnitude larger than those observed for the \mpgadget simulations.

\begin{figure}
    \centering
    \includegraphics[width=0.475\linewidth]{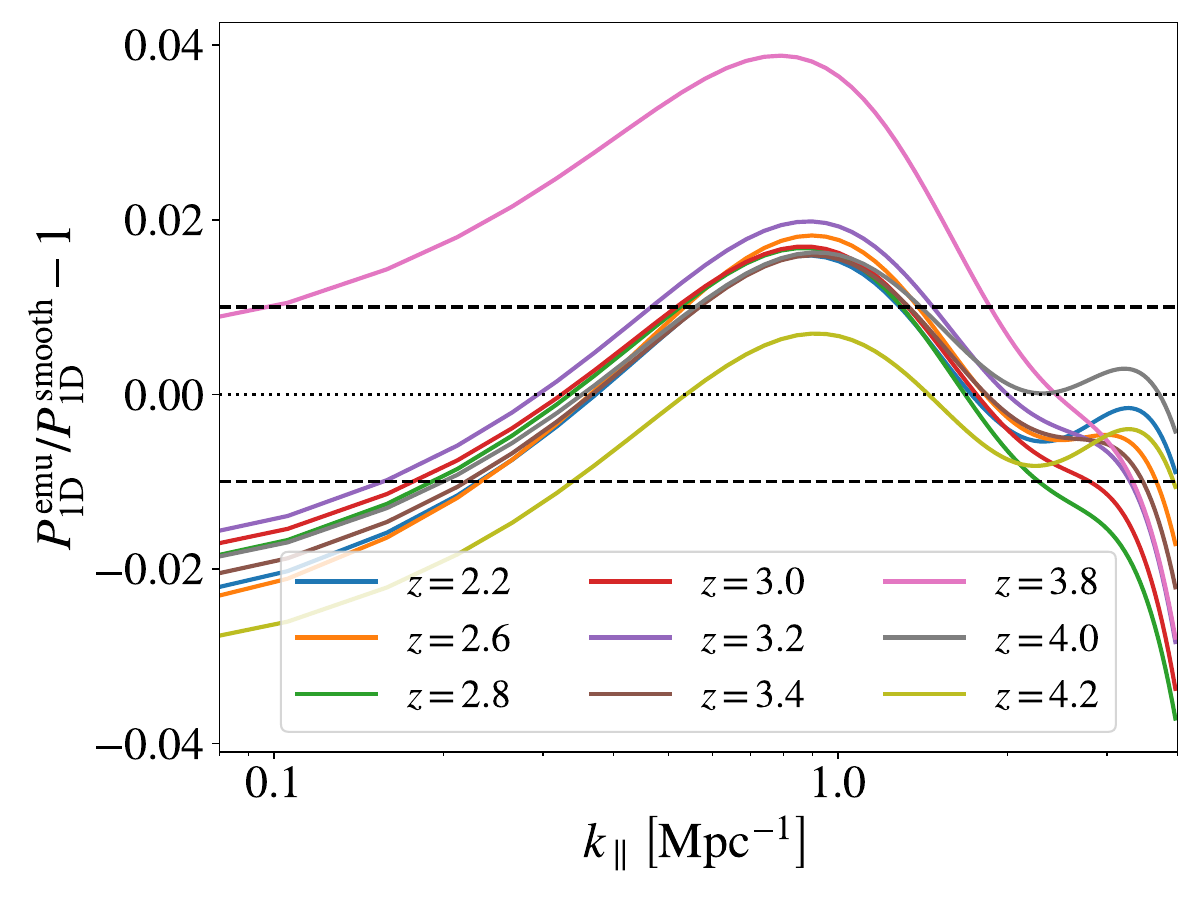}
    \includegraphics[width=0.475\linewidth]{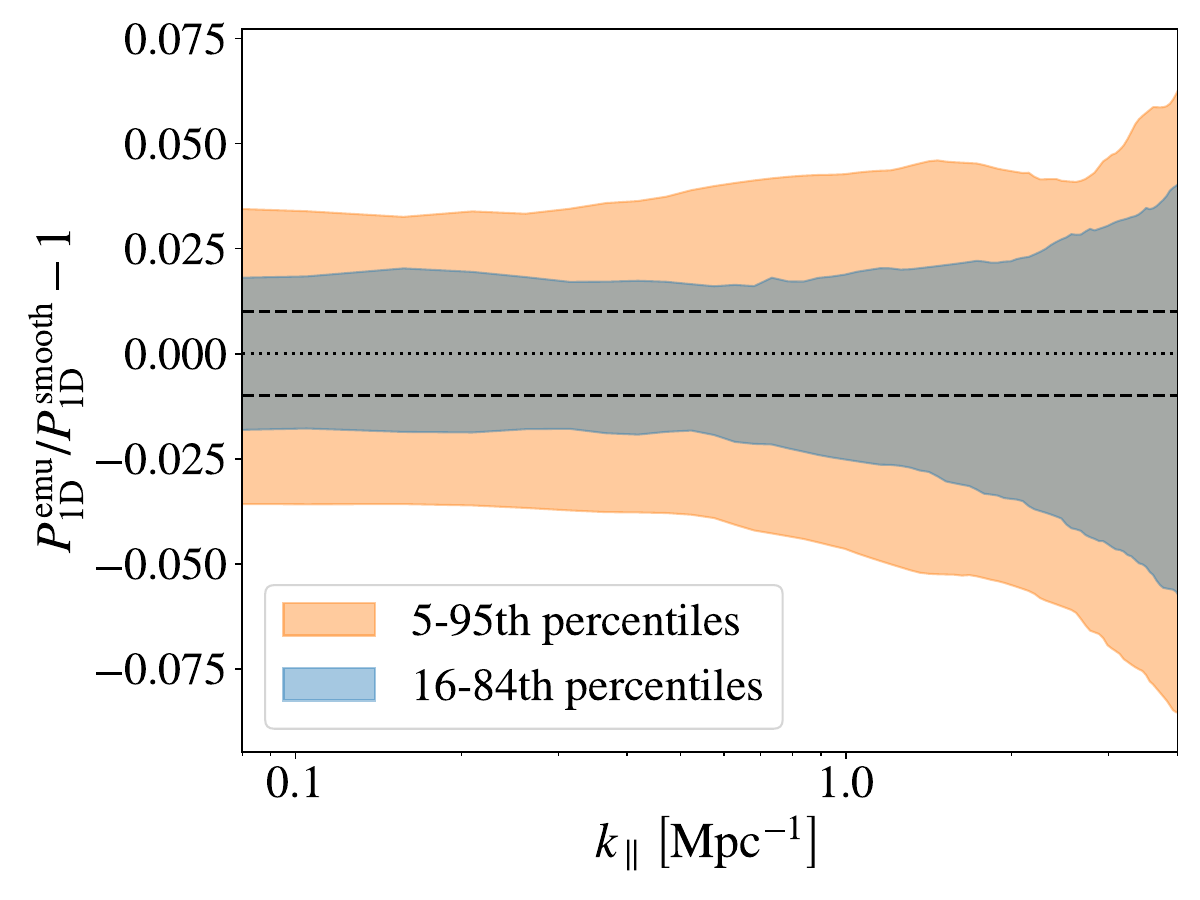}
    \caption{Same as \cref{fig:emulator_performance} but for the \lacelyssa emulator. In the left panel, we show the relative difference between emulator results and smoothed predictions from the \texttt{lyssa-seed} simulation. In the right panel, the blue and orange shaded areas show the 16 to 84th and the 5 to 95th percentile regions, respectively, of 14 leave-one-out tests.
    }
    \label{fig:lyssa_emulator_performance}
\end{figure}

In the left panel of \cref{fig:lyssa_emulator_performance}, we show the performance of \lacelyssa in reproducing \pone predictions from the \texttt{lyssa-seed} simulation. The differences reach up to 4\% and follow the same trend as the discrepancies between \texttt{lyssa-seed} predictions and the best-fitting smooth model to these shown in the left panel of \cref{fig:lyssa_emulator_smooth}. For comparison, the \lacempg emulator reproduces \pone predictions from \texttt{mpg-seed} with an accuracy better than 1.5\% across all scales. This indicates that the impact of cosmic variance on the predictions from the \lacelyssa emulator is nearly three times larger than on the predictions from the \lacempg emulator.

In the right panel of \cref{fig:lyssa_emulator_performance}, we present the results of 14 leave-one-out tests for the \lacelyssa emulator. Three simulations from the training set were excluded from the test because some of the IGM input parameters required for emulation were unavailable. As shown, the \lacelyssa emulator achieves an accuracy better than 2\% at $1\,\sigma$ for most scales, whereas the \lacempg emulator exhibits approximately twice the precision. This is because the number of \mpgadget simulations is approximately a factor of two larger than the number of \lyssa simulations, and the latter sample a much broader range of cosmological parameters. Figure~4 of \cite{Walther2025_lyssa} shows that emulators trained on the \lyssa simulations achieve better than 1\% precision on large scales at most redshifts. However, that analysis employs a separate emulator at each redshift, with an internal parameterization that differs from ours, and the emulator accuracy degrades to worse than 1\% at $z<2.6$ and $z=3.6$. As a result, the effective redshift-independent accuracy may be closer to that of \lacelyssa than these results might initially suggest.

Although the \texttt{lyssa} simulations have a larger volume and better resolution than the \mpgadget simulations, we opted to use the \lacempg emulator in the baseline analysis for the following reasons: 

\begin{itemize}
    \item The impact of cosmic variance on \lacelyssa is substantially larger than the ``interpolation'' errors estimated through leave-one-out tests, whereas this was not the case for \lacempg. Consequently, a precise characterization of emulator errors for \lacelyssa would require incorporating cosmic variance into the emulator covariance matrix. In turn, estimating the cosmic variance uncertainties would necessitate running additional \texttt{lyssa} simulations with different realizations of the initial Fourier phases --- an effort beyond the scope of this work.

    \item The \texttt{lyssa} simulations were initialized using a single transfer function for baryons and dark matter, rather than distinct ones. This difference induces a redshift- and wavenumber-dependent suppression in \pone compared to the two-transfer case that reaches up to 5\% at high redshift and small scales \cite{Walther2025_lyssa}. We correct for this difference during inference; however, this correction was derived from a single simulation \cite{Walther2025_lyssa}, and it remains unclear whether it should depend on cosmology.

    \item The main advantage of the \lyssa simulations is that they have better resolution than the \mpgadget simulations. Nonetheless, in \cref{sec:model_validation} we analyze a \pone mock based on the \texttt{lyssa-central} simulation using the \lacempg emulator, and we recover the true compressed parameters within $1\,\sigma$. This demonstrates that the lower accuracy of the \mpgadget simulations relative to \lyssa simulations does not bias the cosmological constraints.
\end{itemize}

Moreover, we highlight that cosmological constraints obtained from the analysis of DESI measurements using the \lacempg and \lacelyssa emulators are consistent at the $\simeq0.5\,\sigma$ level (see \cref{sec:results_robust_emulator}). 

%% file: app_nuisance_journal.tex
\section{Contaminants and systematics}
\label{app:nuisance}

Throughout the main body of this work, we validate the robustness of our constraints on the compressed parameters using \pone mocks and multiple data analysis variations. However, we do not perform an analogous validation for the IGM, metal, HCD, and resolution parameters. As a result, the inferred values of some of these parameters may be affected by the limited resolution of the simulations or by other systematic effects. With this caveat in mind, we present our constraints on these parameters below.

\begin{figure}
    \centering
    \includegraphics[width=0.6\linewidth]{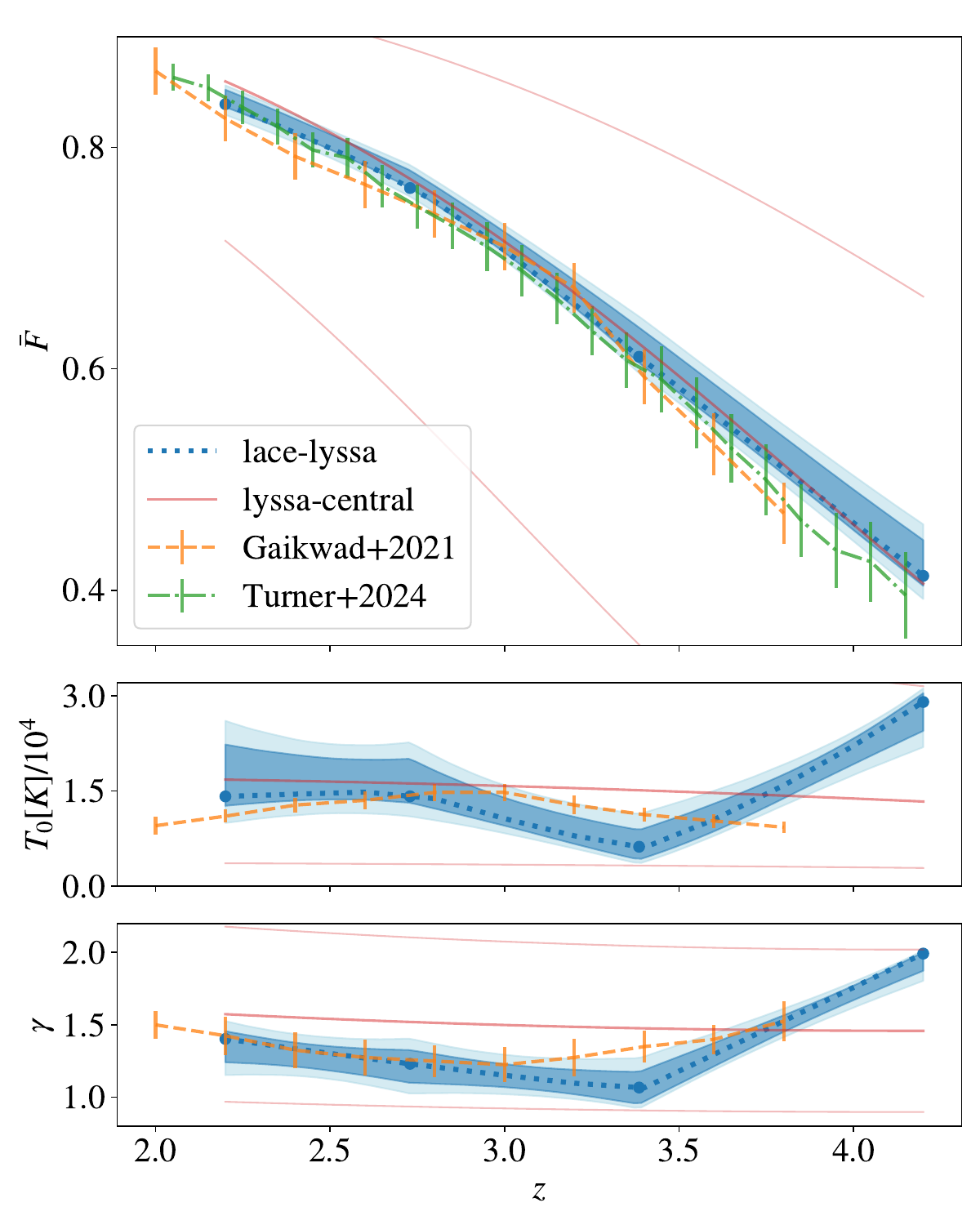}
    \caption{
    Same as \cref{fig:IGM_best}, but when using the \lacelyssa emulator in the analysis. In this plot, the thick red line display predictions from \texttt{lyssa-central} simulation.
    }
    \label{fig:IGM_best_lyssa}
\end{figure}

In \cref{fig:IGM_best}, we present baseline constraints on the redshift evolution of the mean flux and on the amplitude and slope of the temperature–density relation, with the corresponding numerical values reported in \cref{tab:IGM_best}. In \cref{fig:IGM_best_lyssa}, we show the analogous constraints obtained by replacing the \lacempg emulator used in the baseline analysis with the \lacelyssa emulator. We include this comparison because the \lyssa simulations provide a denser and broader coverage of the IGM parameter space than the \mpgadget simulations. The mean-flux constraints derived from this variation are consistent with KODIAQ and LyCAN measurements at low redshift, but are slightly higher than these at high redshift, in contrast to the baseline results. Conversely, the inferred amplitude and slope of the temperature–density relation are consistent with KODIAQ measurements at most redshifts, unlike in the baseline analysis. This difference is likely driven by a combination of the higher resolution of the \lyssa simulations, their improved coverage of the thermal-parameter space, and the larger uncertainties on the best-fitting parameters in this analysis relative to the baseline case.

\begin{table}[]
    \centering
    \begin{tabular}{cccc}
         $z$ & $\bar{F}$ & $T_0[K]/10^4$ & $\gamma$ \\
\hline
2.20 & $0.8147^{+0.0048}_{-0.0040}$ & $1.537^{+0.056}_{-0.098}$ & $1.880^{+0.057}_{-0.102}$ \\
2.40 & $0.7905^{+0.0043}_{-0.0040}$ & $1.329^{+0.061}_{-0.071}$ & $1.682^{+0.044}_{-0.055}$ \\
2.60 & $0.7639^{+0.0046}_{-0.0048}$ & $1.116^{+0.095}_{-0.103}$ & $1.492^{+0.070}_{-0.063}$ \\
2.80 & $0.7306^{+0.0052}_{-0.0057}$ & $0.99^{+0.11}_{-0.11}$ & $1.424^{+0.087}_{-0.077}$ \\
3.00 & $0.6851^{+0.0052}_{-0.0052}$ & $1.016^{+0.105}_{-0.095}$ & $1.563^{+0.057}_{-0.054}$ \\
3.20 & $0.6347^{+0.0058}_{-0.0055}$ & $1.03^{+0.14}_{-0.11}$ & $1.704^{+0.035}_{-0.045}$ \\
3.40 & $0.5804^{+0.0073}_{-0.0066}$ & $1.03^{+0.18}_{-0.15}$ & $1.823^{+0.031}_{-0.058}$ \\
3.60 & $0.5317^{+0.0061}_{-0.0059}$ & $1.05^{+0.15}_{-0.12}$ & $1.663^{+0.031}_{-0.040}$ \\
3.80 & $0.4812^{+0.0049}_{-0.0056}$ & $1.05^{+0.13}_{-0.11}$ & $1.503^{+0.054}_{-0.037}$ \\
4.00 & $0.4297^{+0.0044}_{-0.0066}$ & $1.03^{+0.14}_{-0.13}$ & $1.348^{+0.081}_{-0.052}$ \\
4.20 & $0.3775^{+0.0051}_{-0.0082}$ & $1.00^{+0.17}_{-0.16}$ & $1.198^{+0.109}_{-0.069}$ \\
    \end{tabular}
    \caption{Best-fitting constraints on the redshift evolution of the mean flux and the amplitude and slope of the temperature-density relation from the baseline analysis.}
    \label{tab:IGM_best}
\end{table}

\begin{figure}
    \centering
    \includegraphics[width=\linewidth]{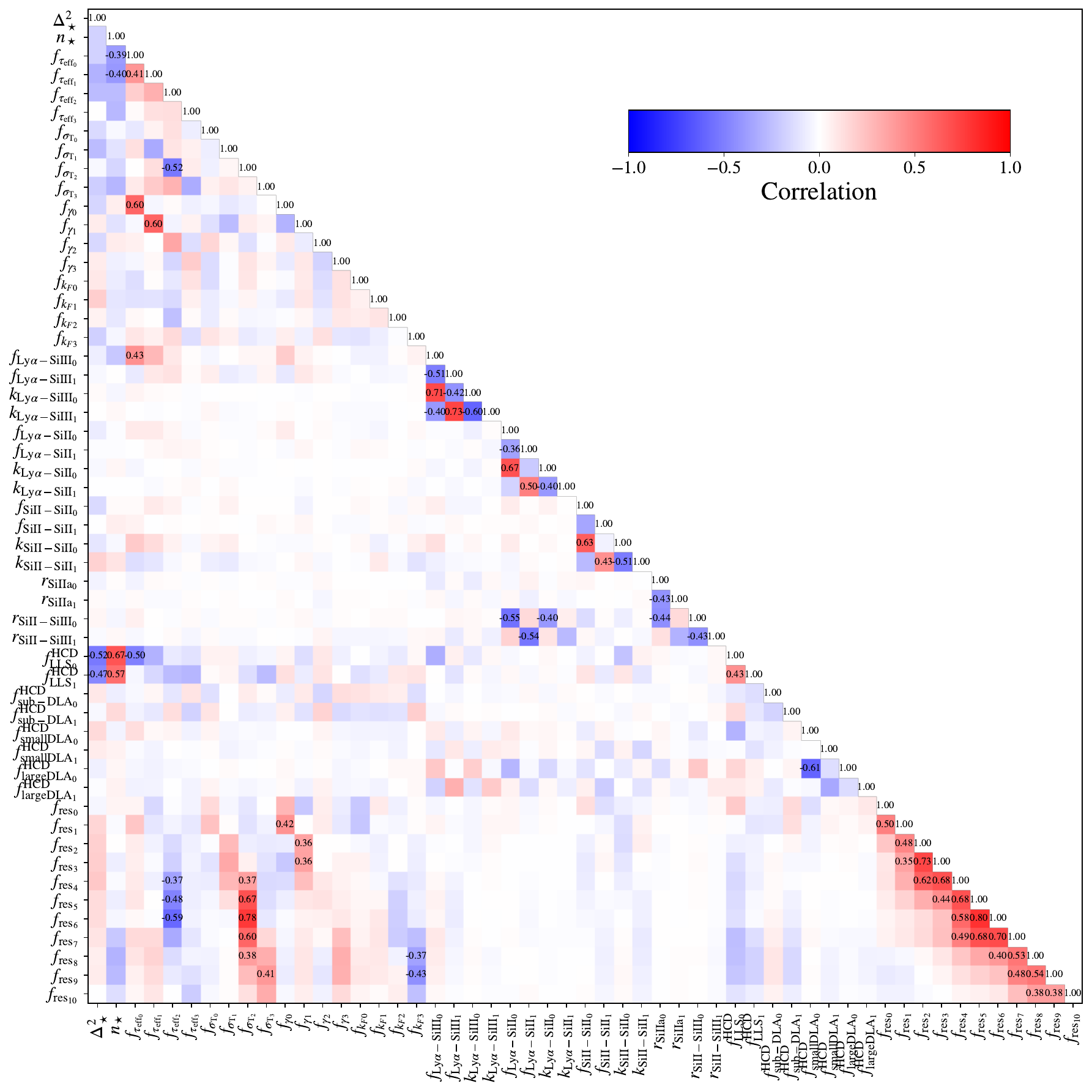}
    \caption{
    Correlation matrix for all model parameters. Red and blue colors indicate positive and negative correlations, respectively. Numerical values are displayed when the absolute value of the correlation exceeds 0.35.
}
    \label{fig:corr_matrix}
\end{figure}

By contrast, the temperature parameters reach their prior bounds at specific redshifts, a behavior also present in the baseline analysis. To investigate the origin of this effect, in \cref{fig:corr_matrix}, we show the correlation matrix of all model parameters from the baseline analysis. The mean flux parameters are anticorrelated (correlated) with the parameters describing the amplitude (slope) of the temperature–density relation, as well as with the resolution parameters. In addition, the resolution parameters are strongly correlated with the amplitude and slope of the temperature–density relation, while being anticorrelated with the pressure parameters. These correlations arise because the IGM parameters affect \pone on intermediate to large scales in a similar manner --- an approximately flat decrease (increase) for $\bar{F}$ and $\gamma$ ($\sigma_\mathrm{T}$ and $k_\mathrm{F}$) --- and because the effects of $\sigma_\mathrm{T}$, $\gamma$, and $f_\mathrm{res}$ are aligned on small scales \cite{cabayol-garcia2023NeuralNetworkEmulator}. Since the sensitivity of \pone to the IGM state and the spectrograph resolution differs on scales smaller than those probed by DESI, a joint analysis combining DESI data with higher-resolution measurements could break these degeneracies and yield more robust IGM constraints.

We also find strong correlations between the parameters controlling the amplitude and damping of metal contamination, as well as between the two parameters describing deviations from the optically thin limit. Additionally, the parameters governing departures from the optically thin limit for \siisiii contamination are correlated with the strength of the \lyasiii component. In \cref{tab:best_params}, we gather the best-fitting value of the parameters describing metal contamination, HCD contamination, and biases in the characterization of the DESI spectrograph resolution from the baseline analysis. Note that the majority of these parameters are constrained at least at the $2\sigma$ level, and their values are not dominated by the flat priors presented in \cref{tab:parameters}.

\begin{table}[]
    \centering
    \begin{tabular}{cc|cr}
    Parameter & Median [68\% CI] & Parameter & Median [68\% CI]\\
    \hline
$f_{{\mathrm{Ly}\alpha-\mathrm{SiIII}}_0}$ & $1.52^{+0.09}_{-0.08}\times10^{-2}$ & $f^\mathrm{HCD}_{{\rm sub-DLA}_0}$ & $1.69^{+5.48}_{-1.00}\times10^{-4}$ \\
$f_{{\mathrm{Ly}\alpha-\mathrm{SiIII}}_1}$ & $0.0183^{+0.0036}_{-0.0030}$ & $f^\mathrm{HCD}_{{\rm sub-DLA}_1}$ & $0.026^{+0.032}_{-0.018}$ \\
$k_{{\mathrm{Ly}\alpha-\mathrm{SiIII}}_0}$ & $7.44^{+0.55}_{-0.50}\times10^{-3}$ & $f^\mathrm{HCD}_{{\rm small DLA}_0}$ & $0.0057^{+0.0039}_{-0.0032}$ \\
$k_{{\mathrm{Ly}\alpha-\mathrm{SiIII}}_1}$ & $0.0076^{+0.0020}_{-0.0016}$ & $f^\mathrm{HCD}_{{\rm small DLA}_1}$ & $0.0043^{+0.0065}_{-0.0030}$ \\
$f_{{\mathrm{Ly}\alpha-\mathrm{SiII}}_0}$ & $0.0252^{+0.0043}_{-0.0039}$ & $f^\mathrm{HCD}_{{\rm large DLA}_0}$ & $0.0091^{+0.0051}_{-0.0045}$ \\
$f_{{\mathrm{Ly}\alpha-\mathrm{SiII}}_1}$ & $0.0076^{+0.0046}_{-0.0029}$ & $f^\mathrm{HCD}_{{\rm large DLA}_1}$ & $0.0096^{+0.0095}_{-0.0058}$ \\
$k_{{\mathrm{Ly}\alpha-\mathrm{SiII}}_0}$ & $3.50^{+0.83}_{-0.66}\times10^{-3}$ & $f_\mathrm{res_{0}}$ & $0.0080^{+0.0057}_{-0.0059}$ \\
$k_{{\mathrm{Ly}\alpha-\mathrm{SiII}}_1}$ & $0.0156^{+0.0170}_{-0.0082}$ & $f_\mathrm{res_{1}}$ & $0.0109^{+0.0062}_{-0.0067}$ \\
$f_{{\mathrm{SiII}-\mathrm{SiII}}_0}$ & $1.49^{+0.16}_{-0.16}$ & $f_\mathrm{res_{2}}$ & $0.0012^{+0.0068}_{-0.0068}$ \\
$f_{{\mathrm{SiII}-\mathrm{SiII}}_1}$ & $2.65^{+1.17}_{-0.91}$ & $f_\mathrm{res_{3}}$ & $-0.0077^{+0.0066}_{-0.0063}$ \\
$k_{{\mathrm{SiII}-\mathrm{SiII}}_0}$ & $0.0121^{+0.0017}_{-0.0017}$ & $f_\mathrm{res_{4}}$ & $0.0018^{+0.0069}_{-0.0067}$ \\
$k_{{\mathrm{SiII}-\mathrm{SiII}}_1}$ & $0.0046^{+0.0021}_{-0.0016}$ & $f_\mathrm{res_{5}}$ & $0.0021^{+0.0080}_{-0.0072}$ \\
$r_{{\mathrm{SiIIa}}_0}$ & $2.29^{+0.60}_{-0.49}$ & $f_\mathrm{res_{6}}$ & $0.0158^{+0.0114}_{-0.0099}$ \\
$r_{{\mathrm{SiIIa}}_1}$ & $1.82^{+1.42}_{-0.80}$ & $f_\mathrm{res_{7}}$ & $0.0217^{+0.0097}_{-0.0087}$ \\
$r_{{\mathrm{SiII}-\mathrm{SiIII}}_0}$ & $1.81^{+0.45}_{-0.38}$ & $f_\mathrm{res_{8}}$ & $-0.0034^{+0.0109}_{-0.0097}$ \\
$r_{{\mathrm{SiII}-\mathrm{SiIII}}_1}$ & $28^{+23}_{-13}$     & $f_\mathrm{res_{9}}$ & $-0.010^{+0.012}_{-0.010}$ \\
$f^\mathrm{HCD}_{{\rm LLS}_0}$ & $0.246^{+0.086}_{-0.074}$ & $f_\mathrm{res_{10}}$ & $0.006^{+0.016}_{-0.015}$ \\
$f^\mathrm{HCD}_{{\rm LLS}_1}$ & $0.218^{+0.095}_{-0.086}$ & & \\
    \end{tabular}
    \caption{Best-fitting value of the parameters describing the metal contamination, HCD contamination, and biases in the characterization of the spectrograph resolution for the baseline analysis. The units of the $k$ terms are s/km, those of the $f_{{\mathrm{SiII}-\mathrm{SiII}}}$ terms are km/s, and all other parameters are dimensionless.
   }
    \label{tab:best_params}
\end{table}

%% file: app_asns_journal.tex
\section{Constraints on the shape of the primordial power spectrum}
\label{app:primodial_pk}

\begin{figure}
    \centering
    \includegraphics[width=0.495\linewidth]{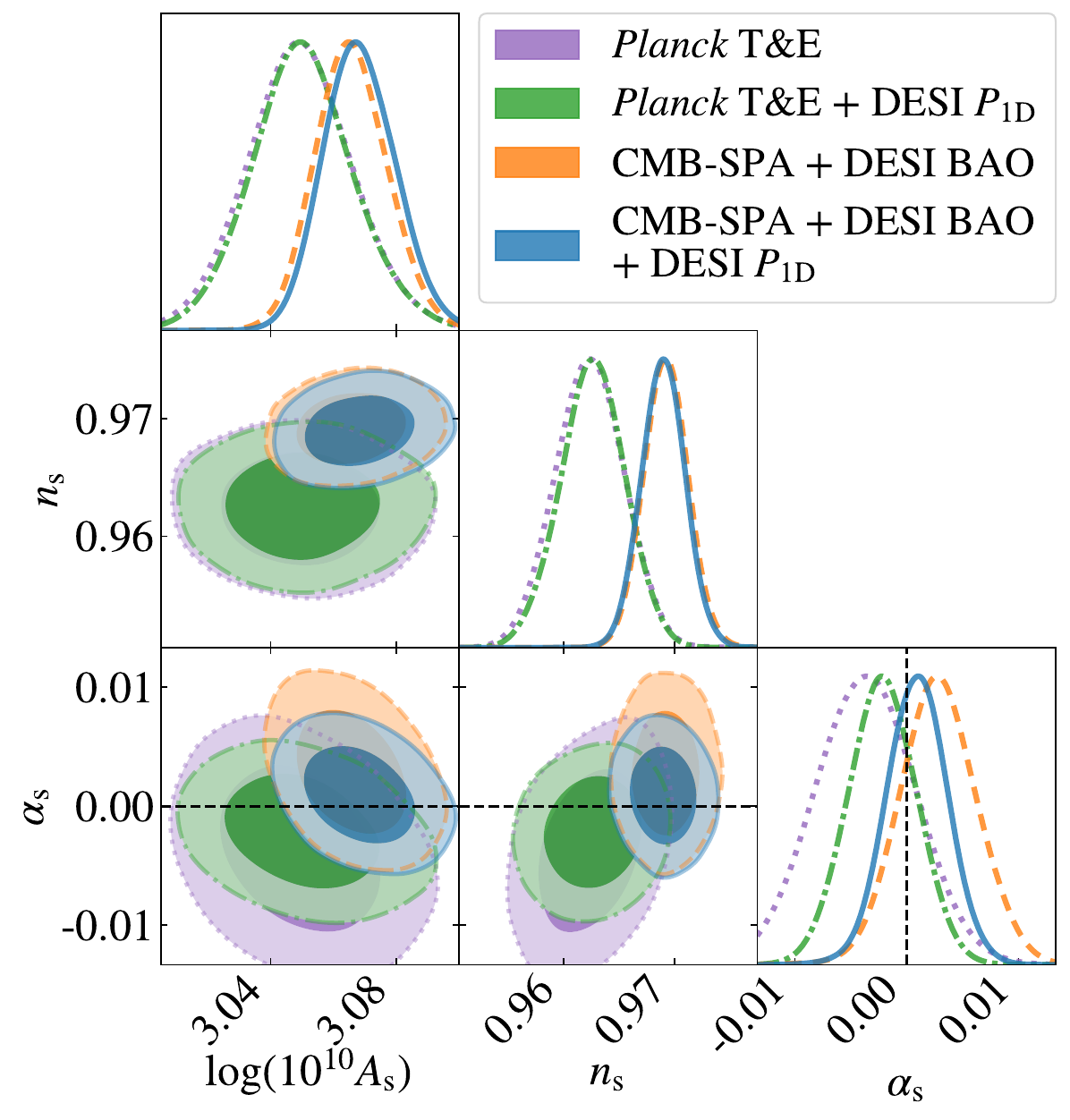}
    \includegraphics[width=0.495\linewidth]{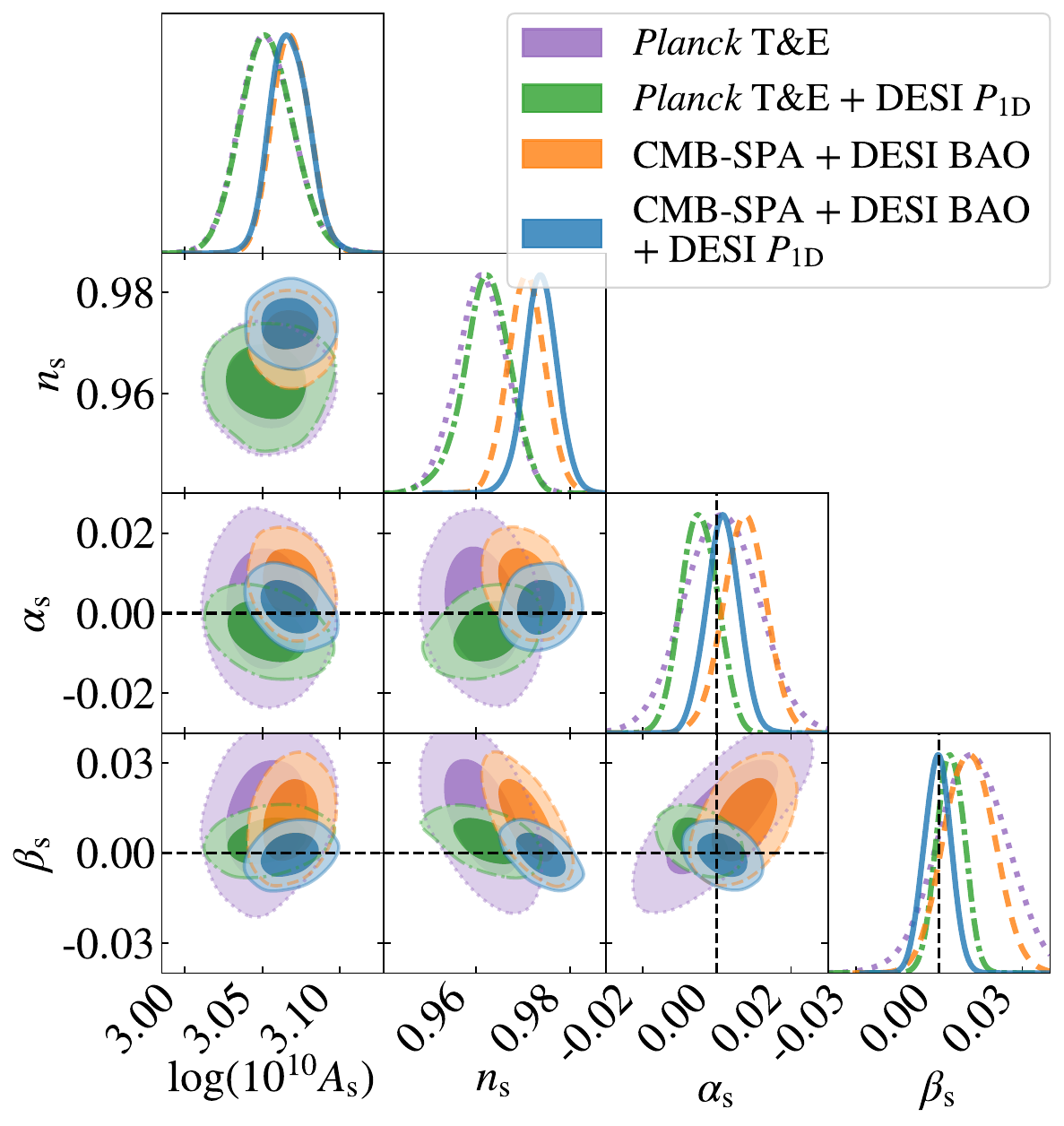}
    \caption{Same as \cref{fig:import_mnu_nnu}, but showing the constraints on parameters defining the primordial power spectrum rather than on the compressed parameters and $\alpha_\mathrm{s}$ and $\beta_\mathrm{s}$.}
    \label{fig:import_nrun_asns}
\end{figure}

In \cref{sec:ext_shape}, we present constraints on the compressed parameters together with two parameters describing deviations of the primordial power spectrum from a pure power law: the running of the spectral index and the running of the running. In \cref{fig:import_nrun_asns}, we show the corresponding constraints on these two last parameters, as well as on the amplitude ($A_\mathrm{s}$) and spectral index ($n_\mathrm{s}$) of the primordial power spectrum at $k_\mathrm{s}=0.05\,\iMpc$. As shown, DESI DR1 \pone measurements do not tighten the constraints on $A_\mathrm{s}$, in contrast to their strong impact on \deltastar. This behavior reflects the fact that \pone measurements probe much smaller scales than $k_\mathrm{s}=0.05\,\iMpc$, and that $A_\mathrm{s}$ is already measured with high precision by CMB experiments. 

On the other hand, \pone measurements improve the precision of $n_\mathrm{s}$ constraints; this is because $n_\mathrm{s}$ influences a much larger range of scales than $A_\mathrm{s}$. When varying both the running and the running of the running, the combination of CMB-SPA, DESI BAO, and DESI \pone yields $n_\mathrm{s} = 0.9738 \pm 0.0036$, improving the precision of the combination of CMB-SPA and DESI BAO constraints by a factor of 1.11. We gather the results from all combinations in \cref{tab:lcdm_shape}.

\begin{table}[]
    \centering
    \begin{tabular}{ccccc}
         & \Planck T\&E & \makecell{\Planck T\&E \\+ DESI \pone} & \makecell{CMB-SPA \\+ DESI BAO} & \makecell{CMB-SPA \\+ DESI BAO \\+ DESI \pone}\\
        \hline
        $\log(10^{10} A_\mathrm{s})$
        & $3.053^{+0.016}_{-0.018}$
        & $3.053\pm 0.017$ 
        & $3.068\pm 0.012$ 
        & $3.068\pm 0.012$\\
        $n_\mathrm{s}$ 
        & $0.9612\pm 0.0053$ 
        & $0.9622^{+0.0052}_{-0.0045}$ 
        & $0.9707\pm 0.0040$ 
        & $0.9738\pm 0.0036$\\
        $\alpha_\mathrm{s}$ 
        & $0.001\pm 0.010$ 
        & $-0.0047\pm 0.0049\;\;\,$ 
        & $0.0075\pm 0.0058$ 
        & $0.0016\pm 0.0045$\\
        $\beta_\mathrm{s}$  
        & $0.012\pm 0.013$ 
        & $0.0041\pm 0.0049$       
        & $0.0110\pm 0.0091$ 
        & $-0.0006\pm 0.0048\;\;\,$\\
    \end{tabular}
    \caption[]{
    Constraints on the shape of the primordial power spectrum when varying both the running and the running of the running of the primordial power spectrum. We use the same datasets as in \cref{tab:lcdm_extensions}.
    }
    \label{tab:lcdm_shape}
\end{table}

%% file: app_pivot_journal.tex
\section{Pivot scale}
\label{app:pivot}

In \cref{sec:results_cosmo}, we report the value of \deltastar and \nstar defined at a pivot scale of \kstarval and redshift $z_\star = 3$, following the convention established by the analysis of SDSS DR2 data \cite{McDonald2006}. However, this pivot scale is not optimal for DESI DR1, as the correlation between \deltastar and \nstar, while small, is nonzero. In order to find the optimal scale for DESI DR1 data, we recomputed the compressed parameter using different values of the pivot scale until minimizing the correlation. We find that the correlation is $r=-0.0011^{+0.0007}_{-0.0006}$ for $\tilde{k}_\star = 0.019\,\ikms$, which is roughly at the center of the wavenumber range probed by DESI DR1 data. The values of the compressed parameters for this pivot scale are
\begin{equation}
    \tilde{\Delta}^2_\star = 0.600^{+0.049}_{-0.052}, \qquad
    \tilde{n}_\star = -2.446^{+0.019}_{-0.019},
\end{equation}
The signal-to-noise ratio of the amplitude is 11.8 at this scale, compared to 11.7 at \kstarval. The resulting improvement is therefore negligible, and for consistency with previous studies, we perform the analysis using the compressed parameters defined at \kstarval.

%% file: acknow.tex
\section*{Data availability}

The \lacempg and \lacelyssa emulators are publicly available in the \texttt{LaCE} repository\footnote{\url{https://github.com/igmhub/lace}}, and the likelihood code in the \texttt{cup1d} repository\footnote{\url{https://github.com/igmhub/cup1d}}, together with all scripts used to perform the analysis and generate the figures and tables. A Cobaya \cite{cobaya_code, cobaya_paper} likelihood enabling to combine our measurements with others can be found \href{https://github.com/igmhub/cobaya_lya_p1d}{here}, while all data points shown in the figures are \href{https://doi.org/10.5281/zenodo.18396025}{here}. We acknowledge that this work made direct use of the following \texttt{python} packages: \texttt{camb}\cite{lewis2000EfficientComputationCosmic, Howlett2012_camb}, 
\texttt{emcee}\cite{foremanmackey13}, 
\texttt{forestflow}\cite{chavesmontero2025forestflow},
\texttt{getdist} \cite{Lewis:2019},
\texttt{matplotlib}\cite{matplotlib}, 
\texttt{mpi4py} \cite{dalcin2005_mpi, dalcin2008_mpi, dalcin2011_mpi, Dalcin2021_mpi, Rogowski2023_mpi}, 
\texttt{numpy} \cite{Harris:2020}, 
\texttt{scipy} \cite{virtanen2020_SciPyFundamentalalgorithms}, and
\texttt{scikit-learn} \cite{Pedregosa2011_scikitlearn}.


\section*{Acknowledgments}

We thank the anonymous referee for their insightful comments and suggestions, Laura Cabayol-Garcia and Chris Pedersen for their early contributions to the development of the \texttt{lace} emulators and \texttt{cup1d}, and Ming-Feng Ho, Vid Ir\v{s}i\v{c}, and other members of the DESI Lyman-$\alpha$ working group for valuable comments and insightful discussions. JCM and AFR acknowledge financial support from the Spanish Ministry of Science and Innovation (MICINN) through the Spanish State Research Agency, under Severo Ochoa Centres of Excellence Programme 2025-2029 (CEX2024-001441-S) and the European Union (ERC Consolidator Grant, COSMO-LYA, grant agreement 101044612). Views and opinions expressed are however those of the authors only and do not necessarily reflect those of the European Union or the European Research Council Executive Agency. Neither the European Union nor the granting authority can be held responsible for them. AFR acknowledges financial support from the Spanish Ministry of Science and Innovation under the Ramon y Cajal program (RYC-2018-025210) and the PID2024-159420NB-C41 project. CGQ acknowledges support provided by NASA through the NASA Hubble Fellowship grant HST-HF2-51554.001-A awarded by the Space Telescope Science Institute, which is operated by the Association of Universities for Research in Astronomy, Inc., for NASA, under contract NAS5-26555. IFAE is partially funded by the CERCA program of the Generalitat de Catalunya. This research used resources of the National Energy Research Scientific Computing Center (NERSC), a Department of Energy User Facility. 

This material is based upon work supported by the U.S. Department of Energy (DOE), Office of Science, Office of High-Energy Physics, under Contract No. DE–AC02–05CH11231, and by the National Energy Research Scientific Computing Center, a DOE Office of Science User Facility under the same contract. Additional support for DESI was provided by the U.S. National Science Foundation (NSF), Division of Astronomical Sciences under Contract No. AST-0950945 to the NSF’s National Optical-Infrared Astronomy Research Laboratory; the Science and Technology Facilities Council of the United Kingdom; the Gordon and Betty Moore Foundation; the Heising-Simons Foundation; the French Alternative Energies and Atomic Energy Commission (CEA); the National Council of Humanities, Science and Technology of Mexico (CONAHCYT); the Ministry of Science, Innovation and Universities of Spain (MICIU/AEI/10.13039/501100011033), and by the DESI Member Institutions: \url{https://www.desi.lbl.gov/collaborating-institutions}. Any opinions, findings, and conclusions or recommendations expressed in this material are those of the author(s) and do not necessarily reflect the views of the U. S. National Science Foundation, the U. S. Department of Energy, or any of the listed funding agencies.

The authors are honored to be permitted to conduct scientific research on I'oligam Du'ag (Kitt Peak), a mountain with particular significance to the Tohono O’odham Nation.

%% file: affiliations.tex
\section*{Affiliations}
\label{sec:affiliations}

$^{1}$ Institut de F\'{i}sica d’Altes Energies (IFAE), The Barcelona Institute of Science and Technology, Edifici Cn, Campus UAB, 08193, Bellaterra (Barcelona), Spain\\
$^{2}$ Institució Catalana de Recerca i Estudis Avançats, Passeig de Lluís Companys, 23, 08010 Barcelona, Spain\\
$^{3}$ Lawrence Berkeley National Laboratory, 1 Cyclotron Road, Berkeley, CA 94720, USA\\
$^{4}$ IRFU, CEA, Universit\'{e} Paris-Saclay, F-91191 Gif-sur-Yvette, France\\
$^{5}$ Center for Astrophysics $|$ Harvard \& Smithsonian, 60 Garden Street, Cambridge, MA 02138, USA\\
$^{6}$ NASA Einstein Fellow\\
$^{7}$ The Ohio State University, Columbus, 43210 OH, USA\\
$^{8}$ Center for Cosmology and AstroParticle Physics, The Ohio State University, 191 West Woodruff Avenue, Columbus, OH 43210, USA\\
$^{9}$ Department of Physics, The Ohio State University, 191 West Woodruff Avenue, Columbus, OH 43210, USA\\
$^{10}$ Department of Astronomy, The Ohio State University, 4055 McPherson Laboratory, 140 W 18th Avenue, Columbus, OH 43210, USA\\
$^{11}$ Universit\'{e} Clermont-Auvergne, CNRS, LPCA, 63000 Clermont-Ferrand, France\\
$^{12}$ University Observatory, Faculty of Physics, Ludwig-Maximilians-Universit\"{a}t, Scheinerstr. 1, 81677 M\"{u}nchen, Germany\\
$^{13}$ Excellence Cluster ORIGINS, Boltzmannstrasse 2, D-85748 Garching, Germany\\
$^{14}$ Department of Physics, Boston University, 590 Commonwealth Avenue, Boston, MA 02215 USA\\
$^{15}$ Dipartimento di Fisica ``Aldo Pontremoli'', Universit\`a degli Studi di Milano, Via Celoria 16, I-20133 Milano, Italy\\
$^{16}$ INAF-Osservatorio Astronomico di Brera, Via Brera 28, 20122 Milano, Italy\\
$^{17}$ Department of Physics \& Astronomy, University College London, Gower Street, London, WC1E 6BT, UK\\
$^{18}$ Instituto de F\'{\i}sica, Universidad Nacional Aut\'{o}noma de M\'{e}xico,  Circuito de la Investigaci\'{o}n Cient\'{\i}fica, Ciudad Universitaria, Cd. de M\'{e}xico  C.~P.~04510,  M\'{e}xico\\
$^{19}$ University of California, Berkeley, 110 Sproul Hall \#5800 Berkeley, CA 94720, USA\\
$^{20}$ Departamento de F\'isica, Universidad de los Andes, Cra. 1 No. 18A-10, Edificio Ip, CP 111711, Bogot\'a, Colombia\\
$^{21}$ Observatorio Astron\'omico, Universidad de los Andes, Cra. 1 No. 18A-10, Edificio H, CP 111711 Bogot\'a, Colombia\\
$^{22}$ Institut d'Estudis Espacials de Catalunya (IEEC), c/ Esteve Terradas 1, Edifici RDIT, Campus PMT-UPC, 08860 Castelldefels, Spain\\
$^{23}$ Institute of Cosmology and Gravitation, University of Portsmouth, Dennis Sciama Building, Portsmouth, PO1 3FX, UK\\
$^{24}$ Institute of Space Sciences, ICE-CSIC, Campus UAB, Carrer de Can Magrans s/n, 08913 Bellaterra, Barcelona, Spain\\
$^{25}$ University of Virginia, Department of Astronomy, Charlottesville, VA 22904, USA\\
$^{26}$ Departamento de F\'{\i}sica, DCI-Campus Le\'{o}n, Universidad de Guanajuato, Loma del Bosque 103, Le\'{o}n, Guanajuato C.~P.~37150, M\'{e}xico\\
$^{27}$ Fermi National Accelerator Laboratory, PO Box 500, Batavia, IL 60510, USA\\
$^{28}$ Department of Astronomy, The University of Texas at Austin, 2515 Speedway Boulevard, Austin, TX 78712, USA\\
$^{29}$ Institut d'Astrophysique de Paris. 98 bis boulevard Arago. 75014 Paris, France\\
$^{30}$ Department of Physics, The University of Texas at Dallas, 800 W. Campbell Rd., Richardson, TX 75080, USA\\
$^{31}$ NSF NOIRLab, 950 N. Cherry Ave., Tucson, AZ 85719, USA\\
$^{32}$ Department of Physics and Astronomy, University of California, Irvine, 92697, USA\\
$^{33}$ Sorbonne Universit\'{e}, CNRS/IN2P3, Laboratoire de Physique Nucl\'{e}aire et de Hautes Energies (LPNHE), FR-75005 Paris, France\\
$^{34}$ Department of Astronomy and Astrophysics, UCO/Lick Observatory, University of California, 1156 High Street, Santa Cruz, CA 95064, USA\\
$^{35}$ Department of Astronomy and Astrophysics, University of California, Santa Cruz, 1156 High Street, Santa Cruz, CA 95065, USA\\
$^{36}$ Departament de F\'{i}sica, Serra H\'{u}nter, Universitat Aut\`{o}noma de Barcelona, 08193 Bellaterra (Barcelona), Spain\\
$^{37}$ Instituci\'{o} Catalana de Recerca i Estudis Avan\c{c}ats, Passeig de Llu\'{\i}s Companys, 23, 08010 Barcelona, Spain\\
$^{38}$ Department of Physics and Astronomy, Siena University, 515 Loudon Road, Loudonville, NY 12211, USA\\
$^{39}$ Instituto Avanzado de Cosmolog\'{\i}a A.~C., San Marcos 11 - Atenas 202. Magdalena Contreras. Ciudad de M\'{e}xico C.~P.~10720, M\'{e}xico\\
$^{40}$ Department of Physics and Astronomy, University of Waterloo, 200 University Ave W, Waterloo, ON N2L 3G1, Canada\\
$^{41}$ Perimeter Institute for Theoretical Physics, 31 Caroline St. North, Waterloo, ON N2L 2Y5, Canada\\
$^{42}$ Waterloo Centre for Astrophysics, University of Waterloo, 200 University Ave W, Waterloo, ON N2L 3G1, Canada\\
$^{43}$ Instituto de Astrof\'{i}sica de Andaluc\'{i}a (CSIC), Glorieta de la Astronom\'{i}a, s/n, E-18008 Granada, Spain\\
$^{44}$ Departament de F\'isica, EEBE, Universitat Polit\`ecnica de Catalunya, c/Eduard Maristany 10, 08930 Barcelona, Spain\\
$^{45}$ Department of Physics and Astronomy, Sejong University, 209 Neungdong-ro, Gwangjin-gu, Seoul 05006, Republic of Korea\\
$^{46}$ CIEMAT, Avenida Complutense 40, E-28040 Madrid, Spain\\
$^{47}$ Department of Physics, University of Michigan, 450 Church Street, Ann Arbor, MI 48109, USA\\
$^{48}$ University of Michigan, 500 S. State Street, Ann Arbor, MI 48109, USA\\
$^{49}$ Department of Physics \& Astronomy, Ohio University, 139 University Terrace, Athens, OH 45701, USA\\
$^{50}$ National Astronomical Observatories, Chinese Academy of Sciences, A20 Datun Road, Chaoyang District, Beijing, 100101, P.~R.~China\\